\documentclass[a4paper]{article}

\newcommand{\cint}{s}

\providecommand{\esymbol}{e}
\providecommand{\esymbolv}{\ev}

\providecommand{\cint}{c}

\usepackage{setspace} 
\usepackage{xr}
\usepackage{wrapfig}
\usepackage{enumitem}
\usepackage{color}
\usepackage{amsmath,amsfonts,amssymb}
\usepackage{epsfig}
\usepackage{graphicx}
\usepackage{colortbl}
\usepackage{url}
\usepackage{lineno}

\definecolor{darkblue}{rgb}{0.05,0.,0.65}
\definecolor{grey}{rgb}{0.9, 0.9, 0.9}

\newcommand{\myparagraph}[1]{\vspace{-3mm}\paragraph{#1}}

\newcommand{\mybox}[1]{\fbox{\parbox{16.5cm}{\begin{center}\parbox{15.5cm}{#1}\end{center}}}}

\newcommand{\coout}[1]{[{\color{magenta} #1}]}
\newcommand{\co}   [1]{[{\it \color{red} #1}]}
\newcommand{\todo} [1]{[{\color{blue} #1}]}

\usepackage{amsmath,amsfonts,amssymb}
\usepackage{relsize}
\usepackage{ marvosym }

\newcommand{\inv}{^{-1}}

\newcommand{\md}{{\rm d}}

\newcommand{\diag}{\mbox{\rm Dg}}

\newcommand{\e}   {\mbox{\rm e}}

\newcommand{\cv}{{\bf c}}
\newcommand{\vv}{{\bf v}}
\newcommand{\yv}{{\bf y}}

\newcommand{\fv}{{\bf f}}

\newcommand{\bv}{{\bf b}}

\newcommand{\ev}{{\bf e}}
\newcommand{\xv}{{\bf x}}

\newcommand{\pv}{{\bf p}}

\newcommand{\qv}{{\bf q}}

\newcommand{\av}{{\bf a}}

\newcommand{\betav}{{\boldsymbol \beta}}
\newcommand{\gammav}{{\boldsymbol \gamma}}

\newcommand{\nuv}{{\boldsymbol \nu}}

\newcommand{\Nmat}{\mathbf{N}}
\newcommand{\Imat}{\mathbf{I}}

\newcommand{\Mmat}{\mathbf{M}}

\newcommand{\Hmat}{\mathbf{H}}

\newcommand{\Emat}{\mathbf{E}}

\newcommand{\Amat} {\mathbf{A}}

\newcommand{\Bmat} {\mathbf{B}}
\newcommand{\Wmat}{\mathbf{W}}
\newcommand{\Lmat}{\mathbf{L}}

\newcommand{\Rmat}{{\bf R}}

\newcommand{\trans}{^{\top}}
\newcommand{\half}{\frac{1}{2}}
\newcommand{\real}{\mbox{Re}}

\newcommand{\re}{\mbox{Re}}

\DeclareMathAlphabet{\mathpzc}{OT1}{pzc}{m}{it}
\DeclareMathAlphabet{\mathcalligra}{T1}{calligra}{m}{n}

\providecommand{\esymbol}{e}

\newcommand{\steady}{^{\rm st}}

\newcommand{\cvsteady}   {\cv\steady}

\newcommand{\RYmat}{\Rmat^{\rm Y}}

\usepackage{etoolbox}

\newtoggle{bookversion}      \togglefalse{bookversion}
\newtoggle{shortbookversion} \togglefalse{shortbookversion}
\newtoggle{anhang}           \togglefalse{anhang}
\newtoggle{hidecomments}      \togglefalse{hidecomments}

\newcommand{\bookco}[1]{}

\definecolor{brown}{rgb}{0.9,0.69,0.34}
\definecolor{samoabrownlight}{rgb}{0.89,0.69,0.4}
\definecolor{samoabrowndark} {rgb}{0.5,0.3,0.15}
\definecolor{cbasamoabrown1}{rgb}{0.87,0.6,0.23}
\definecolor{cbasamoabrown2}{rgb}{0.87,0.6,0.23}
\definecolor{cbabrown1}{rgb}{0.87,0.6,0.23}
\definecolor{cbabrown2}{rgb}{0.87,0.6,0.23}
\definecolor{cbabrown3}{rgb}{0.87,0.6,0.23}
\definecolor{cbabrown4}{rgb}{0.87,0.6,0.23}
\definecolor{cbaecoblue1}{rgb}{0.8,0.8, 1.0}
\definecolor{cbaecoblue2}{rgb}{0.7,0.7, 1.0}
\definecolor{cbaecoblue3}{rgb}{0.87,0.6,0.23}
\definecolor{cbaecoblue4}{rgb}{0.87,0.6,0.23}
\definecolor{cbablue2}{rgb}{0.87,0.6,0.23}
\definecolor{cbapink}{rgb}{.99,0.92,0.75}
\definecolor{cbabeige1}{rgb}{0.86, 0.797, 0.625} 
\definecolor{cbabeige2}{rgb}{0.93, 0.812, 0.56}  
\definecolor{cbabeige3}{rgb}{1.0, 0.97, 0.88}  
\definecolor{cbahelleslila}{rgb}{1.0, 0.99, 1.0}  
\definecolor{cbalightgrey}{rgb}{0.95,0.95,0.95}
\definecolor{cbatablecolor1}{rgb}{0.86, 0.797, 0.625} 
\definecolor{cbatablecolor2}{rgb}{1,1,1}         

\newcommand{\myvalue}      {value}

\newcommand{\gain}         {gain}

\newcommand{\price}        {price}

\newcommand{{\fluxvalue}}  {flux \myvalue}
\newcommand{{\fluxgain}}   {flux \gain}
\newcommand{\metabolicobjective}{metabolic objective}

\newcommand{{\valueflow}}    {value flow}
\newcommand{{\Valueflow}}    {Value flow}

\newcommand{ {\flow}}        {flux profile}
\newcommand{ {\Flow}}        {Flux profile}

\newcommand{\MCA}{MCT}

\providecommand{\esymbol}   {e}
\providecommand{\esymbolv}  {\mathbf{e}}

\providecommand{\cint}      {c^{\rm int}}

\providecommand{\prodrate}  {r}

\newcommand{\intprod}   {\prodrate^{\rm int}}

\newcommand{\rate}{\nu}
\newcommand{\ratev}{\nuv}
\newcommand{\ratelaw}{k}

\newcommand{\ffit}        {{\mathcal F}} 
\newcommand{\fluxbene}    {b}

\newcommand{\gplus}       {q}
\newcommand{\hminus}      {h}

\newcommand{\wsymbol}     {w}
\newcommand{\loadsymbol}  {y}

\newcommand{\Deltar}{\square}

\newcommand{\vvt}{{\tilde{\bf v}}}

\newcommand{\Ffit}   {{\bf F}}
\newcommand{\Fuu}    {\Ffit_{\rm uu}}
\newcommand{\Fee}    {\Fuu}

\newcommand{\fes}    {\fes}
\newcommand{\fel}    {\fel}
\newcommand{\fevs}   {\fevs}

\newcommand{\gev}   {\loadvsymbol_{\rm \esymbol}}

\newcommand{\hminusv}{{\bf \hminus}}

\renewcommand{\u}     {\esymbol}

\newcommand{\hu}      {\hminusv_{\rm \esymbol}}

\newcommand{\Hminus} {{\Hmat}}
\newcommand{\Huu}     {\Hminus_{\rm \esymbol\esymbol}}
\newcommand{\Hee}     {\Hminus_{\rm \esymbol\esymbol}}

\newcommand{\fluxbenev}     {{\bf \fluxbene}}

\newcommand{\fluxbenetotv} {\fluxbenev}
\newcommand{\bvtot}        {\fluxbenetotv_{\rm v}}

\newcommand{\wvsymbol}{{\bf \wsymbol}}

\newcommand{\loadvsymbol}{{\bf \loadsymbol}}

\newcommand{\wint}  {\wvsymbol_{\rm \prodrate}^{\rm int}}

\newcommand{\fluxenzymecostl}  {\Delta \wsymbol_{\intprod_l:}}

\newcommand{\hev}{\mathbf{\hminus}_{\rm \esymbol}}

\newcommand{\opt}{^{\rm opt}}

\newcommand{\et}   {{\tilde e}} 
\newcommand{\evt}   {{\tilde \esymbolv}} 
\newcommand{\xvt}  {{\tilde {\bf x}}} 
\newcommand{\esymbolvt}  {{\tilde {\bf \esymbol}}} 
 
\newcommand{\xt}   {{\tilde x}}

\newcommand{\ys}{z} 
\newcommand{\ysv}{{\bf \ys}}

\newcommand{\ysvd}{\Delta \ysv}

\newcommand{\yy}{y}

\newcommand{\devt}{\Delta {\bf \tilde e}}
\newcommand{\dxvt}{\Delta {\bf \tilde x}}
\newcommand{\devb}{\Delta {\bf  e}}
\newcommand{\dxvb}{\Delta {\bf  x}}

\newcommand{\xb}{\bar x}
\newcommand{\ub}{\bar \u}
\newcommand{\eb}{\bar e}

\newcommand{\pb}{\bar p}

\newcommand{\ab}{\bar a}
\newcommand{\bb}{\bar b}

\newcommand{\evb}{\bar {\bf e}}

\newcommand{\pvb}{\bar {\bf p}}

\newcommand{\vvb}{\bar {\bf v}}

\newcommand{\evref}{{\bf e}^{\rm ref}}
\newcommand{\xvref}{{\bf x}^{\rm ref}}

\newcommand{\qz}{{\bf \gplus}_{\rm \ys}}
\newcommand{\Qzz}{{\bf Q}_{\rm \ys\ys}}

\newcommand{\at}{{\tilde{a}}}
\newcommand{\bt}{{\tilde{b}}}

\newcommand{\pt}{{\tilde{p}}}

\newcommand{\vt}{{\tilde{v}}}

\newcommand{\pvt}{{\tilde{\bf p}}}

\newcommand{\epsilonSmat} {{\bf E}_{\rm c}}
\newcommand{\epsilonXmat} {{\bf E}_{\rm x}}

\newcommand{\epsilonUmat} {{\bf E}_{\rm u}}

\newcommand{\ffitu}{\fv_{\rm u}}
\newcommand{\Ftemp}{\ffit}
\renewcommand{\Fee}{\Ffit_{\rm ee }}
\newcommand{\Fex}{\Ffit_{\rm ex }}
\newcommand{\Fxe}{\Ffit_{\rm xe }}
\newcommand{\Fxtet}{\Ffit_{\rm \tilde{x} \tilde{e} }}
\newcommand{\Fetxt}{\Ffit_{\rm \tilde{e} \tilde{x} }}
\newcommand{\Fetet}{\Ffit_{\rm \tilde{e} \tilde{e} }}

\renewcommand{\RYmat}              {{\bf R}^{\rm \ys}}

  \newcommand{\RYbmat}             {{\bf R}^{\rm \ys}}
  \newcommand{\RYtmat}             {{\bf R}^{\tilde{\rm \ys}}}

  \newcommand{\RYbsecmat}          {{\bf R}^{\rm \ys}}

\newcommand{ \Aux}{{\bf A}^{\rm u}_{\rm x}}

\renewcommand{\Nmat} {{\bf N}}
\newcommand{\NRmat}{\Nmat_{\rm R}}
\renewcommand{\Lmat} {{\bf L}}

\newcommand{\Eper}{\mathring{\bf E}}

\newcommand{\Gex}{\Wmat_{\rm ex}}
\newcommand{\Gee}{\Wmat_{\rm ee}}

\newcommand{\Getet}    {\Wmat_{\rm \et\et}}
\newcommand{\Getxt}    {\Wmat_{\rm \et\xt}}

\newcommand{\hevt} {\tilde{\bf h}_{\rm e}}

\newcommand{\Hetet}{{\bf H}_{\rm \et\et}}

\newcommand{\gevt}{\tilde{\bf w}_{\rm e}}

\newcommand{\pergev}  {\mathring{\bf w}_{\rm e}}
\newcommand{\pergevt} {\mathring{\bf w}_{\rm \et}}
\newcommand{\perwrv}  {\mathring{\bf w}_{\rm \prodrate}}
\newcommand{\perwrvt} {\mathring{\bf w}_{\rm \tilde{\prodrate}}}
\newcommand{\perhev}  {\mathring{\bf h}_{\rm e}}
\newcommand{\perhevt} {\mathring{\bf h}_{\tilde e}}

\newcommand{\wri}   {w_{\prodrate_i}}
\newcommand{\wrti}  {w_{\tilde{\prodrate}_i}}
\newcommand{\dwrv}  {\Deltar \wint}
\newcommand{\dwrl}  {\Deltar w_{\prodrate_l}}

\newcommand{\wrv}   {{\bf w}_{\rm \prodrate}}
\newcommand{\wrvt}  {\tilde{\bf w}_{\rm \prodrate}}

\newcommand{\cintvt}     {\tilde{\cv}}
\newcommand{\cintvb}     {\bar{\cv}}

\renewcommand{\coout}[1]{}

\definecolor{brown}{rgb}{0.9,0.69,0.34}
\definecolor{lightyellow}{rgb}{1,0.99,0.85}

\newcommand{\psfilesrhythms}{ps-files}

\usepackage{CJKutf8}

\renewcommand{\co}[1]{}
\renewcommand{\todo}[1]{#1}
\renewcommand{\myparagraph}[1]{}

\begin{document}

\title{Optimal enzyme rhythms in cells}

\author{Wolfram Liebermeister\\[3mm] 
Universit\'e Paris-Saclay, INRAE, MaIAGE, 78350 Jouy-en-Josas, France}

\date{}

\maketitle

\begin{abstract} 
  Cells can use periodic enzyme activities to adapt to periodic
  environments or existing internal rhythms and to establish metabolic
  cycles that schedule biochemical processes in time.  A periodically
  changing allocation of the protein budget between reactions or
  pathways may increase the overall metabolic efficiency.  To study
  this hypothesis, I quantify the possible benefits of small-amplitude
  enzyme rhythms in kinetic models.  Starting from an enzyme-optimised
  steady state, I score the effects of possible enzyme rhythms on a
  metabolic objective and optimise their amplitudes and phase shifts.
  Assuming small-amplitude rhythms around an optimal reference state,
  optimal phases and amplitudes can be computed by solving a quadratic
  optimality problem.  In models without amplitude constraints,
  general periodic enzyme profiles can be obtained by Fourier
  synthesis. The theory of optimal enzyme rhythms combines the
  dynamics and economics of metabolic systems and explains how optimal
  small-amplitude enzyme profiles are shaped by network structure,
  kinetics, external rhythms, and the metabolic objective. The
  formulae show how orchestrated enzyme rhythms can exploit synergy
  effects to improve metabolic performance and that optimal enzyme
  profiles are not simply adapted to existing metabolic rhythms, but
  that they actively shape these rhythms to improve their own (and
  other enzymes') efficiency.  The resulting optimal enzyme profiles
  ``portray'' the enzymes' dynamic effects in the network: for
  example, enzymes that act synergistically may be coexpressed,
  periodically and with some optimal phase shifts. The theory yields
  optimality conditions for enzyme rhythms in metabolic cycles, with
  static enzyme adaptation as a special case, and predicts how cells
  should combine transcriptional and posttranslational regulation to
  realise enzyme rhythms at different frequencies. \co{reflection of
    time scales; generally, reflection found!}
\end{abstract}


\co{reviewer didier gonze (theoretical chronobiology)}

\co{anfang anstract? anfang intro? discussion?
  "why are biological cycles beneficial? a perturbation theoretical formalism"}

\co{move some text about wiederspiegelung to CBA reg discussion?}

 \co{2nd order approx gilt evtl nicht  an
  bifurktionspunkten (e.g. if there are jumps between reference states)!!}

\co{\textbf{JA! MACHEN!}} \co{title war vorher: ``A theory of optimal enzyme rhythms in cells''?}
 \co{nochmal u und e klaeren, geht durcheinander, zb in box 1 un din supplement proofs}
\co{besseres wort als ``promote'' finden! doch ``favour''?}
\co{JA! gutes wort fuer ``Enzymes that are
  active only in rhythmic states'' .. ``Rhythm-promoted enzymes''; use
  uea} \co{JA! doch metabolic reward hier in metabolic objective
  umbenennen? (oder doch ``metabolic performance''?) dann erklaeren
  und aendern. das gleiche in cba kin? // oder in fn sagen:
  ``metabolic benefit'' entspricht ``metabolic reward'' in cba kin}
\co{rainers preprint
  https://www.biorxiv.org/content/10.1101/2021.05.26.444658v1
  zitieren} \co{kurz durchgehen und korrigieren; dann abbildungen (und
  vor allem ein beispiel mit mehr abbildungen, und aehnliche abb in
  SI?)}  \co{betonen: intro usw ``periodic synergies''} \co{bei abb
  wie fig 4 immer auch heatmap / kurven der variablen zeigen!}
\co{show arrows with discs!}  \co{note here: ``metabolic response''
  denotes physical effects, without enzyme adaptation; ``metabolic
  behaviour'' rather involves adaptation; clarify the difference in
  the introduction} \co{rename synergy effect $\rightarrow$ synergy;
  and synergy: $\rightarrow$ synergy coefficient, JA, gleichzeitig in
  STM} \co{ueberlegen: andere woerter fuer enzyme concentration vs
  activity!  (efficient concentration?)} \co{kommt auch in cba kin:
  enzyme level vs protein level verwirrend! (``enzyme
  abundance''/''concentration''; einmal sagen: ``level'' means
  ``effective active concentration''= explain!) ``activity'' is auch
  nicht gut in diesem zusammenhang .. neues wort praegen, zb
  ``effective enzyme level'' vs ``actual enzyme level''?}  \co{JA!
  show graphics with detailed results (as in website document)}
\co{JA! in matlab implement pointers with circles!  show external
  rhythm by pointer (in same plot: show flux by arrows)!  redo
  graphics} \co{JA! ``at a constant external substrate level'' kommt
  immer wieder: ein wort dafuer einfuhren?: ``in a static
  environment''?}  \co{set x=cext, s=cint, nur hier und cba kin als
  abkuerzung!  but keep rext und rint; also set m = ln c (nur in cba
  opt und model balancing)} \co{wort fuer modelle / fitness functions
  ohne/mit explizite zeitabhaengigkeit: ``time-shift-invariant''?}
\co{Github-Repo anlegen} \co{datentabellen: komplexe
  amplitudenvektoren} \co{ueber hypothese nachdenken: ist ein guter
  rhythmus, aber rueckwaerts, besonders schlecht? NEE: das
  schlechteste (bei externam rhythmus) ist ein guter rhythmus mit
  UMGEDREHTEN VORZEICHEN! (gile auch fuer static adapatation ::
  DISCUSSION!)}
 
\co{\textbf{Weniger wichtig / spaeter}} \co{Journal of Biological
  Rhythms // Phys rev E (IF 2.3); Plos Comp Biol (4.6) // JTB //
  BioSystems // CHAOS waere wohl auch OK.}  \co{Storage /
  ethanol-shuffling- beispiel implementieren!}  \co{FIX calculation
  and plotting of periodic economic potentials} \co{benjamin pfeuty,
  willi: bisher nichts zitierbares?} \co{cite other paper by stefan?
  // check Bartl 2013, \cite{bksl:13}?}  \co{arren: what is the
  optimal benefit in the best possible scenario?}  \coout{bessere
  woerter fuer synergy bzw synergy terms?}  \co{mention/show models
  with NADH / ATP cycles?  eher Opt rhythms II?}  \co{filme zu synergy
  graphs: 1. frequenzen durchfahren. 2. uebergang von netzwerk zu
  zeitkreis fuer alle modelle in reports; fuer verschiedene modelle in
  SI zeigen?}  \co{bilder mit rotierenden pfeilen als
  film?  spaeter}  \co{bei osz .. falls
  problem mit akkumulation .. puffer und ketten!}
\co{oscillationsdilution.speedy: diskrepanz zwischen analytischer und
  numerischer loesung!}  \co{woerter klaeren: oscillation / rhythm /
  cycle / alternation / periodic behavior} \co{check steady state conc
  und amplitudes in linear chain model} \co{die reports schoen machen
  (klare texte, ueberfluessige bilder weg)} \co{schönes matlabskript,
  das mit sbml klarkommt} \co{systematische bifurkationsanalyse fuer
  lin kette mit verduennung .. wann werden enzyme inaktiv, wann wird
  der refzustand instabil?  ein beispiel!; Ein modell mit bifurkation
  besprechen; chain with dilution: 3d-plot mit lambda und omega als
  achsen, principal synergy bzw fitness als funktion} \co{alle
  beispiele neu durchrechnen} \co{implement inequality constraints on
  output variables; implement equality constraints on output
  variables: e.g.~required oscillating flux to be produced!}  \co{Neue
  gute Modelle?; symbole klar?  alle faktoren 1/2 klar?  elast/resp
  klar?  konkretere diskussion von zyklen am beispiel;
  dynamic/economic= physical/physiological; - skaliert oder unskaliert
  klaeren.  Invarianzen zwischen nutzen und kosten usw.}

\co{\textbf{Short description (website) ``Optimal enzyme rhythms''} I
  outline a theory of optimal enzyme rhythms. If metabolic variables
  (fluxes, metabolite concentrations, and enzyme activities) vary around a
  steady reference state, this may impair or improve metabolic
  performance. To quantify the effects on fitness, I describe
  metabolism by kinetic models with a fitness function that scores the
  metabolic performance (e.g.~a high average flux) and punishes enzyme
  cost (e.g.~the cost of enzyme production). Focusing on optimal
  enzyme rhythms, I consider simple sine-wave profiles for enzymes and
  determine their optimal amplitudes and phases across the pathway or
  network. Using this framework, I study how enzyme rhythms can
  provide benefits, whether self-promoted rhythms can emerge, and what
  general principles underlie the optimal amplitudes and phase shifts
  of enzymes, metabolite concentrations, and fluxes.}

\iftoggle{bookversion}
{\section{Rhythms in metabolism}}
{\section{Introduction}}

\myparagraph{\ \\Adaptive rhythms and biological cycles} The lives of
organisms are shaped by biological rhythms such as sleep rhythms or
seasonal flowering.  By adapting to daily or yearly rhythms in the
environment, organisms can perform actions when conditions are best,
and can anticipate changes, e.g.~producing storage compounds for the
night or winter times. Other rhythms like heart beat, breathing, or
menstrual cycle emerge spontaneously. There are also rhythms within
cells (e.g.~the cell division cycle or circadian photosynthesis
\cite{nbas:03}) and specifically in metabolism \cite{kbfm:04,mubk:07},
as widely visible in gene expression data. Yeast cells show autonomous
metabolic oscillations that involve genome-wide periodic  gene expression
\cite{Spe98,Cho98}, as well as periodic production, storage, and
consumption of metabolites.  These oscillations also exist in single
cells \cite{pnwh:17}, \co{more refs} and cell populations may
auto-synchronise to show joint oscillations \cite{wohe:00} in which
cells adapt to varying conditions created by the entire cell
population. They involve large parts of the transcriptome, with
different functional subsystems peaking in different phases. \co{REF
  spellmann rainer usw} Expression oscillations accompany both
``outside'' changes like day-night cycles or cell cycle and
``internal'' oscillations like to repiratory or metabolic cycles in
yeast, which appear to be autonomous, but possibly synchronised with
other cycles.

\myparagraph{Potential functions of metabolic rhythms} While
mechanisms behind biochemical oscillations \cite{gold:96} and the
molecular regulators of genome-wide expression oscillations have been
thoroughly studied, their functions are still debated. Metabolic
oscillations may be a side effect of dynamics, for example of
overshooting gene regulation, but they may also have specific
functions -- in other words: they may provide benefits.  It has been
claimed that ``pumping'', i.e.~periodic concentration and flux
changes, can increase the efficiency of biochemical pathways
\cite{helr:89,hika:16}. More generally, metabolic cycles may allow
cells to arrange their metabolic processes in time and across the
entire metabolic network to optimise the resource usage
\cite{mubk:07}: mutually incompatible processes may be run at
different times to avoid adverse effects, while other biochemical
processes may run in ``concerted actions'' or in a favourable temporal
order.  The yeast metabolic cycle shows a characteristic sequence of
physiological phases that functionally build on each other
\cite{mamu:12}, \co{more details?  show graphics with ``meaningful''
  expression oscillations (like in yin/yang article)?}  which may be
coupled to the cell cycle or not \cite{kbfm:04,tkrm:05}, and
characteristic metabolic changes also occur during the cell cycle and
circadian rhythms \cite{pnwh:17}. \co{refs} \co{konkreter sagen (mit
  abb?): wie beim zellzyklus unf metab cycles, verschiedenes in
  ineinandergreift (oben schon kurz erwaehnt). hypothese: kausale
  notwendigkeit (vorbereitung) spiegelt sich in zeitlich versetzten
  prozessen} \co{mention thermodyn forces} \co{levine liu + elowitz:
  functional role of pulsing in genetic circuits (REFERENCE + comment
  in opt osc paper)}

\co{hier schon ein erstes schemabild, was portraitierung klarmacht?
  kausalitaet (notwendige bedingung) fuehrt zur abfolge (mit / ohne
  auesseren trigger) hier schon bauprojekt als analogie zeigen?
  ``bau'' von proteinen als biol beispiel? metab cycle - rainers
  cluster zeigen und erklaeren; ralfs cyano-aktivierung zeigen?
  spellman cell cycle? dann sagen: das alles koennte auch im steady
  state vor sich gehen, aber anscheinend kommt es unter bestimmten
  umstaenden (ausgeloest durch aussere oszillationen, oder weil es
  guenstig ist) von steady state zu oscillation! das ganze als
  vorbereitung zu ``runder stunde'' mit philosophieabschnitt am ende}

\myparagraph{Enzyme adaptation in large metabolic networks} In this
paper, I study the potential benefits of metabolic rhythms from a
theoretical angle.  \co{briefly discuss ralf's (+ others') FBA-type
  models} If metabolic rhythms are controlled by enzyme activities,
optimal metabolic cycles must involve optimal periodic enzyme profiles
or ``enzyme rhythms''. Based on kinetic models, I ask what enzyme
amplitudes and phases would maximise metabolic efficiency as a proxy
for cell fitness.  Predicting optimal enzyme activity profiles across
the metabolic network and in time, is a difficult problem.  In a
slowly changing environment, cells may adapt their enzyme expression
quasi-statically in every moment \cite{rubs:15}, and slow rhythms in
the environment promote slow, synchronous enzyme rhythms. However, such
a myopic strategy does not allow cells to produce storage compounds
for future usage, and it ignores that metabolism is dynamic: when
external metabolite concentrations or enzyme levels oscillate fast, they
create damped waves in the network, reaching different network
regions at different times.  In this case, a quasi-static adaptation
is logically impossible.  But cells may use existing (e.g.~day-night)
rhythms to their advantage: they may arrange biochemical processes in
time, running each of them when the biochemical conditions are best
(availability of substrates, cofactors, or thermodynamic driving
force).  Plants, for example, produce and store energy compounds
during the day, when light is available, \co{cite ralf? say: in
  contrast to plants, cyanos do not do this kind of storage (unless
  glycogen storage, in resource allocation models, is made much
  cheaper than in reality. this finding actually supports the resource
  allocation hypothesis!)}  and shift some energy-demanding processes
to night hours. Once some of the enzyme levels oscillate, they change
the metabolic dynamics and create incentives for further enzyme
adaptation. This changes again the metabolic dynamics, requiring
further adaptation, and so on. Eventually, this leads to a metabolic
cycle in which different processes are specifically run at different
times, possibly anticipating future demands. In this cycle, enzyme
levels are not just adapted in each moment, but they also produce
compounds for the next phase of the cycle.  In an optimal cycle, the
enzyme profiles must be self-consistent, i.e.~adapted to the dynamics
created by the environment and by the enzyme profile itself.

\begin{figure*}[t!]
 \begin{center}
 \includegraphics[width=16.5cm]{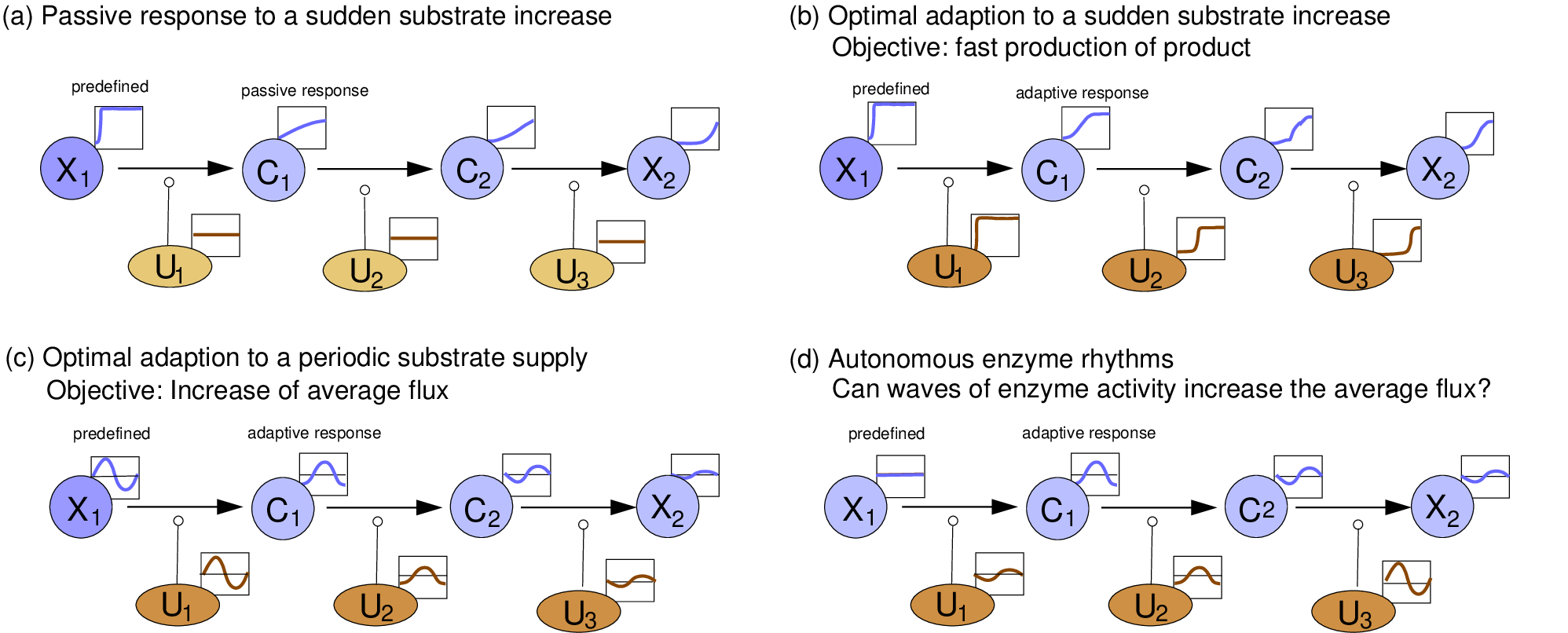}\\[2mm]
 \caption{Metabolic dynamics and enzyme adaptation in a linear
   pathway.  Metabolites are shown by circles (dark blue: external
   substrate with predefined level; light blue: internal metabolites),
   enzymes are shown by ellipses.  Each enzyme produces the substrate
   for the following enzyme. (a) Constant enzyme levels
   (yellow). After a sudden increase in substrate level, metabolite
   levels increase along the chain with different time delays.  (b)
   Optimal temporal enzyme profile (brown).  Following a
   ``just-in-time'' strategy, enzymes are induced sequentially.  This
   sequential activation can speed up the process (i.e.~the time until
   half of the substrate has been turned into product) at a fixed
   total enzyme cost \cite{klhh:02,zmrb:04}. (c) ``Just-in-phase''
   enzyme rhythm as an optimal adaptation to a periodic substrate
   supply.  Substrate oscillations alone would lead to metabolite
   oscillations.  Adaptative enzyme rhythms can modify these rhythms,
   increasing average flux and catalytic rate.  (d) A self-promoting
   enzyme rhythm creates metabolite and flux rhythms even in a
   constant environment. In the article I discuss how externally
   promoted or self-promoting rhythms can be beneficial and what
   determines their optimal phases and amplitudes.}
 \label{fig:examples_linear_chain}
 \end{center}
\end{figure*}

\myparagraph{Promoted rhythms and self-promoting rhythms} Enzyme
rhythms may serve as adaptations to external rhythms or as a way to
drive ``spontaneous'' metabolic cycles. The two cases are closely
related. By ``pumping'', a metabolic cycle can create favourable
biochemical conditions at different times in different parts of the
network. If we look at individual pathways, the effect of these
\todo{self-promoting beneficial changes} is not very different from
(temporally varying) favourable conditions caused by an external
oscillation. Then, the cell can exploit this dynamics oscillations by
investing enzyme wherever conditions are favourable, leading to
periodic enzyme activity changes, phase-shifted along the pathways. In
an autonomous metabolic cycle, each pathway is surrounded by a
periodic environment -- the rest of the network -- to which it needs
to adapt. If the pathway is optimally adapted to a surrounding system,
we can take the surrounding system's behaviour as given, no matter if
we regard it as optimal or as predefined. Therefore, whether rhythms
are promoted by the cell's environment or whether they emerge in the
cell, the incentives for enzyme adaptation are the same on the level
of single reactions or pathways. We may also argue: if cells can take
advantage of oscillations in their environment, which makes metabolic
pathways work more efficiently, they may achieve similar benefits by
enforcing similar oscillations inside the cell. Once a metabolic cycle
has been established (perhaps at some cost), other processes may
become adapted to it, creating benefits in different places that lead,
overall, to a net advantage\footnote{Even if metabolic oscillations in
  cells were beneficial, it may still be the case that their
  synchronisation between cells is harmful, decreasing the metabolic
  efficiency in each cell. This question will not be considered here,
  but it could be studied by extending the present approach.}.
Following these two arguments, I describe both types of rhythms --
promoted and self-promoting rhythms -- by one theory.  To see whether
rhythms can improve metabolic performance, I study dynamic metabolic
models and search for periodic enzyme profiles that optimise the
performance of some biological function.

\myparagraph{Mathematical models of enzyme adaptation} How can optimal
enzyme rhythms be predicted from mathematical models?  Here we
consider an optimal control problem based on metabolic efficiency. We
describe the system under study (a metabolic pathway or network) by a
kinetic model and ask how enzyme oscillations (as opposed to static
enzyme levels) can improve the performance of the system, e.g.~the
average flux per average enzyme amount. Mathematically, metabolite
concentrations and rates are described by a kinetic model an enzyme
profiles are optimised. This approach has been applied to static
enzyme levels \cite{hekl:1996}, static enzyme adaptation
\cite{lksh:04} and temporal enzyme profiles
\cite{klhh:02,zmrb:04}. For example, imagine a metabolic pathway in
which the substrate level switches from zero to some constant positive
value. How should the enzymes be activated in time? If all enzyme
activities were constant, the metabolite concentrations would slowly
increase one after the other (Figure \ref{fig:examples_linear_chain}
(a)). By choosing time-dependent enzyme profiles, the chemical
conversion can be accelerated at a constant enzyme
investment. Optimisations of enzyme profiles \cite{klhh:02} or enzyme
regulation mechanisms \cite{zmrb:04} predicted a sequential induction
of enzymes (Figure \ref{fig:examples_linear_chain} (b)), a behaviour
observed in amino acid biosynthesis pathways \cite{zmrb:04}.  In
\cite{klhh:02}, a sequential induction of enzymes was predicted: each
enzyme became active only when enough of its substrate had
accumulated. The resulting sequential expression was dubbed
``just-in-time production'' because enzymes are expressed only when
needed.  Here I ask the same question for periodic states.  If the
concentration of a pathway substrate changes periodically, can
synchronous enzyme rhythms increase the average flux and thereby
the catalytic rate?  Can we expect wave-like activity patterns that
``push'' or ``guide'' metabolites along a pathway (Figure
\ref{fig:examples_linear_chain} (c))?  And can autonomous enzyme
waves increase metabolic efficiency, even in a static environment
(Figure \ref{fig:examples_linear_chain} (d))?

\myparagraph{A theory of optimal enzyme rhythms} Here I propose a
theory of optimal enzyme rhythms in kinetic metabolic models,
applicable from small pathways to entire metabolic networks.  Enzyme
rhythms can be understood in two ways: causally, through physical
mechanisms, and functionally, through economic demands.  To combine
these two views, I consider kinetic metabolic models with optimal
enzyme profiles. In the models, optimal enzyme activities are not
determined by \emph{regulation} (e.g.~transcriptional gene
regulation), but by their \emph{function}, i.e.~by benefits they
provide. The focus on optimal profiles rather than regulation
mechanisms is reflected in my terminology. If an external rhythm
provides an incentive for enzyme rhythms, we say that it
``\emph{promotes}'' the enzyme rhythm. Likewise, if an enzyme rhythm
provide a fitness advantage just by itself, in a static environment,
this rhythm is called \emph{self-promoting}.  Whether and how these
optimal rhythms can be realised by cells is a separate question and
should be discussed separately.

\myparagraph{Computing optimal enzyme rhythms} To define optimality
problems for enzyme rhythms, I assume a given environment (e.g.~static
or periodic external substrate levels) and search for periodic enzyme
profiles that lead to a maximal fitness.  Then, to obtain tractable
formulae, I apply a perturbation theory: oscillations are described by
small sine-wave oscillations around an enzyme-optimised steady
reference state, and all oscillating variables (enzyme activities,
metabolite concentrations, and fluxes) are represented by amplitudes and
phases (``curve parameters''). To relate the amplitudes and phases of
different metabolites, enzyme, and fluxes, periodic response
coefficients from Metabolic Control Theory ({\MCA}) are used
\cite{lieb:2005}.  With all these approximations, finding optimal
enzyme profiles becomes a quadratic optimality problem with linear
constraints (see Figure \ref{fig:OptimalityProblem} in appendix).  If
none of the constraints are hit, the adaptations to perturbations at
different frequencies are additive and the adaptations to
non-sine-wave perturbations can be obtained by Fourier synthesis.  The
search for optimal enzyme rhythms resembles a search for optimal
static enzyme adaptations, which has been addressed in
\cite{lksh:04}. Optimal static enzyme adaptations can be computed from
metabolic response coefficients and curvatures of the fitness
function.  Similarly, to compute optimal enzyme rhythms, we replace
real-valued static adaptations by complex-valued oscillation
amplitudes, and static response coefficients by their complex-valued,
periodic counterparts. A new finding is that periodic enzyme
variations can improve an already optimal state, even if static enzyme
shifts cannot provide an advantage.

\myparagraph{Overview of the article} In this article I extend the
theory of optimal enzyme adaptation from \cite{lksh:04} into a theory
of optimal enzyme rhythms. The text is structured as follows.  I first
consider a single reaction and show how a given substrate rhythm can
provide an incentive for enzyme rhythms. Then I study optimal enzyme
rhythms across an entire network and derive formulae for enzyme
amplitudes and phases in externally promoted or self-promoting
rhythms.  Known formulae for static enzyme adaptation \cite{lksh:04}
are reobtained as a special case. Finally, I ask how enzyme rhythms
should be realised by combining gene expression and posttranslational
modifications. As expected, transcriptional regulation is preferably
used for slow changes, while posttranslational regulation is
preferably used for faster changes. \co{:REF?} While this article shows simple
example models, the theory applies to networks of any size.
Mathematical details are given in the appendix and in the
supplementary material. More examples are shown at
\url{www.metabolic-economics.de/enzyme-rhythms/}.

\begin{figure*}[t!]
  \mybox{
    \textbf{Box 1: Beneficial enzyme oscillations in a single reaction}\\[3mm]

    \centerline{\includegraphics[width=15cm]{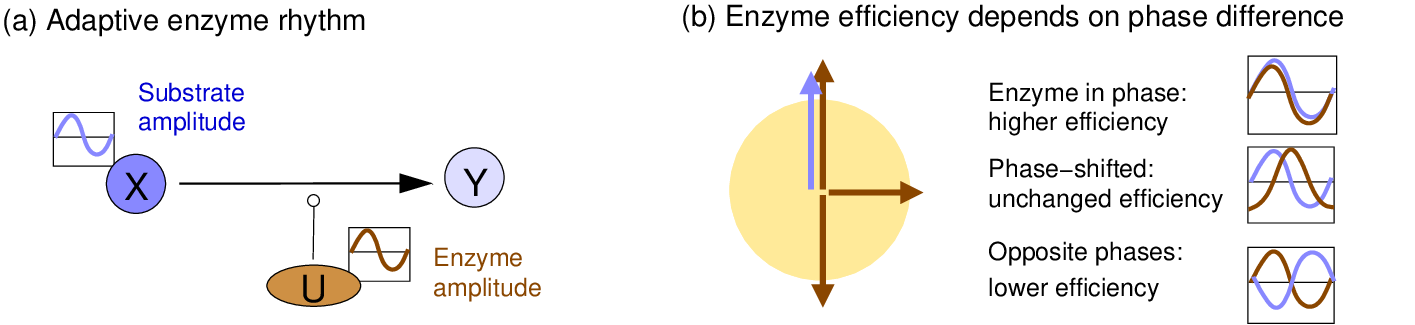}}\ \\

    Figure (a) shows a reaction X $\rightarrow$ Y with an irreversible
    mass-action rate law $\rate(u,x) = u \, k\, x$ and a given
    substrate rhythm $x(t)$. The enzyme activity $u(t)$ describes the
    concentration of active enzyme.  We consider the reaction rate $v$
    in mM/s, enzyme activity $u$ in mM, substrate level $x$ in mM,
    rate constant $k$ in mM$\inv$ s$\inv$.  What is the enzyme rhythm
    $u(t)$ (with a given average value) that maximises th average
    flux?  We assume sine-wave profiles\footnote{Sine-wave
      oscillations can be conveniently described by complex
      exponential functions (see Figure \ref{fig:OptimalityProblem}).
      The real part $\real(\e^{i\,\omega\,t}\,\et)$ is a cosine
      function with real amplitude $|\et|$ and a phase shift given by
      the phase angle of $\et$. \co{FN WO? was ist mit negativen
        frequenzen? die sollten ja fuer realle werte sorgen; discuss
        this in the beginning of the paper. eigentlich könnte man
        immer beide frequenzen betrachten und das ergebnis
        aufsummieren}}
\begin{eqnarray}
x(t) = \xb + \real(\e^{i \omega t}\,\xt), \qquad 
u(t) = \ub + \real(\e^{i \omega t}\,\et)
\end{eqnarray}
 with mean values $\xb$ and $\ub$ and complex amplitudes $\xt$ and
 $\et$.  Circular frequency $\omega$, frequency $f$, and period 
 $T$ are related by $\omega = 2\,\pi\,f = 2\,\pi/T$.  By inserting the profiles 
 $x(t)$ and $u(t)$ into the rate law, we obtain the time-dependent reaction
 rate
\begin{eqnarray}
\label{eq:fluxsplitting}
  v(t)
 &=& \underbrace{k\,\ub\,\xb}_{\bar v} 
 + \underbrace{k\,\real(\e^{i \omega t}\,\et) \xb}_{\langle \cdot \rangle_{t} = 0}
 + \underbrace{k\, \ub\,\real(\e^{i \omega t}\,\xt)}_{\langle \cdot \rangle_{t} = 0}
 + \underbrace{k\,\real(\e^{i \omega t}\,\et)\, \real(\e^{i \omega t}\,\xt)}_{\frac{k}{2} \real(\e^{i
  2 \omega t}\,\et\,\xt) + \frac{k}{2} \real(\et^{*}\, \xt)},
\end{eqnarray}

a sum of five terms: a static reference flux $\bar v$, two periodic
terms caused by linear effects of the periodic parameters, and two
synergy terms (see SI \ref{sec:proofFourTerms}). The synergy term
$\frac{k}{2} \real(\e^{i 2 \omega t}\,\et\,\xt)$ describes an
oscillation of frequency $2\,\omega$, while the synergy term
$\Delta \bar v = \frac{k}{2} \real(\et^{*}\, \xt)$, with a star for
the complex conjugate, describes a shift of the average flux
\cite{lieb:2005}. \coout{cite stefan's paper on jensen's inequality
  (in ``literature subfolder''); did he cite me?}
This flux shift is the benefit  we are interested in. It is given by
\begin{eqnarray}
\label{eq:fluxsplitting2}
\Delta \bar v =  \langle \Delta v\rangle_{t} = \frac{k}{2}\, |\xt|\,|\et|\, \cos(\Delta \varphi)
\end{eqnarray}
and can be positive or negative depending on the phase
difference  $\Delta \varphi = \varphi(\xt)-\varphi(\et)$
  (Figure (b)). Instead of a mass-action rate law, we may
consider  more general nonlinear rate laws $\rate = u\,\ratelaw(x)$. 
We assume small oscillation amplitudes and use the
linear approximation
\begin{eqnarray}
\label{eq:fluxsplitting2a}
\Delta \bar v =  \langle  \Delta v(t) \rangle_{t} = \frac{E_{\rm x}}{2 \ub}\, |\xt|\,|\et|\, \cos(\Delta \varphi)
\end{eqnarray}
with the unscaled elasticity $E_{\rm x} =\partial \rate/\partial x$
(see SI section \ref{sec:nonlinear}).  For fully saturated enzymes
(with $E_{\rm x}=0$), the oscillation has no effect on the average
flux. As shown in the graphics above (Figure (b)), the average flux
$\langle v \rangle_{t} = \bar v+\Delta \bar v$ depends on the phase
shift between substrate and enzyme. If substrate and enzyme vary in
phase ($\Delta \varphi = 0$), the flux increases by
$\frac{E_{\rm x}\,|\et|\,|\xt|}{2 \ub}$. The overall efficiency
$\langle v\rangle_{t} / \langle u\rangle_{t}$ -- the average rate,
divided by the average enzyme level -- increases while average enzyme
activity and average ratio
$\langle v / u \rangle_{t} = k \langle x \rangle_{t} = k\,\xb$ remain
unchanged.  In contrast, if $x(t)$ and $u(t)$ oscillate with opposite
phases, catalytic rate and average flux may decrease.}
\end{figure*}

\co{wo? ``enzyme rhythm'' alone refers mostly to single enzymes;
  ``orchestrated'' or ``global'' enzyme rhythms refers to the entire
  pattern in the network}

\section{Optimal enzyme profiles}

\subsection{Periodic enzyme levels can increase the catalytic rate}
\label{sec:localeffects}

\myparagraph{\ \\Periodic adaptation of a single enzyme} Let us see
how a periodic redistribution of enzyme resources can improve
metabolic performance. If a reaction's substrate level varies
periodically, a synchronous periodic variation of enzyme activity (at
a constant average enzyme level) can be beneficial. Whenever the
substrate level is high, the enzyme acts more efficiently: shifting
enzyme investments to these moments will increase the average flux per
average enzyme activity\footnote{The distinction between enzyme
  concentration $p(t)$ (concentration of enzyme molecules) and enzyme
  activity $u(t)$ (concentration of enzyme molecules in the ``active''
  protein modification state) does not matter at this point. However,
  it is important when enzyme activities are shaped by expression
  changes and posttranslational modification simultaneously, as
  described further below.}: the oscillations can increase average
fluxes ``for free''! Box 1 shows an example. The fact that
oscillations can provide an advantage challenges a basic assumption in
metabolic modelling, the assumption that optimal metabolic states need
to be steady states\footnote{Steady states are commonly assumed in
  models.  If we assume that cells are \emph{really} in steady state
  in a given environment, the rate laws must hold between the
  steady-state concentrations and fluxes. However, if the steady state
  holds only on average (e.g.~in a metabolic system that oscillates
  around a hypothetical steady state), the average metabolite concentrations
  and fluxes need not satisfy the rate laws precisely
  \cite{rere:16}.}.

\myparagraph{Allosynchrony} Here I will call this phenomenon -- 
synergisms due to synchronised substrate and enzyme rhythms --
``allosynchrony''. Unlike allosteric regulation, which acts in each
moment in time, allosynchrony concerns the overall time profile and
describes the effect of synchronous fluctuations on the average
metabolic state. Similar to channelling, which increases the local
substrate concentration ``seen'' by the enzyme, allosynchrony
increases the substrate concentration ``locally in time'', in moments
when most enzyme is most abundant. This efficiency increase is
achieved by metabolic dynamics, but it also affects cell economy,
namely the economical usage of enzymes.  A higher metabolic efficiency
(average flux per average enzyme activity) leads to a higher economic
efficiency (i.e.~metabolic benefit per enzyme investment). When enzyme
profiles are compared at a given cost (e.g.~identical average enzyme
levels, and assuming a linear cost function), high cost efficiency
implies a high {\metabolicobjective}.  If a substrate rhythm allows
cells to increase their average fluxes at a given enzyme cost, there
will be an incentive for such enzyme rhythms -- in other words, they
will be promoted.

\myparagraph{Dynamic propagation of enzyme perturbations} In metabolic
pathways or networks, enzyme resources can be reallocated not only in
time, but also across reactions.  If the periodic metabolite concentrations
were known, we could determine optimal enzyme rhythms reaction by
reaction, as described above. However, in kinetic models the internal
metabolite concentrations are not predefined, but depend on the enzyme levels!
Thus, our optimality problem contains various degrees of freedom: all
enzyme rhythms must be optimised simultaneously, taking into account
their dynamic effects on metabolite concentrations. All this makes the problem
much harder.  For example, each oscillating enzyme has an adverse
effect on its own substrate: whenever an enzyme level is high, the
substrate level tends to go down, when the enzyme level is low, the
substrate accumulates.  \co{kurz anhaeufen diskutieren;
  selbst-''inhibition''; buffering usw} \co{gibts noch die rechnung
  dazu im SI?}  Due to this adverse effect, enzyme rhythms tend to
decrease an enzyme's average efficiency \cite{aosk:17}! \co{alam et al
  noch lesen!  see subfolder ``literature''} To obtain a beneficial
overall rhythm, these adverse effects need to be overcompensated by
beneficial effects, e.g.~by creating phases in which certain enzymes
have a higher thermodynamic efficiency \cite{hika:16}. Moreover,
enzymes may also be coupled through costs. For example, if there is a
fixed overall enzyme budget, but investments can be reallocated
between enzymes in every moment, the increasing one enzyme level
implies decreasing another one\footnote{Like a fixed enzyme budget,
  also non-linear cost functions can lead to coupled enzyme rhythms.}
Mathematically, rhythms in a networks resemble rhythms in a single
reaction, but with some additional complications.  First, only some of
the metabolite profiles are predefined, while all others are
determined by the system dynamics and dependent on enzyme
profiles. Second, perturbations propagate through the networks
dynamically, and in finding optimal enzyme amplitudes and phase
patterns, we need to account for this.  Third, all enzyme profiles are
optimised simultaneously and must be adapted to the effects of other
enzyme rhythms, which are optimally adapted as well. The result is a
self-consistent, optimal metabolic state.

To model how enzyme rhythms shape metabolic states and are
shaped by them at the same time, we need to first consider metabolic dynamics and understand how enzyme levels act on metabolite concentrations and
fluxes. Then we 
turn its logic around, interpreting   the possible  \emph{effects} of an
enzyme change as an \emph{incentive}, or teleological cause, for this  change. \co{see
  freud} When a reaction rate is perturbed, the perturbation propagates
through the network, reaching different parts of the network with
different delays: the details of this forward propagation -- where perturbations
arrive, with what delays, and how they are damped -- are themselves reflected in the necessary or optimal
enzyme profiles. In the following sections, we
develop a theory of  optimal enzyme rhythms in three steps: (i) we define an
optimality problem for periodic metabolic states; (ii) we compute
synergy effects between oscillating model parameters
(external parameters and enzyme levels), which depend on oscillation
amplitudes and phase shifts; (iii) and we determine an optimal network-wide
enzyme rhythm that maximises the sum of all synergy effects.

\begin{figure*}[t!]
 \begin{center}
 \includegraphics[width=16.5cm]{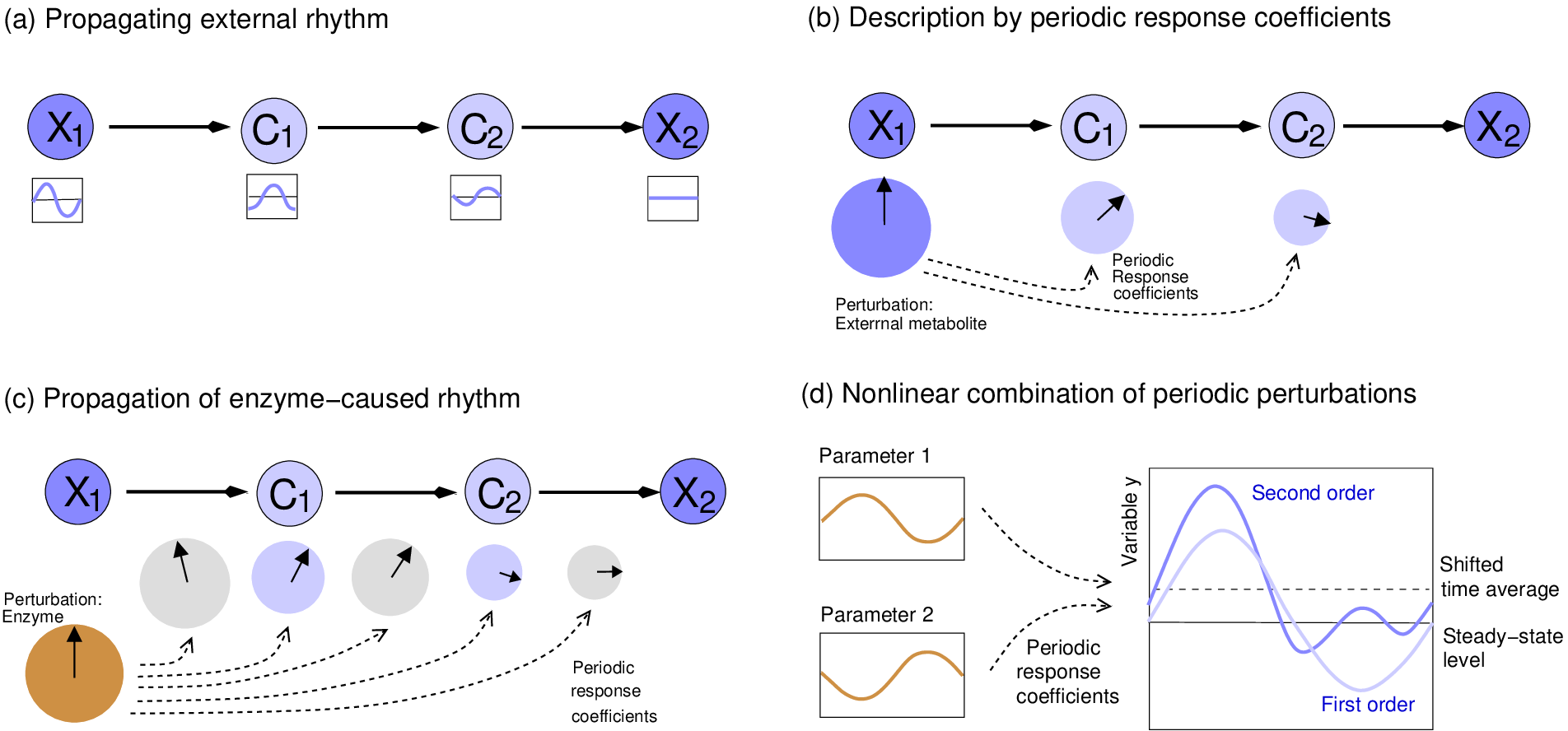}
 \caption{Forced oscillations in a metabolic pathway (schematic
   example).  (a) Metabolic pathway with external substrate
   rhythm. The periodic substrate level $x_{1}(t)$ (blue curve) causes
    metabolite and flux variations that propagate down
   the pathway as damped waves.  (b) Forced oscillations in a model with linearised
   rate laws
   $\Delta \vv(t) \approx \epsilonUmat \devb(t) + \epsilonSmat \Delta
   \cv(t) + \epsilonXmat \dxvb(t)$.  Amplitudes and phases are
   described by complex amplitudes (pointer lengths and angles).  The
   amplitude of metabolite $i$ is obtained by multiplying the
   amplitude of the perturbed parameter (index $m$) with the
   spectral response coefficient $\tilde R^{c_{i}}_{p_{m}}(\omega)$
   between parameter $p_{m}$ and concentration $c_i$.  (c) Metabolite
   and flux oscillations enforced by enzyme rhythms.  (d) Synergistic
   parameter rhythms.  A pair of  periodic parameters (with circular
   frequency $\omega$) leads to oscillations in a state variable $\ys$;
   in a second-order approximation, the forced oscillation consists of
   an average shift (second-order effect), a sine-wave oscillation of
   frequency $\omega$ (first-order effect), and a second harmonic
   (frequency $2 \omega$, second-order effect). }
 \label{fig:forced}
 \end{center}
\end{figure*}

\subsection{Dynamics of enzyme rhythms in pathways and networks}
\label{sec:DynamicsInNetworks}

\myparagraph{\ \\Metabolic states under periodic pertubations} To
predict beneficial enzyme rhythms, we first need to model the
metabolic dynamics, i.e.~the effect of enzyme rhythms on metabolite
levels and fluxes. To do so, we consider a kinetic model with internal
concentrations $c_{i}$ and reaction rates $v_{l}$ as state
variables. The rate laws depend on metabolite concentrations $c_i$,
external metabolite concentrations, $x_{j}$, and enzyme
activities\footnote{In fact, the distinction between $\xv$ and $\esymbolv$
  in the models below is not biological, but mathematical: we assume
  that $\xv$ decribes environment variables, which are externally
  defined and uncontrollable, while $\esymbolv$ describes control variables
  that need to be optimised.  Aside from external metabolite
  concentrations, the parameters $x$ may also describe other
  quantities that affect reaction rates. For example, the ATP synthase
  in plants relies on a proton gradient that affects the equilibrium
  constant of the ADP $\rightarrow$ ATP conversion. At night, when the
  gradient is low, the ATP synthase should be shut down. Otherwise it
  would start running in reverse and degrade ATP.  To model this, the
  proton gradient may be treated as an external parameter $x$, and the
  ATP synthase activity as a control variable $u$ to be optimised. In
  general, besides enzyme activities, the control variables $\esymbol_l$ may
  also represent any other quantities that influence the reaction
  rates and can oscillate.} $e_{l}$. \co{note that these would be
  called el in other papers; or change it here?} If enzyme activities
oscillate, they evoke oscillations in internal metabolite
concentrations and fluxes. The amplitudes and phases depend on network
structure, rate laws, enzyme amplitudes, and oscillation frequency. If
oscillations are small, amplitudes and phases can be predicted from
spectral response coefficients \cite{inga:04,lieb:2005} (see Figure
\ref{fig:forced}).  In the calculation, we start from a steady
reference state with constant external parameters and optimal enzyme
activities (which must be a stable steady state). Then we consider
sine-wave oscillations of external parameters $x_{j}(t)$ and enzyme
activities $\esymbol_l(t)$ around this state (with circular frequency
$\omega$ and complex amplitudes $\xt_{j}$ and $\et_{l}$). In addition,
we may apply shifts $\Delta \xb_{j}$ and $\Delta \ub_{l}$ of the
average values.  Then we apply perturbation theory: the state variable
profiles are approximated by the leading terms of a Taylor
expansion. In a linearised model, sine-wave perturbations lead to
sine-wave metabolite and flux oscillations of the same frequency, and
the amplitude profiles $\cintvt$ and $\vvt$ of the state variables
depend linearly on the amplitude profiles $\xvt$ and $\esymbolvt$ of the
perturbed parameters. We can write the dynamics as
$\cintvt = \tilde \Rmat^{\rm S}_{\rm \tilde x}\, \xvt + \tilde
\Rmat^{\rm S}_{\rm \tilde u}\, \esymbolvt$, with expansion coefficients
given by the spectral response coefficients\footnote{Response
  coefficients resemble reaction elasticities, but refer to an entire
  network (instead of a single reaction).  In an isolated reaction
  with known periodic metabolite concentrations and enzyme activities,
  the amplitude of the reaction rate can be computed, to first order,
  with the help of periodic elasticities.  Similarly, in a network,
  external metabolite and enzyme rhythms lead to oscillations of
  internal metabolite concentrations and fluxes, each with a different
  amplitude and phase shift.  These rhythms can be computed using
  periodic response coefficients. The synergy coefficients
  (second-order response coefficients) describe synergy effects of
  enzymes or external concentrations on state variables. In a single
  reaction, the second-order enzyme elasticity $E_{\rm uu}=0$
  vanishes, so enzyme rhythms alone have no second-order effects.}
(in matrices $\Rmat^{\rm S}_{\rm \tilde x}$ and
$\tilde \Rmat^{\rm S}_{\rm \tilde u}$), and an analogous formula holds
for flux amplitudes $\vvt$ \cite{lieb:2005}.  Similarly, average
parameter shifts $\dxvb$ and $\devb$ lead to average shifts
$\Delta \cintvb = R^{\rm S}_{\rm x}\, \dxvb + R^{\rm S}_{\rm u}\,
\devb$ with static response coefficients as prefactors. Generally,
linearised metabolic models act as low-pass filters: at low
frequencies, there is a quasi-static response, while at high
frequencies perturbations are strongly damped. At intermediate
frequencies, dynamic resonance may occur \cite{lieb:2005}.  To handle
combined perturbations, involving multiple perturbations and
frequencies, we can split them into sums or integrals of basic
perturbations (with single perturbations and frequencies), and sum
over the dynamic responses.  This linear superposition works in the
first-order approximation only. At larger perturbation amplitudes, the
approximation becomes unreliable and we need to include higher-order
effects: in a second-order approximation, the synergistic interactions
between enzymes lead to second harmonics at frequency $2\,\omega$ and
to shifts in the average concentrations and fluxes (see appendix
\ref{sec:forcedOscillations} and SI \ref{sec:periodicdynamics}).

\subsection{Optimal rhythms in metabolic pathways and networks}

\label{sec:EconomicsInNetworks}

\myparagraph{\ \\Cost-benefit problem for enzyme profiles} If we know
how to simulate enzyme rhythms and their effects on metabolic fluxes,
we can also turn this around and ask: what enzyme profiles are needed
to obtain a certain desired system output?  For example,  given an 
external metabolite rhythm $\xvt$, which enzyme profiles could
realise a desired flux oscillation at a minimal cost?
Such inverse problems can be hard, but our small-amplitude
approximation makes them tractable (see SI
\ref{sec:achievepredefineddynamic}).  To formulate cost-benefit
problems for metabolic networks (see Figure \ref{fig:optimality}), we
combine  the internal metabolite concentrations $c_{i}$ and fluxes $v_{l}$ into a
state vector $\ysv$ and define a {\metabolicobjective} function
$\yy(\ysv)$. The vector $\ysv$ may also include other state variables
such as pH values, membrane potentials,  or organelle volumes.  In our optimality problem, we
consider  two types of parameters: external parameters $x_{j}$
imposed by the environment (here usually external metabolite concentrations), and
control variables $\esymbol_l$ to be optimised (here usually enzyme activities $e_{l}$).  As a 
fitness function, we consider  
the difference $\ffit(\esymbolv,\xv) = \gplus(\esymbolv,\xv)-\hminus(\esymbolv)$ of a {\metabolicobjective}
$\gplus(\esymbolv.\xv) = \yy(\ysv(\esymbolv,\xv))$ and an enzyme cost
$\hminus(\esymbolv)$ \cite{lieb:18theory,lieb:14a}. Typically, the
{\metabolicobjective} $\yy(\cv,\vv)$ requires high production
fluxes and low metabolite concentrations, while the cost $h$
increases with the enzyme activities $\esymbol_l$.

\begin{figure*}[t!]
\begin{center}
 \includegraphics[width=16.5cm]{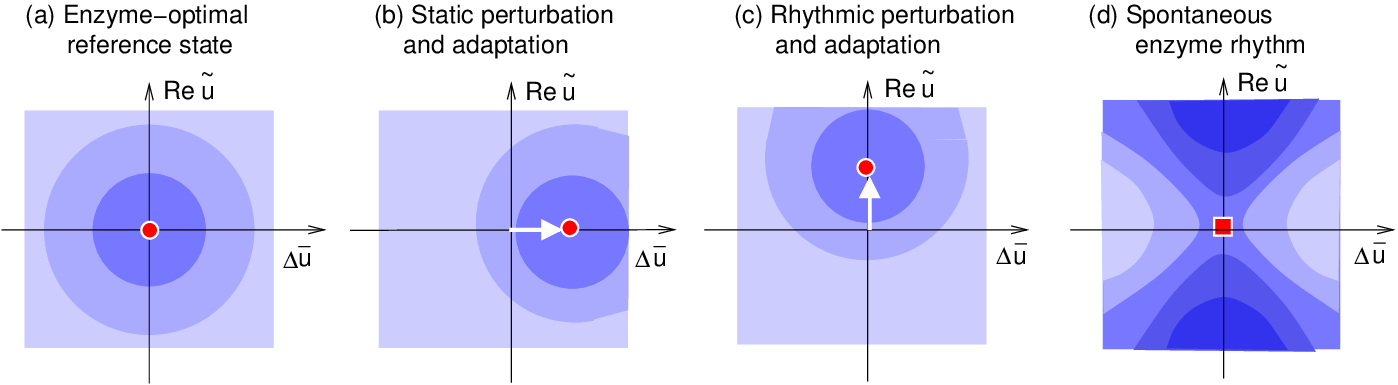}
 \caption{\coout{aehnliches bild wie in (a) und (b) fuer CBA adaptation?}
   Prediction of optimal enzyme profiles.  A periodic enzyme profile,
   described by a static shift $\Delta \ub$ and complex-valued
   amplitude $\et$ for a single enzyme, influences the behaviour of a
   metabolic system (compare Figures \ref{fig:OptimalityProblem} (a)
   and \ref{fig:optimality}). The resulting fitness of the metabolic
   system can be described by a fitness function $F(\Delta \ub, \et)$
   in the space of enzyme profiles. The different panels show fitness
   landscapes under different external metabolite perturbations
   $\Delta \xb$ or $\xt$. A red dot marks the optimal state.  (a)
   Unperturbed fitness landscape ($\Delta \xb= \xt= 0$, leading to
   $\Delta \ub= \et= 0$; fitness contours shown in blue; dark blue
   represents high fitness).  The fitness maximum lies in the
   reference state.  Negative fitness curvatures for $\Delta \ub$ and
   $\et$, described by fitness synergies $F_{\rm uu}$ and
   $F_{\rm \tilde{u}\tilde{u}}(\omega)$ in this point, show that any
   change $\Delta u$ or $\et$ would decrease the fitness.  (b) An
   external parameter shift $\Delta \xb$ changes the landscape,
   displaces the optimum point, and promotes a static shift
   $\Delta \ub^{\rm opt}$ of the optimal enzyme level. (c) An external
   rhythm with amplitude $\xt$ displaces the optimum and promotes an
   enzyme rhythm with amplitude $\et^{\rm opt}$.  (d) The reference
   state is a saddle point of the fitness landscape (red square),
   fitness-unstable against enzyme rhythms. Any static enzyme change
   $\Delta u$ would decrease the fitness, but some oscillations
   ($\et \ne 0$) increase the fitness even in the absence of external
   oscillations. These beneficial oscillations are called
   ``self-promoting''. }
 \label{fig:Guulandscape}
 \end{center}
\end{figure*}

\myparagraph{Fitness in periodic metabolic states} To study optimal
enzyme rhythms, we start from a steady reference state with given
external concentrations $x_{j}^{\rm ref}$ and static enzyme
levels\footnote{In the optimisation, enzyme profiles that lead to
  unstable steady states are discarded.} $\esymbol_l^{\rm ref}$ that are
already optimised.  As an optimality condition, $\ffit(\esymbolv,\xv)$ must
have negative curvatures with respect to $\esymbolv$ (except for those
enzymes that are not used in this optimal state).  Based on this state
we study the adaptation of the enzyme profile to external
perturbations.  To define our optimality problem, we use a functional
that assigns fitness values to metabolic time courses.  We consider
two possibilities, the \emph{state-average fitness}
$\Ftemp=\yy(\langle \ysv \rangle_{t})-\hminus(\langle \esymbolv
\rangle_{t})$ and the \emph{fitness-average fitness}
$\Ftemp=\langle \yy(\ysv) -\hminus(\esymbolv) \rangle_{t}$. Both of them are
based on time averages: in one case the static fitness function
$\ffit$ is applied to the average metabolic state, in the other one
the fitness is evaluated in every moment and then averaged over time,
which means that variations around the average value can have a
fitness effect.  Using our fitness functional, we can study the
fitness changes caused by perturbations and adaptations. If a static
parameter change $\dxvb$ is applied to our reference state, the
optimal enzyme adaptation $\devb$ follows from the second-order
metabolic response coefficients \cite{lksh:04}.  With periodic
external perturbations, the calculation works similarly, but using
periodic response coefficients \cite{lieb:2005} (see SI
\ref{sec:periodicdynamics}).  Let us see how this works. We start from
our optimal reference state, apply periodic external profiles
$\xv(t) = \xvref + \dxvb + \re(\e^{i\,\omega\,t}\,\xvt)$ and enzyme
profiles $\esymbolv(t) = \evref + \devb + \re(\e^{i\,\omega\,t}\,\esymbolvt)$, and
evaluate the resulting periodic state (Figure \ref{fig:Guulandscape}
(b)).  The fitness can be written as a function
$\Ftemp(\devb,\dxvb,\esymbolvt,\xvt)$ of the shifts $\devb$ and $\dxvb$ and
of the complex amplitude profiles $\esymbolvt$ and $\xvt$ (see Figure
\ref{fig:Guulandscape}).  Near the reference state, a second-order
approximation yields
\begin{eqnarray}
  \label{eq:deltafexpansion}
   \Delta \Ftemp(\devb,\esymbolvt, \omega)  \approx 
   \underbrace{\left[ \dxvb\trans\,\Fxe \devb + \frac{1}{2} \devb\trans \,\Fee \devb \right]}_{\mbox{static}}
  + \underbrace{\left[ \real(\xvt^{\dag}\,\Fxtet(\omega)\, \esymbolvt)  + \frac{1}{2} \esymbolvt^{\dag}\,\Fetet(\omega)\, \esymbolvt \right]}_{\mbox{periodic}}.
\end{eqnarray}
The shape of the the fitness landscape near the reference state
depends on the local curvatures of $\Ftemp$ with respect to $\devb$,
$\dxvb$, $\esymbolvt$, and $\xvt$, contained in the fitness synergy matrices
($\Fee$ and $\Fex$ for static variations, $\Fetet$ and $\Fetxt$ for
periodic variations) as components.  The synergies depend on the
kinetic model, on cost and benefit functions, and on the type of
fitness functional used (state-average or fitness-average fitness).
If the reference state is known, the curvature matrices can be easily
computed with formulae from {\MCA} (see appendix
\ref{sec:AppFitness})\footnote{The synergy matrices $\Fetet$ and
  $\Fetxt$ are frequency-dependent. If we assume slow oscillations
  ($\omega \rightarrow 0$) and a fitness-average fitness functional (see appendix \todo{B}),
  the matrices can be approximated by $\Fetet(\omega=0) = \half \Fee$ and
  $\Fetxt(\omega=0) = \half \Fex$, based on the static synergy
  matrices.}. What can we learn from the fitness expansion in
Eq.~(\ref{eq:deltafexpansion})? The first bracket term describes the
effect of static perturbations (and enzyme adaptation), while the
second bracket describes the effect of parameter rhythms (and
adaptative enzyme rhythms) at frequency $\omega$.  In the formulae,
irrelevant terms have been omitted: terms that depend only on $\dxvb$
and $\xvt$ do not matter for the choice of the enzyme profiles; terms
linear in $\devb$ vanish because of the optimality condition
$\fv_{\rm u}=0$ in the reference state; and terms linear in $\esymbolvt$ do
not lead to any shifts on time average.  The expansion formula
Eq.~(\ref{eq:deltafexpansion}) holds for sine-wave perturbations of
frequency $\omega$.  Due to the second-order approximation (and
assuming a time-independent fitness function), there are no synergies
between perturbations of different frequencies, or between static and
periodic perturbations.  To compute the effects of mixed perturbations
with multiple parameters and multiple frequencies, we can sum or
integrate over the adaptive responses at different
frequencies\footnote{A linear superposition is only possible if
  higher-order terms (beyond the quadratic expansion) are neglected,
  if our fitness function is not explicitly time-dependent, and if our
  combined solution does not violate any constraints (see SI
  \ref{sec:HigherHarmonics}).}.

\myparagraph{Synergies between periodic parameters.}  To summarise,
the fitness effect of an enzyme profile can be computed via the
synergy matrices $\Fee$ and $\Fex$ (for static adaptation) or $\Fetet$
and $\Fetxt$ (for adaptive rhythms).  The matrix elements represent
three sorts of synergies: synergies between external parameters and
enzymes (elements of $\Fex$ and $\Fetxt$), synergies between different
enzymes (off-diagonal elements of $\Fee$ and $\Fetet$), and enzyme
self-synergies (diagonal elements of $\Fee$ and $\Fetet$).  All three
effects together determine the total fitness effect of an enzyme
profile: if it is positive, the enzyme rhythm provides an advantage
over the unperturbed steady state.  Enzyme self-synergies are usually
negative because of dynamic self-inhibition or because of the effect
of diminishing returns (due to fitness-average fitness functionals
with non-linear cost functions\footnote{Negative self-synergies may
  also be caused by a frequency-dependent enzyme cost function. I briefly
  discuss this below.}).  To make enzyme rhythms beneficial this
negative effect must be overcompensated by beneficial synergy effects
between external parameters and enzymes or between different enzymes.

\subsection{Promoted and self-promoting enzyme rhythms}
\label{sec:OptimalEnzymeProfiles}

\myparagraph{\ \\Optimally adapted enzyme profiles} If an external
parameter shift $\Delta \xb$ or an external rhythm $\xt$ is applied to
our reference state, how should the  (static or periodic)
enzyme profile be optimally adapted?  The expansion formula (\ref{eq:deltafexpansion})
helps us answer this question.  To compute the profile, we consider
the fitness landscape $F(\devb,\esymbolvt)$, approximate it by  Eq.~(\ref{eq:deltafexpansion}), consider
an external perturbation $\xv(t)$, and determine the optimal curve
parameters (in vectors $\devb$ and $\esymbolvt$) that
\begin{eqnarray}
  \label{eq:simpleOptimalityProblem}
   \mbox{Maximise} \quad \Delta \Ftemp(\devb,\esymbolvt, \omega)
\end{eqnarray}
under the constraint $\devb \ge - \evb$. Without external
perturbation, the system should stay in the reference state (with
$\devb=\esymbolvt=0$) because this state is a local fitness
optimum\footnote{Here we make two extra assumptions (which we will
  drop below): first, we assume that the reference state is
  economically stable against spontaneous enzyme oscillations of any
  frequency; and second, that all enzymes in the reference state are
  active; that is, the reference state is an interior optimum with
  respect to static enzyme activities $\evb$ (Figure
  \ref{fig:Guulandscape} (a)).  A more general case, reference states
  with inactive enzymes, will be discussed below.}.  With a static
perturbation $\dxvb$, the fitness landscape changes and the optimum
moves. If the new optimum $\evb$ respects all constraints (e.g.~if all
enzyme levels remain positive), the displacement $\devb^{\rm opt}$
yields the optimal enzyme adaptations (Figure \ref{fig:Guulandscape}
(b)).  Optimal adaptations to periodic perturbations are computed
similarly: an external rhythm with amplitude profile  $\xvt$ displaces the
optimum by $\esymbolvt^{\rm opt}$ (in the space of amplitude profiles
$\esymbolvt$), the vector of amplitudes and phases of the optimal enzyme
profile.  In both cases, the optimal enzyme profile follows from
maximising Eq.~(\ref{eq:deltafexpansion}) with the perturbation
$\dxvb$ or $\xvt$.  If there are no no further active constraints,
each of the bracket terms (for static and periodic shifts) can be
optimised separately. A static perturbation $\dxvb$ promotes a static
adaptation 
\begin{eqnarray}
 \label{eq:optstaticvector}
 \devb^{\rm opt} &\approx& - \Fee\inv\, \Fex\, \dxvb
\end{eqnarray}
while a periodic perturbation profile $\xvt$ promotes a periodic adaptation
with amplitude profile
\begin{eqnarray}
 \label{eq:optperiodic}
 \esymbolvt^{\rm opt} &\approx& - \Fetet\inv\, \Fetxt\, \xvt.
\end{eqnarray}
If a mixed perturbation is applied (a sum of static perturbations and
periodic perturbations with multiple frequencies), the optimal
adapatation is given by the sum of the single optimal adaptations
(again, under the conditions mentioned above that allow for linear
superposition).

\co{hier einmal eine box mit kompletter analyse! weitere analysen zu
  beispielmodellen ins supplement

 \includegraphics[width=10.5cm]{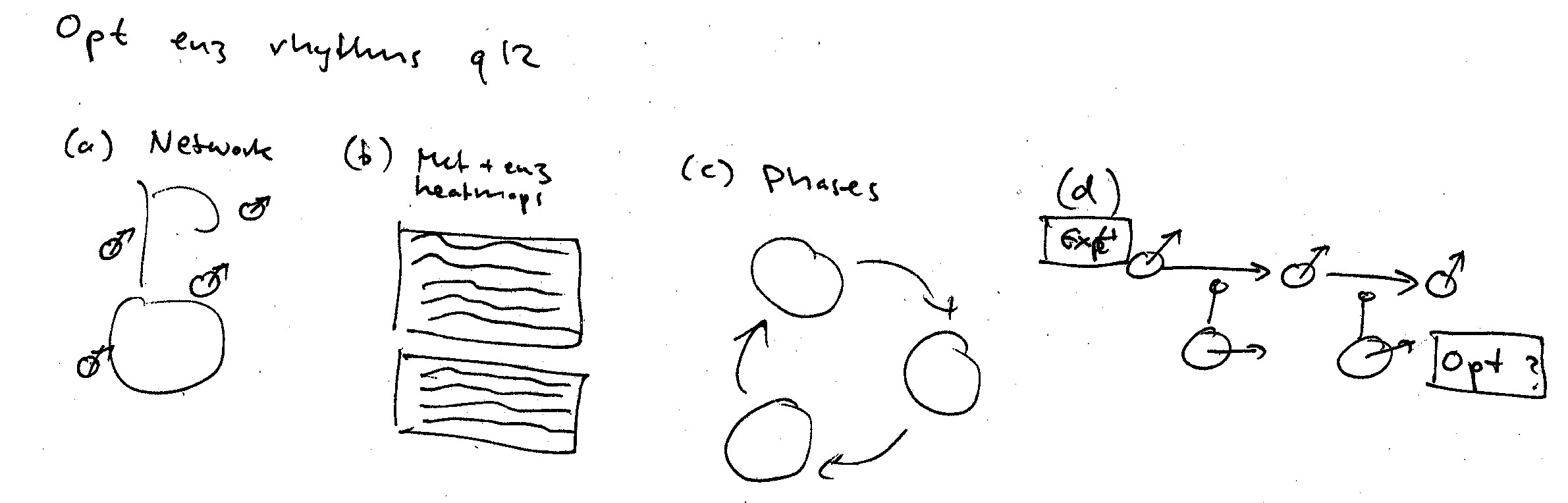}
 
 \co{WRITE ME; Overview graphics depicting metabolic rhythms in different ways}
}

\begin{figure*}[t!]
  \begin{center}
      \begin{tabular}{lll}
        (a) Forced oscillations caused by  substrate rhythm &&   (b) Adaptive enzyme rhythm \\[4mm]
        External substrate rhythm &&        Enzyme rhythm \\[1mm]
        \includegraphics[width=7.5cm]{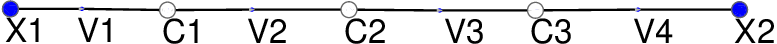} &&
        \includegraphics[width=7.5cm]{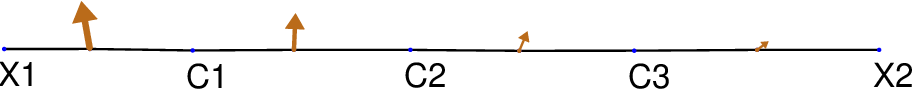}\\[2mm]
         Responsive metabolite rhythm &&   Metabolite rhythm \\[1mm]
        \includegraphics[width=7.5cm]{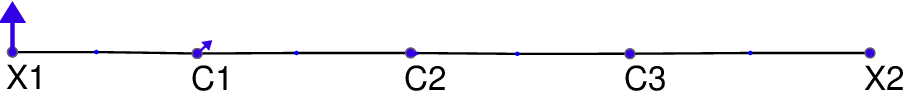}&&
        \includegraphics[width=7.5cm]{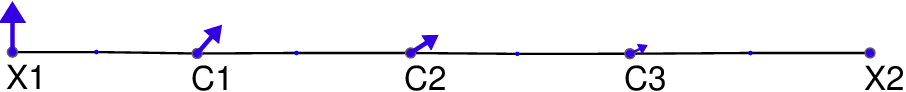}\\[2mm]
        Responsive flux rhythm &&     Flux rhythm \\[1mm]
        \includegraphics[width=7.5cm]{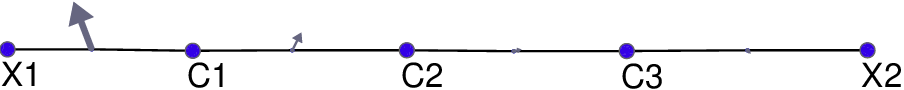}&&
        \includegraphics[width=7.5cm]{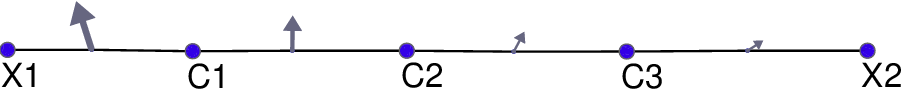}
      \end{tabular}
      \caption{\co{show arrows with discs! a show flux by thick
          arrows} Enzyme rhythm in a metabolic pathway.  (a) Forced
        metabolic oscillations at constant enzyme levels.  Top:
        pathway structure (metabolites shown as circles, X1 and X2 are
        external metabolites). Fluxes are shown as arrows. Centre:
        sine-wave oscillations of the external substrate (frequency
        $f=0.25$ s$\inv$) cause metabolite oscillations that propagate
        as damped waves.  Amplitudes and phase angles shown by arrows.
        Bottom: oscillating fluxes caused by the substrate rhythm.
        (b) Dynamics with optimally adapted sine-wave enzyme
        rhythm. Top: the first enzyme is almost in phase with the
        substrate, while the others show larger phase shifts.  The
        resulting metabolite and flux rhythms are shown in the centre
        and bottom panels.}
 \label{fig:linearChain}
 \end{center}
\end{figure*}

\myparagraph{self-promoting rhythms} We saw how enzyme rhythms
can be promoted by external rhythms.  How can we describe self-promoting
rhythms, that is, enzyme rhythms that provide benefits  in
static environments?  Again, we consider  a
reference state that cannot be improved, at least not by  static enzyme
changes. However,  there may still be  enzyme rhythms that lead to improvements even
in the absence of external rhythms, there is an incentive for such
rhythms.  As mentioned above, I call such rhythms ``self-promoting''.
To increase fitness by enzyme rhythms alone (positive
$\Delta F = \esymbolvt^\dag\,\Fetet\,\esymbolvt$), the matrix $\Fetet$ must have a
positive eigenvalue (see Figure \ref{fig:Guulandscape} (d)). To test
this, we compute the maximal eigenvalue of $\Fetet(\omega)$, called
principal fitness synergy $\sigma(\omega)$.  If this value is
positive, the corresponding eigenvector describes a self-promoting
enzyme rhythm, the most beneficial rhythm at a given amplitude
$||\esymbolvt||\le\esymbolvt^{\rm max}$. In contrast, if $\sigma(\omega)$ is
negative for all frequencies $\omega$, our system is stable against
self-promoting enzyme rhythms.

\begin{figure*}[t!]
  \begin{center}
\parbox{16.5cm}{
  \parbox[t]{8cm}{\vspace{-3.2cm}
    External substrate rhythm \\[4mm]
    \includegraphics[width=7.5cm]{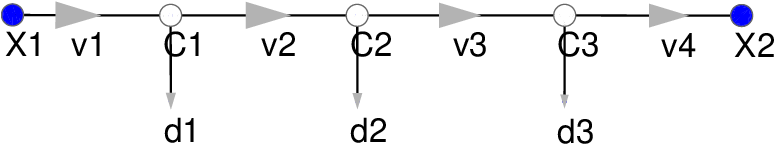}
    Principal synergy \\[4mm]
    \includegraphics[width=4cm]{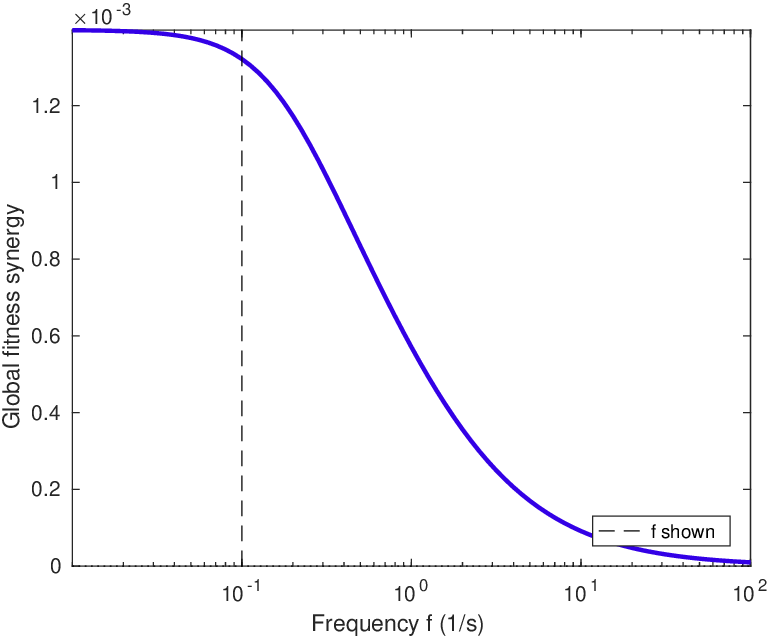}
}\hspace{5mm}
\parbox[t]{8cm}{
  \begin{tabular}{l}
    Enzyme rhythm \\[2mm]
    \includegraphics[width=7.5cm]{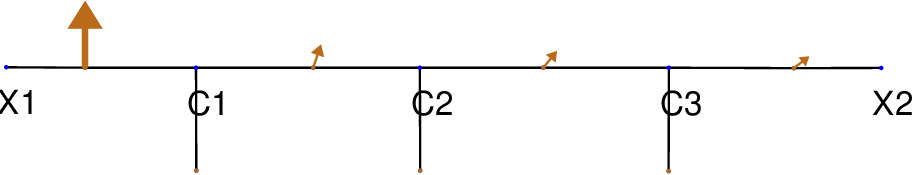}\\
    Metabolite rhythm \\[2mm]
    \includegraphics[width=7.5cm]{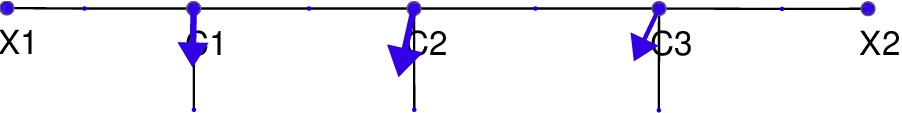}\\
    Flux rhythm \\[2mm]
    \includegraphics[width=7.5cm]{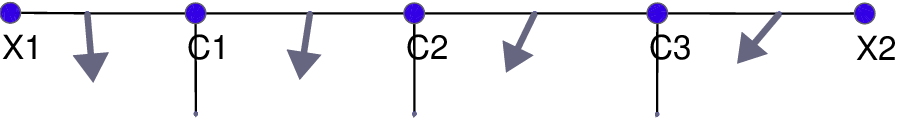}
\end{tabular}
}}
\caption{Self-promoting enzyme rhythm in a linear pathway.  Starting
  from the model in Figure \ref{fig:linearChain}, we now assume that
  metabolites can disappear by dilution, by non-enzymatic degradation,
  or by passive diffusion through cell membranes. Enzyme rhythms can
  reduce this loss and increase the production flux, even in the
  absence of external rhythms. We can see this from the positive
  principal synergy at low frequencies. Like in Figure
  \ref{fig:linearChain}, we assume that enzyme levels can oscillate
  without any extra cost.}
  \label{fig:linearChainDil}
 \end{center}
\end{figure*}

\begin{figure*}[t!]
  \begin{center}
      \begin{tabular}{lll}
        (a) Network structure & (b) Principal fitness synergy &
        (c) Enzyme and  metabolite rhythm \\[2mm]
        \includegraphics[height=4cm]{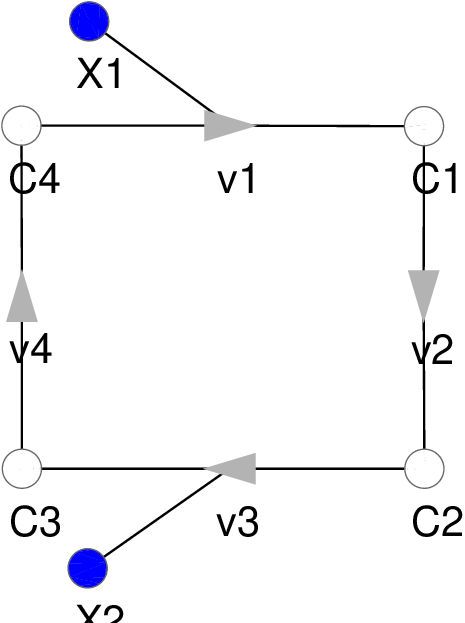}&
                                                                                              \includegraphics[width=5cm]{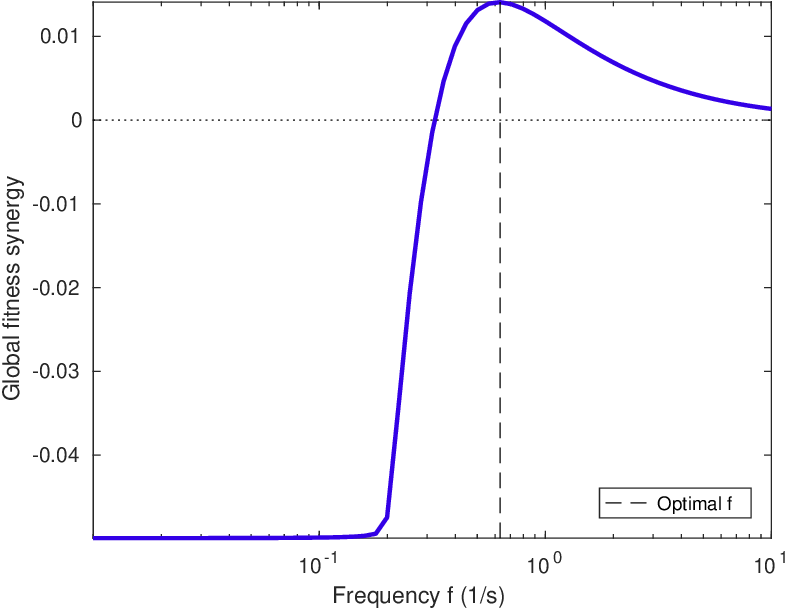}&
                                                                                                                                                                                              \raisebox{3mm}
    {\includegraphics[width=3cm]{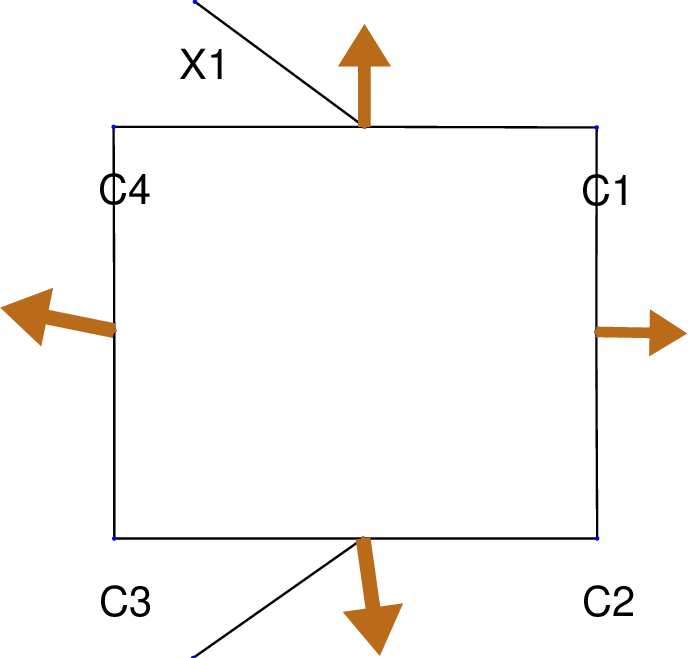}
        \hspace{2mm}
        \includegraphics[width=3cm]{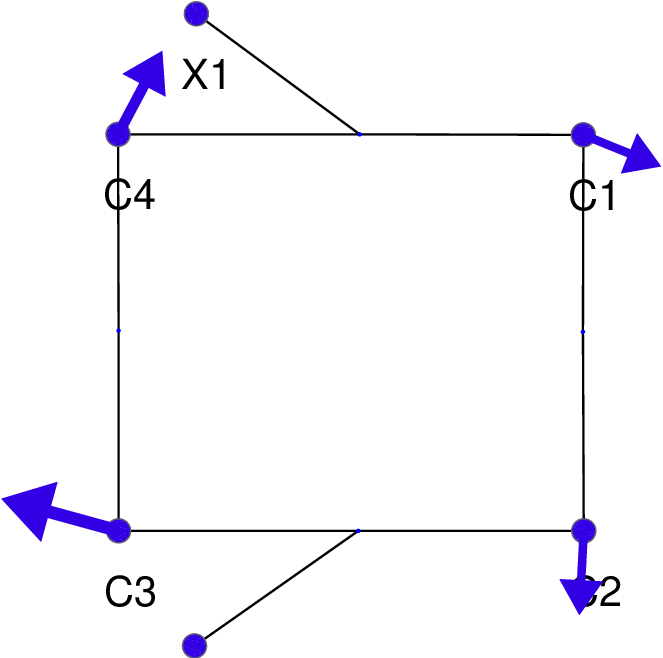}}
      \end{tabular}
      \caption{\co{axis font groesser} Self-promoting rhythm in a cyclic
        pathway.  (a) Network structure and reference state. A circle
        flux is driven by an inflow from metabolite X1 and an outflow
        to metabolite X2.  (b) Principal synergy (i.e.~maximal fitness
        curvature) as a function of frequency.  The local maximum with
        a positive value shows an incentive for self-promoting
        oscillations.   (c) Enzyme and metabolite rhythm.  The
        optimal self-promoting enzyme profiles (brown arrows) are
        phase-shifted along the pathway. Blue arrows show the
        metabolite rhythm.}
 \label{fig:metabolicLoop}
 \end{center}
\end{figure*}

\myparagraph{Example: linear chain with dilution} Self-promoting
rhythms may involve the entire metabolic network like the metabolic
cycle in yeast.  However, to see the reasons for self-promoting
oscillations more generally, let us consider a simple examples.  In
the pathway in Figure \ref{fig:linearChain}, a given substrate level
and optimally adapted enzyme activities define a static reference
state.  A quasi-static substrate increase would promote, i.e.~favour,
an increase in enzyme activities.  What behaviour will periodic
substrate rhythms promote?  At constant enzyme activities, a substrate
rhythm would lead to damped concentration and flux waves along the
pathway (Figure \ref{fig:linearChain} (a)).  But if a wave-like enzyme
rhythm is added\footnote{In this model, enzyme activities can vary
  fast, with a characteristic time of 1 s. This unrealistic assumption
  will be given up below (see Figure
  \ref{fig:linearChainPostTrans}).}, orchestrated with the substrate
rhythm, the average flux can be increased by allosynchrony. In other
words: the substrate oscillation \emph{promotes} a wave-like enzyme
rhythm (Figure \ref{fig:linearChain} (b)).  Can an enzyme rhythm
alone, without substrate rhythm, be beneficial?  In the example, the
principal synergy is negative at all frequencies, so there is no
incentive for spontaneous enzyme waves.  However, this changes if the
intermediate metabolites are non-enzymatically degraded\footnote{Fast,
  non-enzymatic metabolite degradation is not very likely in cells.
  However, some relevant processes can be described in this way:
  diffusion of gaseous compounds such as H$_2$S or acetaldehyde
  through the cell membrane; \co{example from yeast met osc} dilution
  (relevant for molecules with a slow turnover rate compared to
  dilution rates, e.g.~DNA); and chemical damage by free radicals
  (e.g.~during DNA elongation)}. \co{im moment dilution mu =0.1;
  aendern?}  In a model with uncontrollable degradation, spontaneous
enzyme rhythms have the potential to increase the average flux at
constant external substrate level and are therefore beneficial. (Figure
\ref{fig:linearChainDil}).  As a variant of the model, we consider a
loop-shaped pathway with input and output compounds (Figure
\ref{fig:metabolicLoop}).  Also here, self-promoting enzyme waves can
be beneficial: the principal synergy (Figure \ref{fig:metabolicLoop}
(b)) has a positive maximum at a finite frequency, caused by a dynamic
resonance of the metabolic system, while slower or faster
self-promoting oscillations lead to a lower benefit.  Regulation
systems that generate enzyme rhythms at the optimal frequency will
provide a selection advantage.  self-promoting oscillations can be
seen as an example of spontaneous pattern formation, that is, a
pattern formation in time \co{cite temporal crystals?} that is not
based on dynamics alone, but on incentives and optimal choices (for
details, see \url{www.metabolic-economics.de/enzyme-rhythms/}).

\iftoggle{bookversion}
{\section{Enzyme rhythms that combine transcriptional  and posttranslational regulation}}
{\subsection{Enzyme rhythms that combine transcriptional and posttranslational regulation}}
\label{sec:AmplitudeConstraints}

\co{say: particularly important bcs transcript wave in yeast can create almost no enzyme levels changes!}

\myparagraph{\ \\Optimal enzyme rhythms under constraints} We saw how
optimal enzyme rhythms can be computed from synergies given in the
matrices $\Fetxt$ and $\Fetet$. However, so far we ignored all
possible side constraints.  These constraints may prevent, for
example, that enzyme amplitudes exceed the enzyme's average value
(which would lead to negative values), a problem that may occur if an
enzyme is inactive in the reference state and starts oscillating.  For
example, storage reactions are useless in a strictly static
environment: a steady conversion of glucose into glycogen and back
would be futile and the reactions should be inactive. \co{show example
  in figure? ja, waere gut! extraabschnitt nach diesem abschnitt!}
However, in a periodic environment, storage and release of glycogen
may buffer the fluctuation inside the cell and provide a relatively
constant glucose supply, which can be beneficial.  Finally, amplitude
constraints are also important when desired amplitudes cannot be
realised by gene expression at higher frequencies (see SI
\ref{sec:geneexpression}). What are the relevant constraints for a
profile
$\esymbolv(t) = \evb + \Delta \evb + \real(\e^{i\,\omega\,t}\,\esymbolvt)$?
Bounds on the amplitudes $\esymbolvt$, and constraints between $\evb$ and
$\esymbolvt$, are obtained as follows.  First, to prevent negative enzyme
activities, the shifts $\Delta \ub_{l}$ must not go below $-\ub_{l}$
and the amplitude $|\et_{l}|$ must not exceed the average level
$\ub_{l} + \Delta \ub_{l}$. This means that the average level of any
oscillating enzyme must be positive. Moreover, if enzyme rhythms are
caused by rhythmic mRNA profiles, the enzyme amplitudes at high
frequencies become very small and, conversely, there is a bound on the
possible enzyme amplitudes (see appendix \ref{sec:computingoptimal}
and SI \ref{sec:geneexpression}).  If we impose such constraints, the
solutions $\ub_{l}$ and $\et_{l}$ from Eqs (\ref{eq:optstaticvector})
and (\ref{eq:optperiodic}) may not be valid any longer.  Instead, we
obtain the optimality problem
\begin{eqnarray}
\label{eq:OptimalityProblem}
 \mbox{Maximise}\;\;\Delta F(\devb, \esymbolvt) \quad \mbox{subject to} \quad
 |\esymbolvt| \le \pvt^{\rm max}(\evb, \omega)
\end{eqnarray}
with the bound $\devb \ge - \evb$, where $\evb = \evref+\devb$.  The
vector $|\esymbolvt|$ contains the absolute values from $\esymbolvt$ and
$\pvt^{\rm max}(\evb, \omega)$ describes some frequency-dependent,
real-valued amplitude bounds as described above.  If a solution
$(\devb, \esymbolvt)$ satisfies all constraints, the constraints can be
ignored and we can use the simple formula
Eq.~(\ref{eq:simpleOptimalityProblem}). This typically holds for
adaptive rhythms with small perturbation amplitudes $\xvt$, and
assuming that all enzymes were active in the reference state.  In all
other cases, i.e~if constraints are active, they affect the solution;
if an enzyme hits a constraint, there will be secondary effects on
other enzymes and eventually on the entire optimal state (see SI
\ref{sec:RhythmConstraintsLagrange}).

\begin{figure*}[t!]
 \begin{center}
   \begin{tabular}{ll}\\
     \parbox{5cm}{ (a) Amplitude constraints \\[2mm]
       \includegraphics[width=4.5cm]{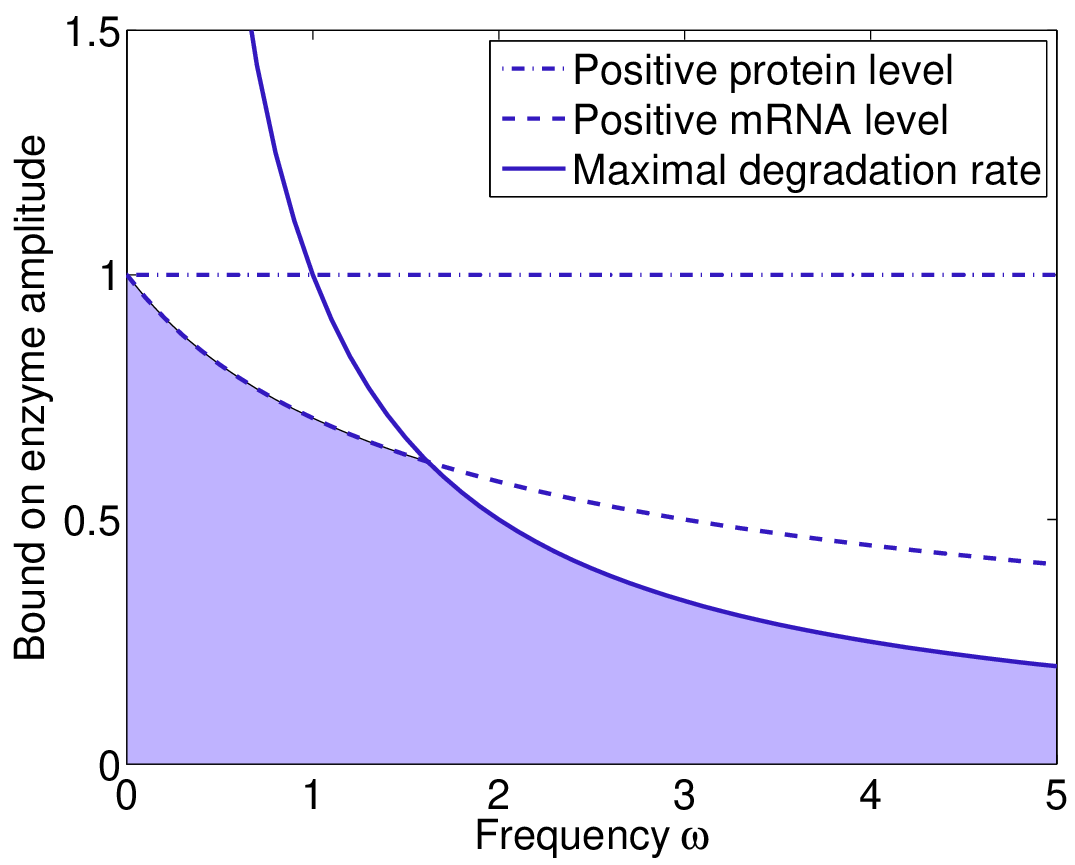}\\[2.7cm]}&
     \parbox{11cm}{\includegraphics[width=11cm]{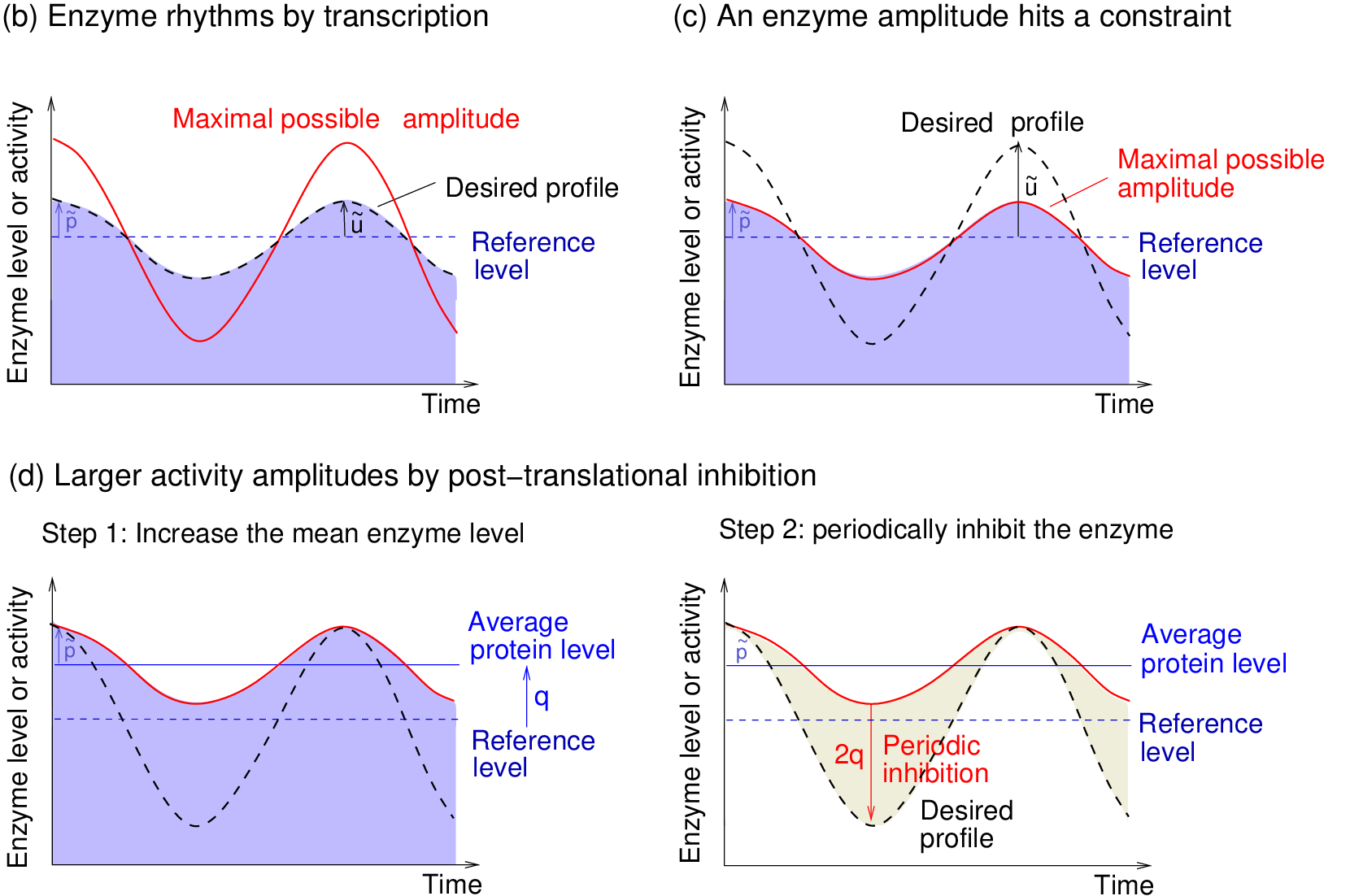}}
   \end{tabular}
   \caption{\co{a axis labels groesser} Bounds on enzyme amplitudes
     create an incentive for posttranslational regulation.  (a) Bounds
     on protein amplitudes (with average concentration $\pb$ and
     protein degradation constant $\kappa = 1$). The blue region shows
     possible protein amplitudes
     $\tilde p_l^{\rm max}(\omega)/\bar p_l$ at different frequencies.
     (b) In this case, an enzyme activity profile $u(t)$ respecting
     the amplitude constraints can be realised by periodic expression.
     (c) If the desired amplitude is larger, the profile cannot be
     realised by periodic gene expression alone, and posttranslational
     regulation needs to be added to achieve the desired
     amplitude. (d) To increase the activity amplitude while keeping
     the average activity unchanged, a cell may increase the average
     enzyme concentration (left) and periodically inhibit the enzyme
     by posttranslational modification (right). The inhibition
     amplitude and the increase in average enzyme level, needed for
     compensation, are both given by $q=|\et|-\pt^{\rm max}$, the
     difference between desired activity amplitude and allowed protein
     amplitude. The extra cost for the higher protein level is
     proportional to the inhibition amplitude and can be regarded as a
     cost of the periodic inhibition.}
 \label{fig:amplitudeConstraint}
 \end{center}
\end{figure*}

\myparagraph{Rythms based on posttranslational modifications} \co{hier
  oder discussion?  interesting to mention as "non-optimality" case in
  oscillations: (or alternative to phosphorylation, proposed in opt
  enz) Rhythmic Degradation Explains and Unifies Circadian
  Transcriptome and Proteome Data Sarah Lück, 1,2 Kevin Thurley, 1,2
  Paul F. Thaben, 1 and Pål O. Westermark 1, Lück et al., 2014, Cell
  Reports 9, 741–751 [in verzeichnis literature] what could be the
  advantage of this?  was sind die kosten? evl noch teurer als
  inhibition by phosphorylation!  } \co{where i say that expression
  response is slow (oder in discussion?): cite Bennett MR, Pang WL,
  Ostroff NA, et al. Metabolic gene regulation in a dynamically
  changing
  environment. Nature. 2008;454(7208):1119–1122. doi:10.1038/nature07211;
  from abstract: ``We find that the metabolic system acts as a
  low-pass filter that reliably responds to a slowly changing
  environment, while effectively ignoring fluctuations that are too
  fast for the cell to mount an efficient response.''
  (zweckoptimismus!) } Until now, we assumed that changes in enzyme
activities are caused by changing enzyme amounts that is, by changes
in gene expression. However, at high frequencies gene expression
changes alone will not suffice to realise large enzyme amplitudes,
because the dynamics of transcription and translation will damp any
changes. To generate larger enzyme activity amplitudes, a cell could
periodically inhibit its enzymes by posttranslation modification, such
as phosphorylation or acetylation\footnote{The same effect can be
  obtained by allosteric regulation, but in our models, allosteric
  regulation is included in the rate laws and thus modelled as a
  mechanistic fact rather than a feature to be optimised. However,
  this is just a modelling choice. To optimise allosteric regulation,
  we could treat in the same way as regulation by posttranslational
  modifications.}. What is the cost of such posttranslational
regulation mechanisms? And how should the two mechanisms be combined?
With expression changes alone, and with a linear enzyme cost function
(or a state-average fitness functional), the cost of a periodic
enzyme profile depends directly on the average enzyme levels, and the
periodic fluctuations around it are cost-neutral.  Posttranslational
activity changes, in contrast, allow us to bypass the amplitude
constraint, but now the enzyme rhythms are, effectively, costly.  We
can see this as follows: if a desired enzyme amplitude $|\et|$ exceeds
the maximal possible protein amplitude $\pt^{\rm max}$, the difference
$|\et|-\pt^{\rm max}$ can only be realised posttranslationally,
i.e.~by increasing the average enzyme level and applying rhythmic
posttranslational inhibition. In this way, the desired amplitude is
realised while keeping the average activity unchanged (see Figure
\ref{fig:amplitudeConstraint}). The extra cost is proportional to the
amplitude difference.  To account for this cost in a numerical
optimisation, we can introduce an auxiliary variable $q$ for each
enzyme, describing the difference between desired enzyme activity
amplitude and maximal protein amplitude: this amplitude difference is
the amount by which the average enzyme concentration must be increased
(see appendix \ref{sec:extending}).  The resulting model allows for
fast high-amplitude oscillations, but since enzyme modification
rhythms are costly, they need to be beneficial!  In Figure
\ref{fig:linearChainPostTrans}, we consider again the linear pathway
from Figure \ref{fig:linearChain} with optimal enzyme rhythms realised
by expression changes and posttranslational regulation. As expected,
slow enzyme rhythms (on the time scale of hours) are realised by
expression changes, while fast oscillations are mostly realised
posttranslationally.  Notably, expression and posttranslational
modification rhythms must be in phase (i.e.~posttranslational
inhibition should be strongest when enzymes expression is low) because
otherwise enzyme resources are wasted.

\begin{figure*}[t!]
\begin{center}
\begin{tabular}{l}
(a) Enzyme amplitudes (assuming that fast activity changes are possible) \\[1mm]
  \includegraphics[width=16cm]{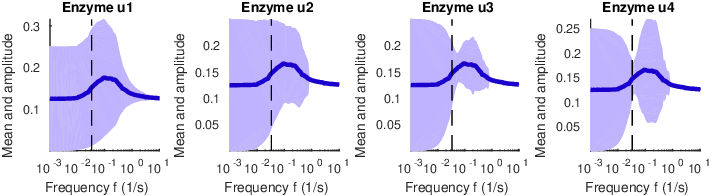}\\[3mm]
(b) Enzyme amplitudes (combination of slow gene expression changes and fast covalant modification)\\[1mm]
  \includegraphics[width=16cm]{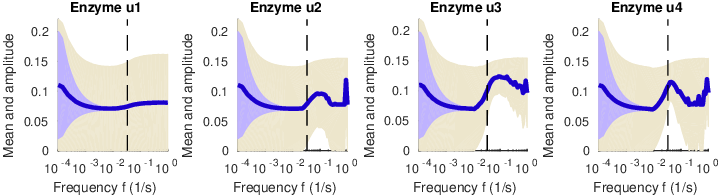}
\end{tabular}
\end{center}\ \\[-10mm]
\caption{\co{ueberschriften aus matlab nicht dick!}  Optimal enzyme
  rhythms, realised by gene expression and posttranslational
  regulation. (a) Enzyme amplitudes in a linear pathway (see Figure
  \ref{fig:linearChain}) as a function of frequency
  $f = \omega/(2\, \pi)$. In the upper panels, we make the
  (unrealistic) assumption that any enzyme rhythms can be realised
  transcriptionally, even at high frequencies. Each plot shows to one
  of the enzymes.  Curves show the average enzyme activities; blue
  bands indicate amplitudes (phase angles not shown). The dashed line
  denotes the oscillation frequency $f = 0.25\, \mbox{s}\inv$ from
  Figure \ref{fig:linearChain}.  (b) In the lower panels we assume
  that gene expression can only generate limited amplitudes and that
  higher amplitudes need to be realised by additional protein
  modification. Blue areas show expression amplitudes, grey areas show
  amplitudes by posttranslational modifications. For optimality
  reasons, both oscillations must be in phase (see Figure
  \ref{fig:amplitudeConstraint}).  Their total enzyme activity
  amplitude, is shown by the outer envelope. Colours correspond to
  amplitudes in Figure \ref{fig:amplitudeConstraint}.  Even a simple
  model like this shows complicated regulation profiles.  At
  low-frequency perturbations, the enzymes oscillate in phase with the
  external perturbation (phases not shown); at high frequencies,
  enzymes 3 and 4 oscillate in opposite phase. At intermediate
  frequencies ($\approx$ 0.5 \,s$\inv$ for enzyme 3, or $\approx$ 0.1
  \,s$\inv$ for enzyme 4), the phases of these two enzymes flip (not
  shown) and their amplitudes become very small. \co{mention or fix
    numerical errors}}
  \label{fig:linearChainPostTrans}
\end{figure*}

\myparagraph{Amplitude constraints and their consequences} To
summarise, constraints on enzyme amplitudes leads to more realistic
predictions of enzyme rhythms and to new types of predicted behaviour.
Constraints are especially important if enzymes are not (or weakly)
expressed in the reference state or if there is an incentive for fast
oscillations (that is, oscillating much faster than the typical time
scale of protein dynamics, which for growing cells is typically given
by the cell-cycle period).  In the first case, an increasing enzyme
amplitude may require an increase in the average value. For example,
if an enzyme in the reference state is inactive (e.g.~the enzymes
catalysing glycogen storage and consumption), then in order to
oscillate, it must increase its average level from zero to a value as
big as the enzyme amplitude. This increase is costly (see appendix
\ref{sec:extending}), so it needs to be justified by a benefit.  In
the second case, protein amplitudes may be limited due to the slow
dynamics of protein concentrations. At higher frequencies, gene
expression changes will lead to very small amplitudes, and
posttranslational modification (e.g.~phosphorylation) have to be used
instead.

\section{Design principles for enzyme rhythms}

\subsection{Optimal rhythmic  enzyme profiles reflect pairwise synergies}
\label{sec:RhythmsAreShaped}

\begin{figure*}[t!]
\begin{center}
\fbox{\parbox{16.5cm}{
    \begin{center}
      \small 
\parbox{15.5cm}{
  \textbf{Box 2: Dynamics and economics of enzyme rhythms\\[2mm]}

  \textbf{Metabolic dynamics}\\
  
  Metabolic models with constant enzyme levels can describe the
  mechanistic effects of external perturbations (e.g.~external
  metabolite concentrations).  If we start from a stable steady state and
  assume a (static or periodic) parameter perturbation, the system's
  response -- the change of steady-state concentrations and fluxes --
  can be approximated with the help of metabolic control or response
  coefficients.  Metabolic steady states can be dynamically unstable:
  if a parameter change destabilises a stable state, this is called a
  dynamical bifurcation.
  \co{formel fuer matrixinversion zeigen!}\ \\

  \includegraphics[width=15cm]{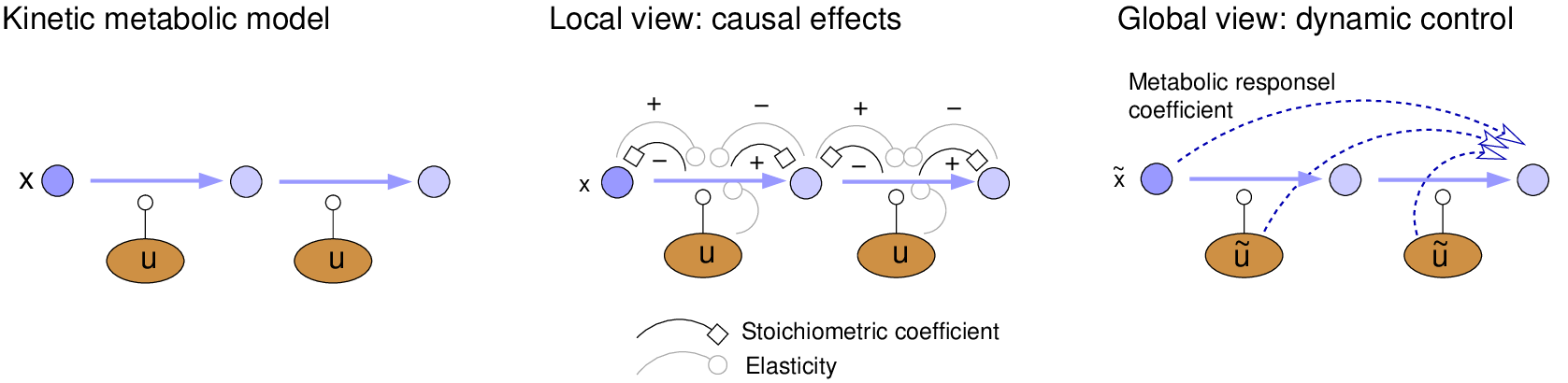}

  \co{``Local causal effects'' ``Network-wide metabolic control''}
  Left: A metabolic pathway described by reactions, metabolites, and
  enzymes.  Centre: In a kinetic model, metabolite concentrations determine
  reaction rates, and reaction rates lead to changes of metabolite
  levels. The causal connections arise from elasticities $E_{li}$
  (from compounds to reactions) and stoichiometric coefficients
  $n_{il}$ (from reactions to compounds), ginving rise to the Jacobian
  matrix $\Mmat = \Nmat\,\Emat$ for metabolite concentrations. Right:
  Metabolic response coefficients translate local perturbations (of
  enzyme activities and external substance levels)
  into global responses of all state variables (metabolite concentrations or fluxes).\\[2mm]

  \textbf{Enzyme economics}\\
  
  On top of metabolic dynamics, we may consider optimal enzyme
  adaptations, described by optimality problems for enzyme
  levels. Starting from a (dynamically and economically stable) state,
  the optimal adaptation to small perturbations can be described by
  adaptation coefficients (assuming a linear approximation).  \co{[def
    in SI! REF here!]}  An optimal steady state may be economically
  unstable against enzyme oscillations, and  transitions from
  economically stable to an economically unstable states can be
  described as economic bifurcations.  \co{explain economic
    bifurcation?}
  \co{formel fuer matrixinversion zeigen!}\ \\
  
  \includegraphics[width=15cm]{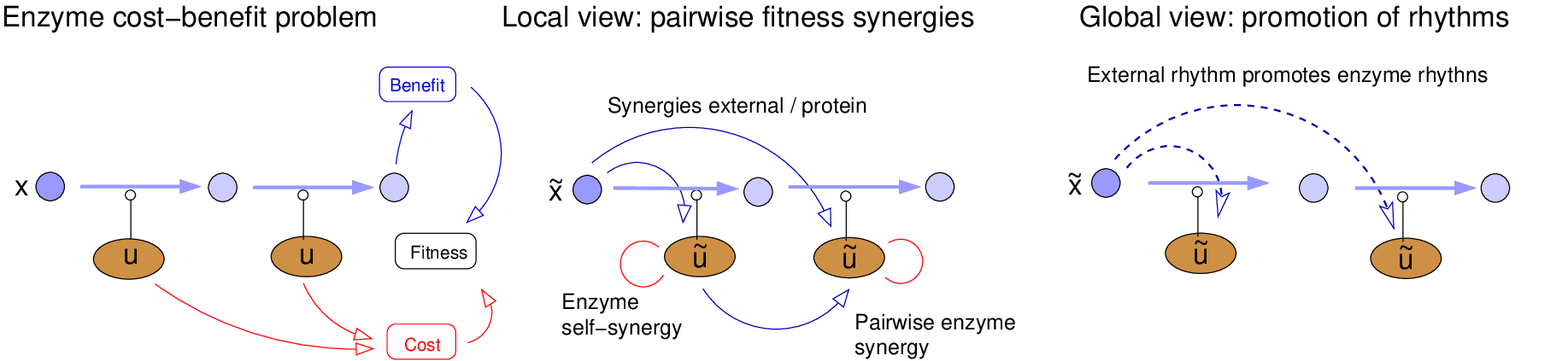}\\

  \co{Cost-benefit problems for enzymes; Pairwise fitness synergies;
    Network-wide view: promotion of enzyme rhythms} Left: Enzyme
  profiles lead to {\metabolicobjective} (via the state variables) and
  enzyme cost.  Centre: Fitness effects of rhythmic perturbations. In
  a second-order approximation, fitness depends on pairwise synergies
  between oscillating parameters (blue arrows), where arrow heads
  denote beneficial phase relations (pointing towards the element that
  should peak later).  Negative self-synergies are shown by red
  arcs. Right: Under the optimality principle, an external rhythm
  promotes a network-wide enzyme rhythm (i.e.~it provides an incentive for a
  rhythm with specific amplitudes and phases). In the example, the
  second enzyme will show a smaller amplitude and a larger phase
  shift. The adaptive enzyme profile is shaped by metabolic network
  structure, {\metabolicobjective} function, and the specific
  perturbation applied.

}
\end{center}}}
\end{center}
\end{figure*}

\begin{figure*}[t!]
  \begin{center}
    \begin{tabular}{llll}
      (a) Network and fluxes & 
      (b) Synergies on network & 
      (c) Synergy phase plot \\[2mm] 
      \raisebox{.5cm}{\includegraphics[width=3.2cm]{\psfilesrhythms/four_cycle_speedy_network_external.eps}}&
      \includegraphics[width=3.9cm]{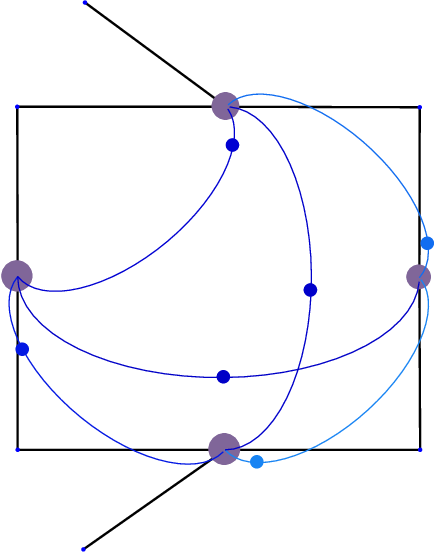}&
      \raisebox{.8cm}{\includegraphics[width=3.6cm]{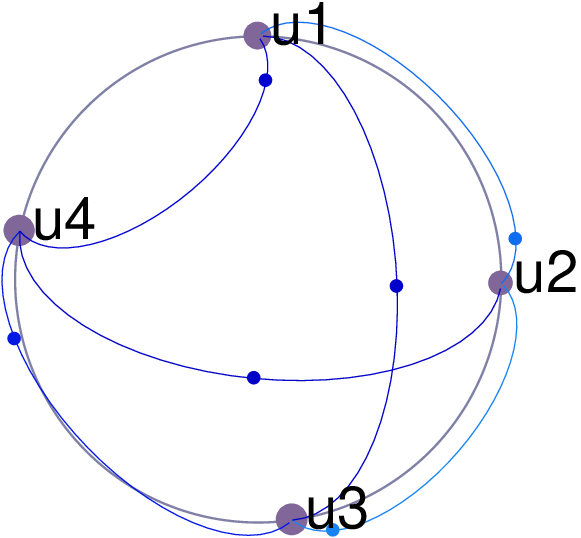}}\\[-1cm]
    \end{tabular}
  \end{center}
  \caption{\co{optimal solution (with arrows), wiederholung von abb 6c
      (enzym)} Synergies in a loop-shaped pathway (see Figure
    \ref{fig:metabolicLoop}).  (a) Network structure and reference
    flux (arrows).  (b) Synergies between enzyme rhythms.  Enzymes are
    shown as nodes, enzyme synergies as arcs with dots.  Absolute
    values are indicated by arc colors (dark: large); phase shifts are
    marked by dots.  Dots do not show actual oscillation phases but
    \emph{desired} phase shifts that would maximise fitness. A dot's
    position on the arc represents the phase shift in radians (units
    of $2\,\pi$): a dot near the end of an arc indicates a small phase
    shift and shows that the nearby node should peak after the more
    distant node (e.g.~the arrow head between $u_1$ and $u_2$ shows
    that $u_1$ should peak slightly before $u_2$). A dot in the middle
    of an arc indicates that two enzymes should have opposite phases
    (like, e.g.~$u_1$ and $u_3$).  Self-synergies are not shown. (c)
    Synergy phase plot. Enzymes are placed on a circle, indicating
    their optimal phases. Arcs are drawn as in (b).}
 \label{fig:synergy_plot}
\end{figure*}

\myparagraph{\ \\Reasons for rhythmic synergies} We saw how optimal
enzyme rhythms in a given metabolic model can be computed.  The
formulae are summarised in the appendix and examples are shown at
\url{www.metabolic-economics.de/enzyme-rhythms/}.  However, can these
formulae also explain general phenomena such as coregulation of
enzymes or their sequential activation in pathways?  To understand
optimal enzyme profiles, and how they are shaped by fitness
objectives, external conditions, metabolic dynamics, and enzyme
constraints, numerical simulations alone are not enough -- we need
general laws that explain how enzyme rhythms -- assuming
small-amplitude, sine-wave oscillations -- are shaped by fitness
synergies and constraints on individual enzyme amplitudes.  There are
three types of synergies: environment/enzyme synergies ($\Fxtet$),
enzyme/enzyme synergies ($\Fetet$ off-diagonal elements), and enzyme
self-synergies ($\Fetet$ diagonal elements). \co{JA! ENTWURF SIEHE
  LETZTER AUSDRUCK! draw pictures of the four possibilities explained,
  and move text in figure caption!} \todo{Synergies often reflect the
  fact that the actions of two parameters converge towards a common
  objective.} They can emerge in different ways, for example: (i)
\emph{Synergy via supply}.  A rhythm (in the environment or of an
enzyme) leads to oscillations in a metabolite somewhere in the
network.  Since an enzyme that consumes this metabolite becomes more
efficient in certain moments in time, there is an incentive to
synchronise enzyme and substrate, thus coordinating the enzyme with
the original oscillating parameter with the right phase delay.  (ii)
Synergy through a common target: for example, two enzymes have control
over the two substrates of a reaction, creating a synergy effect;
different ``temporal distances'' from the target reactions lead to
synergies with different phase shifts.\co{generally define ``phase
  distance'' between processes, according to periodi control coeffs!
  compare to delay equations!  und: variant of (ii), with ext par +
  enzyme!)}  (iii) \emph{Synergy via avoidance.}  Metabolic rhythms
can contain time windows in which certain processes would be
inefficient or even dangerous (for instance, phases in which reactive
oxygen species (ROS) levels are high, which makes DNA replication
problematic).  These processes should be shifted to other phases. (iv)
\emph{Synergy via cost}. If we assume a fixed total enzyme amount in
each moment, then upregulating an enzyme implies a simultaneous
downregulation of others.  Similarly, in models with nonlinear enzyme
cost functions, the upregulation of an enzyme may increase the cost
pressure on other enzymes, creating an incentive for their
downregulation.  In both cases, a rhythm in one enzyme will promote
rhythms in others, typically rhythms of opposite phase. Of course,
these are just examples: there are many other scenarios that lead to
rhythmic parameter synergies, and therefore promote a network-wide rhythmic
dynamics. \co{note that all these reasons also apply to ext/enyme
  synergies and ext/ext synergies}

\co{WO? Examples like in text (i) supply coupling:
  delay/synchronisation example; (ii) cost coupling example; keine
  formeln, aber bilder! // Rhythms coupled through downstream
  reactions and enzyme costs. KUERZEN; SORT // show more of the small
  examples from website; EIN GUTES BSP FUER JEDES PRINZIP; SHOW IN SI}
\co{(moegliche erklaerung fuer negative kopplung: gesamtproteinmenge
  is stabiler als einzelne proteinmengen!)}
  
\myparagraph{From enzyme synergies to optimal enzyme profiles} Optimal
enzyme rhythms as a whole must be self-consistent: each enzyme curve
must be optimally adapted not only to external substrate rhythms, but
also to all other enzyme curves and to the resulting metabolite
rhythms. If all the synergies are known, how can we determine the
optimal pattern of oscillations?  What are the optimal amplitudes and
phase shifts of all enzymes?  A network-wide enzyme rhythm entails
positive and negative synergies between all pairs of variables. Each
periodic synergy defines a optimal magnitude and an optimal phase
shift, according to this single synergy term. Knowing the actual
parameter amplitudes and phase shifts in given periodic state, we can
translate this into positive or negative, strong or weak synergy
effects.  The optimal network-wide enzyme rhythm is determined by the
sum of all these effects: maximising this sum (see Box 2) requires an
optimal compromise, possibly under constraints, between the synergy
terms.  Computing an optimal network-wide rhythm from known pairwise
synergies is like computing a network-wide dynamic state from known
local interactions in a dynamical system.  If a dynamical system,
e.g.~a kinetic metabolic model, is perturbed by external oscillations,
the shape of the resulting internal oscillations depends on the
effects of the external perturbations and on the system's internal
dynamics, a dynamics that also becomes visible, e.g.~in damped
oscillations after a short perturbation.  In dynamical systems, global
dynamics arises from local interactions. Mathematically, the local
interactions between metabolites are described by a sparse Jacobian
matrix $\Amat$, and the global behaviour following a perturbation
$\devb$ (e.g.~propagating waves or steady-state changes) can be
derived from this matrix by matrix inversion\footnote{For a linear
  dynamical system
  $\frac{\md}{\md t}\dxvb=\Amat\,\dxvb + \Bmat\,\devb$, control theory
  provides some standard solutions: a static response, where
  $\frac{\md}{\md t}=0$, a perturbation $\devb$ leads to a static
  response $\dxvb=-\Amat\inv \Bmat\,\devb$. In contrast, a periodic
  perturbation $i\omega\dxvt=\Amat\,\dxvt + \Bmat\,\devt$ yields a
  periodic response $\dxvt=-(\Amat-i\omega\Imat)\inv\Bmat\,\devt$.} In
kinetic models, enzymes, metabolites, and reactions are directly
linked by stoichiometric coefficients and reaction elasticities, which
form the Jacobian matrix.  The \emph{inverse} Jacobian, a non-sparse
matrix, determines the periodic response coefficients, which describe
the global behaviour (including the propagation of perturbations).
The transition from local interactions to network-wide behaviour also
exists in enzyme optimisation: we start from pairwise fitness
synergies, the ``local'' economic effects between single enzymes
(which are, nevertheless, outcomes of the ``global'', network-wide
metabolic dynamics). Then we compute the optimal enzyme profile,
i.e.~the orchestrated pattern of all enzyme curves, by a matrix
inversion.

\myparagraph{Synergy plots} Synergies and optimal phase shifts in
metabolic systems can be shown in a \emph{synergy plot} (see Figure
\ref{fig:synergy_plot}). In this plot, oscillating parameters
(external metabolite and enzyme activities) are displayed on the
network and pairwise synergies are represented by arcs. Similarly, in
a \emph{synergy phase plot} the enzymes are arranged in a circle,
sorted by their optimal phase shifts. While these phase shifts do not
maximise each of the synergy effects individually, at least each
enzyme peaks in the optimal moment, given the other enzyme curves. In
a ``good'' metabolic cycle, arcs point in clockwise direction,
indicating that the phase shifts agree with the phase shifts promoted
by individual synergy terms.

\myparagraph{Optimal orchestrated behaviour as a series of
  adaptations} How can we understand or compute optimal network-wide  enzyme
rhythms?  It is hard to predict them based on intuition alone. One
reason is their required self-consistent behaviour, which is much
harder to depict mentally than simple cause and effect. A way to
understand self-consistent enzyme profiles is to view them as the
result of a hypothetical planning process in a which we start with a
first tentative solution which is then iteratively refined. An
external parameter oscillation causes metabolic oscillations in the
network. We can first naively assume that each enzyme level is adapted
``locally'' to its periodic substrate and product levels, as described
in section \ref{sec:localeffects}. This yields a periodic profile for
each enzyme.  Of course, these enzyme profiles will cause a change in
the metabolite profiles, these require further enzyme adjustments, and
so on. If we consider this infinite number of adjustements, and if
their infinite sum converges, we obtain a self-consistent, optimal
enzyme rhythm. Mathematically, the matrix inversion\footnote{In
  metabolic control theory (see Box 2), control matrices (for global,
  long-term influences) are computed from the Jacobian matrix (for
  local, short-term interactions) by a similar matrix inversion.} in
Eqs (\ref{eq:optstaticvector}) and (\ref{eq:optperiodic}) is an
effective way to compute this infinite sum (see SI
\ref{sec:SIiterative}). The same idea -- describing a self-consistent
enzyme adaptation strategy as an infinite sum of adjustments -- can be
applied to compute self-consistent static adjustments \cite{lksh:04}.

\myparagraph{Enzyme rhythms in time and frequency space} \co{move to
  SI?} Generally, oscillations can be described by changes in time or
by modes in frequency space. \co{Discuss the perspectives in time and
  in Fourier components: what do we learn?  jetzt mehr intuitiv!}
\co{We saw that describing the propagation of periodic perturbation in
  time, based on a causal reasoning, is difficult, and if we turn the
  problem around and consider optimal control rather than causal
  effects, things become even more complicated.}  In time, the most
simple perturbations are peak-like perturbations of single
reactions. In a linearised metabolic dynamics, their propagation is
described by the pulse-response function. The effects of general
time-dependent perturbations can be described by convolution
integrals.  In this convolution, all the propagating effects are
overlaid. If the perturbation itself is a sine wave the result (in a
linear approximation) is a simple sine-wave dynamics as we saw
before. \todo{Assuming sine waves as an ansatz for enzyme rhythms,
  allows us to describe rhythms simply by amplitudes and phases.  In
  Fourier space, this corresponds to a multiplication of
  Fourier-transformed perturbation function with the
  Fourier-transformed pulse-response function.} \co{WD: We can
  describe this directly in frequency space (where, instead of
  computing a convolution, we simple multiply the perturbation
  amplitude with the periodic response coefficients, the Fourier
  transforms of the pulse-response function.} Thus, we can see the
metabolic system as a filter, e.g.~a low pass filter that damps
high-frequency oscillations and lets slow oscillations and static
changes pass through \cite{inga:04,lieb:2005}. \co{die refs weiter
  nach oben?}

\myparagraph{Computing non-sine-wave rhythms by Fourier synthesis} For
non-sine-wave perturbations, dynamic responses can be obtained by
Fourier synthesis \cite{lieb:2005}.  Based on our sine-wave approach,
the dynamic (linearised) response can be computed by Fourier synthesis
as an infinite sum over sine-wave responses of different
frequencies. To do so, the input oscillations are approximated by a
Fourier series of sine-wave oscillations with frequencies
$0, \omega, 2\,\omega, 3\,\omega, ..$.  In practice, considering only
the lowest harmonics often yields a good approximation. So how can we
predict optimal adaptation in time? With our quadratic approximation
Eq.~(\ref{eq:deltafexpansion}) (symmetric against time shifts), and
without constraints, the optimal enzyme amplitudes
Eq.~(\ref{eq:optperiodic}) are linear in the perturbation amplitudes,
and independent between different frequencies: these solutions are
completely additive.  If desired, sine-wave adaptations to different
sine-wave perturbations can be linearly combined, so we can apply
Fourier synthesis: we split the perturbation profile into Fourier
components, compute the respective optimal adaptations, and sum over
them to obtain the optimal adaptation to the original, non-sine-wave
perturbation (see SI \ref{sec:HigherHarmonics}). \coout{say this in
  FN?: A practical application here, for optimal ryhthms: think about
  higher harmonics, show proof (ideen von blatt aufschreiben)
  momentane idee: in der entwicklung (2.ordnung) um den ungestörten
  zustand (und unter bedingungen für zeittrnslinvarianz) sind die
  fitnessterme für verschiedene frequenzen ungekoppelt: d.h., ohne
  constraints und mit aeußerer störung (für mehrere frequenzen!)  kann
  alles pro frequenz optimiert und dann aufsummiert werden. hat die
  äußere störung nur eine frequenz, gibt es auch nur bei dieser
  frequenz eine adaptation.  mit constraints muss alles zusammen unter
  constraints optimiert werden.  (was mit den delta-peaks los ist, ist
  mir nicht klar .. vielleicht ist das ein effekt von mehr als
  2. ordnung?)}  \co{beispiel für nicht-harmonische loesung zeigen?
  verschiedene frequenzen: enzyme können folgen / können nicht folgen!
  abschnitt ueber hoehere harmonische schreiben} \co{is that safe, or
  is a mathematical proof required?} In contrast, if our optimality
problem contains constraints (e.g.~on enzyme amplitudes), this may not
work: in this case, we need to solve Eq.~(\ref{eq:OptimalityProblem}),
whose solutions may not be additive. In this case, blindly applying
Fourier synthesis would lead to wrong results. \co{mention the
  mathematical connection between oscillations and dilution} \co{JA!!
  welche ergebnisse sind zu erwarten?  tiefpass-filter?)}  \co{say
  with Fourier synthesis we're still limited to the range of the
  Taylor approximation!}  \co{wie siehts aus mit fouriersynthese fuer
  spontane rhythmen? ergibt das ueberhaupt sinn in der approximation,
  und ohne weitere constraints?}  \co{modules as example?  forced
  oscillations; synchronisation of oscillations between fuzzy network
  oscillators} \co{clarify duality between dynamics and economics
  (write economics as quasi-dynamics; solve both at the same time)}

\subsection{Periodic economic potentials}

\todo{\myparagraph{\ \\Local incentives for enzyme rhythms} Optimal
  metabolic cycles require a self-consistent choice of the enzyme
  profiles, with the right amplitudes and well-synchronised peaks in
  time. While such arrangements look plausible once we see them, they
  may be hard to predict from intuition alone. The calculation via a
  matrix inverse in Eq.~(\ref{eq:optperiodic}) is possible, but hard
  \todo{to grasp intuitively}, and the entire network needs to be
  known. In our one-reaction example (see Box 1), things were much
  easier: the optimal enzyme profile was directly obtained from a
  \emph{local} variable, the time-dependent catalytic rate, which
  results from the known substrate and product profiles.  Can we find
  a similar local description for the entire network, one in which
  each enzyme adjusts is locally adjusted to substrate availabilty and
  product demand, described as ``economic values''?  Of course, these
  values would change periodically and would be coupled across the
  network. For example, if the product P of a pathway is useful in a
  certain cell cycle phase, it should peak in this phase.  Therefore,
  the producing enzyme, and the enzyme's substrate, should also
  oscillate, but peak a bit earlier. If we continue with this
  reasoning, \todo{and translate the ``should'' again into values}, we
  find that each metabolite, and each enzyme, has a value that
  oscillates in time, and that all these values are interdependent. If
  the periodic values of metabolites (i.e.~the incentives to produce
  them) are known, we may expect that an enzyme level should peak when
  substrate is cheap and product is pricy. With this concept, enzymes
  would adapt themselves locally, not to the concentrations but to the
  values of their reaction substrates and products, showing the
  highest investments when the value difference is large.  Can we
  translate this idea into mathematical definitions and laws?  This is
  in fact possible.  Metabolic value theory   \cite{lieb:14a} translates
  network-wide fitness objectives into economic proxy variables
  (assigned to individual reactions, metabolites, and enzymes, and
  describes these variables by local balance equations. In
  \cite{lieb:18lagrange}, economic values for metabolite production,
  called economic potentials, have been defined for steady and
  periodic states. For details, see \co{appendix \co{?} and} SI section
  \ref{sec:AppPeriodicEconomicPotentials}. To describe oscillating
  states, we can consider complex-valued economic potentials, with
  phase angles describing when a metabolite's value is highest.}
\co{The periodic economic potentials characterise the marginal benefit
  of oscillations in individual metabolites. The difference in
  periodic economic potentials along a reaction is related to the
  periodic demand for the enzyme.} \co{The periodic economic
  potentials represent an idea of \co{note the paradox!}
  ``anticipatory just-in-time production''. In an optimised metabolic
  cycle, enzyme rhythms must be adapted to support processes that are
  distant both in the network and in time. This contradicts the idea
  of a ``just-in-time production'', where enzyme expression is
  directly triggered by substrate supply or demand for product. How
  can the two views be combined?  Could globally optimal enzyme
  rhythms be realised by direct feedback from local substrates and
  products? A theoretical possibility would be to implement such a
  mechanistic feedback based on economic potentials. \co{REF TO CBA
    regulation!}} \co{In reality, regulation by substrate and product
  levels; better: regulation by substrate and product values! (which
  can also anticipate fitness requirements in time!)} \co{wo? analogie
  zu dilution kurz erwaehnen.}

\co{\textbf{Periodic economic potentials and periodic economic balance
    equation} PROBLEM: HIER KOMMT KEINE EINZIGES BEISPIEL VOR! ES
  sollte zumindest eine abbildung und ein echtes beispiel geben. WENN
  das schon in CBA lagrange kommt, dann hier kurz erwaehnen; intuition
  fuer bedeutung der oek variablen geben; } \co{Was lernen wir?  for a
  single enzyme, these variables can be used to answer the question:
  what would be the ``local fitness gain'' from a certain rhythms in
  this enzyme? \co{SHOW similarity and differences to local
    description in section 2.1} in fact, balance between benefit and
  cost for each enzyme! KURZ sagen, dass periodischer bedarf null sein
  muss (falls schwingungen preisneutral sind) oder mit periodischem
  preis ausbalanciert sein muss; beziehungen zu per oek pot (im
  ungestoerten referenzzustand; im optimalen gestoerten zustand)}

\section{Principles of optimal adaptation}

\subsection{Metabolic control, optimality, and portrayal relations}

\myparagraph{\ \\Causal and functional effects} To model optimal
enzyme rhythms, we combined dynamic metabolic models with optimality
problems for enzyme profiles. \co{(see again Box 2)} Metabolic models, on
the first level, imply a logic of cause and effect: enzyme
perturbations affect reaction rates, changes in reaction rates affect
metabolite concentrations, metabolite concentrations affect other
reaction rates, and so on.  \co{NOETIG HIER?  In a metabolic pathway
  with irreversible reactions, perturbations will propagate along the
  chain.}  On the second level, in the search for optimal enzyme
profiles, we turn this logic around.  Starting from an objective (at
the end of our causal chains), we go back to metabolic states that
support this objective and enzyme profiles that support these states.
Requiring an \emph{optimal} objective value and going back from
objective to enzyme profiles, we trace all effects in reverse, from
effect to cause, along pathways and in time.  This inverse causation
is called ``promoting'': ``promoting a rhythm'' does not mean
``causing a rhythm machanistically'', but ``creating the incentive for
a rhythm''. \todo{Any dynamical ``forward'' effects can be reflected
  in ``reverse'' incentives.} For example, if an oscillating external
nutrient is said to ``promote'' a periodic expression of a
transporter, this simply mean that the transporter works more
efficiently if expressed periodically in phase with its substrate.
Any process that is influenced by an enzyme (either directly or
indirectly) can promote adaptations of that enzyme. Importantly, for
an enzyme A to promote enzyme B, no causal connection between them is
needed. It suffices that A and B influence a third, fitness-relevant
process C synergistically.  Thus, metabolic value theory can be seen as
an inversion of metabolic control. While {\MCA} consideres enzymes and
predicting their effects in forward direction, we start from desired
effects and ask which enzyme adapatations might realise this effect in
an optimal way\footnote{A similar logic is used in hierarchical
  regulation analysis, \co{REF} with two main differences.  (i)
  Hierarchical regulation analysis does not concern optimal choices,
  but traces the dynamics of metablic changes. (ii) It considers only
  \emph{direct} effects and how they are partitioned.}.  This is not
new: it has been used to predict adaptation of steady states
\cite{lksh:04}. When we study optimal rhythms, this logic of a causal
inversion seems even more apt because metabolite waves caused by an
oscillating enzyme propagate visibly through the network and can be
traced back along pathways and in time.  \todo{Therefore, any
  ``forward'' features of dynamics may be reflected ``in reverse'', in
  the choice of optimal enzyme strategies.}

\co{zurueeckkommen auf these: kausale beziheungen, zeitlich versetzt!
  jetzt aber: oekonomische beziehungen, zeitlich versetzt! in intro
  noch vorgestellt als diskrete phasen; jetzt koennen wir das alles
  kontinuierlich verstehen}

\myparagraph{Coordinated enzyme rhythms reflect network structure and
  metabolic tasks} \co{MOVE THIS TO CBA reg: just copy the most
  important sentences back to here!}  \todo{Our theory of optimal
  enzyme profiles relates the metabolic effects of enzymes (i.e.~their
  function) to their own regulation (e.g.~regulation mechanisms that
  can realise optimal enzyme adaptations).}  \todo{\co{KEEP THIS HERE}
  Biochemically, it would be conceivable that protein expression
  patterns are completely unrelated to protein function (for example,
  assuming a hypothetical genome in which the regions are
  shuffled). However, it is known (and physiologically important) that
  protein expression patterns reflect the way in which proteins act in
  metabolic pathways or other cellular systems. The components of
  protein complexes are often coexpressed, while their order of
  expression in time reflects the order of complex assembly
  \cite{kaal:04}. Likewise, enzymes involved in a metabolic pathway
  have been shown to be expressed according to their order along the
  pathway \cite{zmrb:04}.  This coordinated activation reflects -- or
  as it were, ``portrays'' -- the metabolic network topology, even if
  metabolic network and enzyme regulation have evolved separately, and
  there is no reason for them to show similar patterns.}  \co{FN: In
  practice this coregulation is often achieved by transcriptional
  feedback regulation (from a pathway product to pathway enzymes,
  through operons, through dedicated transcription factors, and
  through timing mechanisms such as feed-forward loops). But why are
  all these mechanisms in place? What is the (optimal) benefit that
  their synchronisation provides, and what are the best
  synchronisation patterns? And how come they ``portray'',
  structurally and dynamically, the metabolic systems that they
  control?}

In an engineered control system, a ``portrayal
relationship'' between the controller and the system under its control
would not be surprising at all: for instance, in predictive control,
the controller employs a model -- a kind of ``mental representation''
-- of the systems to be steered, to make its choices.  For reliable
predictions, the internal model needs to resemble the real system in
all relevant aspects, and for good outcomes, the controller needs to
employ a meaningful objective function. In this sense, the controller
must ``portray'' both the system to be steered (and possibly its
uncertain environment) and the optimality task.  \co{mention kalman
  filters)} But how can such a portrayal relation emerge in biological
networks that are not engineered, but ``tinkered'' by evolution?
\co{ref jakob; uri} \co{philosophischer exkurs: allgemein frage nach
  der Repraesentation der aussenwelt (environment task) in cell; hier:
  repraesentation der aussenwelt (rest der zelle) in protein
  regulation function. am ende eher .. in allem, was zur regulation
  des proteins beitraegt ... dieses repraesentationsproblem
  (``bewusstsein der zelle'') viel klarer machen. infotaxis erwaehnen?
  ``bewusstsein, gedaechtnis'', in infotaxis .. biologie fragt nicht
  nach repraesentation, sonder nach dem funktionieren .. aber das kann
  gerade zu repraesentation fuehren!}  Whether cells have a ``mental
representation'' of their environment is something that cannot be
shown by scientific means. However, an evolutionary adaption leading
to similar outcomes is well conceivable. The enzyme profiles in a cell
(and the regulation systems behind them) may end up ``portraying'' the
metabolic network, its tasks, and its environment to an extent to
which this provides a selection advantage. Thus, a portrayal relation,
``implicit'' in the behaviour shown by the control system, may evolve
because of a fitness advantage that it provides.  \co{just like
  knowledhe in an artifical neural network is implicit in what the
  network performs} \co{while this is quite general, below we will see
  that a more specific optimality-based explanation, for enzyme
  rhythms, arises from our synergy formulae and from {\MCA}!}

\co{speziell nochmal erwaehnen (``besonders gut zu sehen ist das an ..''): match von zeitskalen und (expression/posttransl) mechanismen!}

\myparagraph{Portrayal relationship arising from optimality} \co{cite
  goethe? gabor filter?} \co{mention operons?} The fact that enzyme
expression reflects enzyme function is not surprising if cells behave
economically: if the components of a complex are not expressed in the
right proportions, or if enzymes in a pathway are not expressed in the
right temporal order, resources are wasted.  The portrayal relation
between function and regulation emerges because of an evolutionary
advantage, and we can study this by optimality models. \co{as in REFs
  nochmal edda; uri; wolf}  The theory of optimal enzyme profiles
explains why enzymes in pathways should be co-expressed and that their
regulation patterns should reflect their own flux control.
\todo{\co{KEEP THIS HERE} Portrayal relations between enzyme profiles
  (our control variables) and metabolism (the system to be controlled)
  have been found for other cases of metabolic control. For example,
  Klipp and Heinrich have studied flux maximisation in metabolic
  models with a fixed total enzyme amount and have shown that the
  optimal (i.e.~flux-maximising) enzyme activities are proportional to
  the enzymes' scaled flux control coefficients, i.e.~to the enzyme's
  relative effect on the steady-state flux \cite{klhe:99}.  Similar
  result was obtained for adaptation of static enzyme levels
  \cite{lksh:04}: optimal adaptation profiles reflect fitness
  synergies ( second-order metabolic response coefficients), which
  reflect metabolic network structure.  According to this prediction,
  the more strongly an enzymes affects relevant state variables, and
  the cheaper its adaptation would be, the more strongly it should be
  adapted.  In sequential activation strategies, as predicted in
  \cite{klhh:02,zmrb:04}, enzyme activation reflect the sequence of
  enzymes in a metabolic pathway.

  The theory of optimal enzyme rhythms extends this to periodic
  behaviour and provides analytical formulae that suggest a portrayal
  relationship between network structure, a meaningful order of
  processes, and optimal enzyme phases and amplitudes.  Now we found
  similar phenomena for enzyme rhythms: optimal rhythms show
  coordinated patterns, and their phase shifts portray the order of
  metabolic reactions in the network.  \co{mention example
    ``repressilator'': portrayal of dynamic resonance in regulation?}}

\co{WO? Any enzyme change in a metabolic system evokes a dynamic
  metabolite profile, which provides opportunities for further enzyme
  adaptation; this additional enzyme adaptation will change the
  metabolite profile again, and so on. An activity peak of one enzyme
  creates time-shifted peaks in downstream metabolites (and possibly
  drops in upstream metabolites). during these peaks, other enzymes
  can work more efficiently. (PEAK, and so on ..) The same things
  holds for a sequence of enzyme peaks or for a sine-wave enzyme
  rhythm.  Whenever cells can improve their fitness by some
  coordinated enzyme behaviour, there is an incentive to realise this
  behaviour.}  \co{WO? case of Portrayal (similar to uri's / edda's
  substrate step case): (dynamics forward: peaks follow! leads to
  protrayal!  however, this is obscured by global optimality (infinite
  series!)}

\myparagraph{Reasons for the portrayal relationship} \co{make it clear
  that the mathematical connection is through enzyme synergies (which
  are thererfore a key concept for understanding the portrayal
  relation} How does the ``portrayal relationship'' between a
metabolic system and enzyme profile arise mathematically?  For
example, can we find a link between enzyme profiles and metabolic
network structure in our mathematical formulae?  The shape of an
optimal enzyme rhythm depends on three aspects: on periodic
perturbations present in the environment, on the way external and
enzyme-driven perturbations propagate in the network, and on the
importance of different state variables for cell fitness.  In our
formulae, the second and third aspect determine the synergy matrices.
The second factor, how perturbations propagate, is described by
periodic response and control coefficients. These coefficients are
closely related to network structure. On the one hand, they follow by
matrix inversion from the Jacobian matrix, which portrays the network
by describing neighbourhood relations between metabolites. On the
other hand, stoichiometric matrix, elasticity matrix, control
coefficients are related through summation and connectivity theorems
\cite{hesc:98}, which directly links network structure, metabolic
control (in ``forward direction''), and the shape of optimal enzyme
rhythms (arising from incentives propagating in ``reverse
direction'').  Thus, enzyme rhythms (and, more generall, temporal
enzyme profiles) portray network structure \todo{as a result of two}
matrix inversions.  The Jacobian matrix reflects the network topology;
by inverting it, we obtain metabolic control coefficients, which play
a role in defining the enzyme synergies; by inverting this synergy
matrix, we obtain the optimal enzyme profiles.  \co{so why does
  topology information remain??  answer this :)} \co{auspassen, dass
  nicht oben irgendwelche direkteren zusammenhaenge behauptet werden!
  was kann man ueber vorwaerts- bzw rueckwaertslaufen von (dynamischen
  und oekonomischen) wellen lernen?}

\co{WO? wichtig: es is NICHT so, dass eine metab welle existiert und die
  enzyme dann gleich schnell folgen; die enzyme MACHEN die welle, und
  die zeitskala kann deshlab viel langsamer sein. weil die
  akkumulation durch enzymaenderungen langsamer sein kann als der
  (metabolische) turnover der metaboliten!. konntrolle und dynamik
  sind nicht das gleiche! (aber beide gehen auf die gleiche
  netzwerkstruktur zurueck - deshalb wiederspiegelung der topologie
  (nicht der zeitskala; in steady-state adaptation ist die zeitskala
  sowieso nicht sichtbar!)) }

\subsection{Principles of optimal periodic behaviour}

\myparagraph{\ \\Temporal strategies of organisms} Should organisms
 in periodic environments behave homeostatically, or should they
use existing environmental variations as opportunities, running different
processes at different times? If biochemical processes follow an  optimal temporal order,
which of them should be synchronised or be separated in time?  How should a metabolic
cycle proceed?  From an optimality perspective, organisms should do
what improves their evolutionary fitness. For example, many trees
lose their leaves in winter and many animals hibernate, consuming the
storage compounds they produced in the warmer seasons.  Wood
production in trees varies between seasons, and also {\flow}ering is
season-dependent. We can think of such yearly cycles  as a whole, as processes
under a selective pressures.  We
can think of daily rhythms (e.g.~the varying glucose supply and demand
in our body and its management of by the liver) in a similar way.  The
behaviour in each phase should be adapted to the environment and to
the behaviour in the past and future phases.

\myparagraph{Resource allocation in time, and anticipation} \co{wird
  das umdrehen in der zeit / entlang wegen / von ableitungen gut
  erklaert? oder kam das schon oben?}  To predict optimal metabolic
\todo{behaviour} in time, we may postulate different control
strategies: ``myopic'' strategies (``In each moment, adapt optimally
to the current conditions as if they wer static''), ``scheduling''
strategies (``Shift processes to times that provide the best external
conditions''), or ``self-consistent'' strategies (``Arrange all
processes optimally in time, such that no rearrangement could provide
an advantage).  Self-consistent strategies imply anticipation
\todo{and a shaping of future conditions}: the enzyme activities are
not adapted ``passively'' to existing substrate levels (as suggested
by ``just-in-time production'' or implemented by substrate
activation).  Instead, they \emph{actively} shape the metabolite
profiles, to create conditions under which enzymes act more
efficiently. We can see anticipation at work in the optimal activation
patterns in linear metabolic pathways \co{WD mit eben:}(which was
described as ``just-in-time'' adaptation): if all enzymes were
activated immediately, most of them would have little substrate and
would therefore act inefficiently, so converting most of the substrate
into product would take very long.  A sequential activation, in
contrast, creates high levels of internal metabolites and high
concentration gradients at specific times, and provides time windows
in which individual enzymes are very efficient.

\iftoggle{bookversion}{
\section{Biological role of enzyme rhythms}}
{\section{Discussion}}

\myparagraph{\ \\Enzyme rhythms can be useful} \co{Why should we model
  not only metabolic dynamics, but also metabolic optimality?  It may
  certainly improve our understanding of biological regulation.}
Whether metabolic rhythms are a by-product of overshooting regulation
or whether they have biological functions (e.g.~to improve metabolic
efficiency) is an open question.  In order to argue for biological
functions, we need to show that oscillations can provide an
advantage. Here I adopted this view and asked, specifically, whether
enzyme rhythms can provide metabolic benefits (e.g.~an improved
metabolic performance at a constant overall enzyme investment) that
cannot be realised by static enzyme changes alone.

\co{DEN FOLGENDEN ABSCHNITT GANZ UMSCHREIBEN!}  \co{JA! oder lieber
  sagen: einerseits kinetische ansaetze mit opt. reg; andererseits
  FBA-ansaetze mit opt verhalten aber ohne kinetik. ich genau
  dazwischen.} \co{CITE willi?  others cite in intro} \co{mention
  periodic FBA and dFBA!}  \co{Mention temporal and periodic FBA,
  e.g. \cite{rubs:15} holzhuetter/bockmayr // ralf steuer}\co{direkt
  gegensatz zu FBA mit phasen erklaeren. CITE?}  \co{explain other two
  ways (evolution of allost reg; fba) clearly} In models, we may study
this question by optimising either the enzyme profiles themselves or
the regulation mechanisms behind them.  Previous studies were focused
on regulation (e.g.~optimising the kinetic constants in allosteric or
transcriptional regulation): \co{REF?} the optimisation resembles a
simulated evolution in which mutations can introduce additional
regulation edges, destroy edges, or modify their regulation strengths,
and allow us to quantify the selection pressures on such mutations.
These studies have shown that the resulting rhythms can improve
metabolic performance. However, the exact functional reasons for
enzyme rhythms -- why certain enzyme rhythms are beneficial and how
they should be orchestrated -- remain unclear. To study how benefits
arise precisely, the present theory predicts enzyme profiles
independently of how they are realised, which makes it more general.
Once these profiles are known, possible ways to realiss them by
regulation mechanisms can be studied separately.

\co{How such rhythms may emerge mechanistically is beyond the scope of
  this paper. The comparison to experiments (e.g.~yeast metabolic
  oscillations) can be done once suitable models have been
  established. \co{Briefly note that regulation must reflect forward
    dynamics in order to be optimal}}

\co{here, the approch was slightly different: instead of constructing
  a (beneficial) series of actions to be performed, I start from a
  steady state and ask whether it can be improved by slight periodic
  modifications, in the sense of a Taylor expansion. If this is the
  case, it indicates that also larger oscillations may be
  beneficial. The advantage is that this approach is suited to take
  metabolic dynamics (enzyme kinetics and the possible accumulation of
  metabolites, a)}

\myparagraph{The economics of enzyme rhythms} A theory of optimal
enzyme profiles, as developed here, has multiple benefits.  First, it
highlights the importance of enzyme synergies. \co{FN:
  \textbf{Generality of the approach} recall that external variables
  and control variables were used to describe external concentrations
  and enzyme activities, but that they could also have other
  meanings!}  The synergy matrices show why there can be incentives
for oscillations, even if the static enzyme activities have already
been optimised. Second, the formulae tell us how transcriptional and
posttranscriptional rhythms should be combined at different
frequencies, and makes the logic behind this more transparent than
numerical simulations would do. \co{Hier auch: kurz was zur anpassung
  an auessere rhythmesn sagen, matchen von zeitskalen, reflexion von
  zeitskalen in mechanismen} \co{fourth, explains portrayal relation?}
\co{another advantage of the frequency representation: idea of
  periodic economic potentials} Third, the theory shows us that
promoted and autonomous enzyme rhythms are closely related, and why.
A metabolic system may benefit from self-promoting metabolic cycles:
whenever there is a positive principal synergy at a finite frequency,
the reference states will be fitness-unstable against certain enzyme
rhythms. If enzyme rhythms can improve the system's performance even
in a constant environment, there is an incentive for self-promoting
oscillations. Other systems may be fitness-stable against
self-promoting enzyme rhythms, and require external rhythms to promote
rhythmic enzyme adaptation.  In fact, a single model may show both
types of behavior depending on parameter choices.  For example, a
linear pathway with strong dilution may show self-promoting rhythms,
while the same pathway with weak dilution shows only adaptive rhythms.
This parameter-dependent switch between strategies can be seen as a
bifurcation. However, it is not a bifurcation in the usual dynamical
sense, but an ``economic'' bifurcation concerning the optimal choice
of strategies\footnote{Phase transitions in classical thermodynamics
  (e.g.~between water and ice) resemble ``economic bifurcations'' if
  the physical laws are formulated as extremal principles,
  e.g.~minimising the system's Gibbs free energy.}.

\myparagraph{Choice of the fitness function \co{and consequences of
    this choice (write more about this!)} } The theory of optimal
enzyme rhythms does not presuppose specific metabolic objectives, but
allows modeller to define arbitrary objectives, as functions of
metabolite concentrations and fluxes. For wild-type cells, a fitness
function may score relevant output variables such as biomass
production.  In biotechnological applications, the objective may
represent a production rate to be maximised. In any case, we choose a
metabolic objective that scores steady state variables such as biomass
production rate and translate it into a functional for scoring time
courses.  Here I proposed two types of fitness functionals.  In the
\emph{state-average fitness}, the metabolic state is first averaged
over time and the static fitness function is then applied to the
time-averaged state: in this case enzymes must shift the average
concentrations and fluxes to have an effect on fitness.  With a
\emph{fitness-average fitness} (and using a nonlinear static fitness
functions), the overall fitness depends not only on the average state,
but also on temporal fluctuations around it\footnote{In some models
  the enzyme cost term does not matter: in models with linear enzyme
  cost functions or state-average fitness functionals, enzyme rhythms
  are cost-neutral and fitness changes are only due to the
  {\metabolicobjective}.}. Both of the functionals are computed from
time averages, but they reflect different assumptions about cellular
time scales\footnote{We may assume that the metabolic state affects
  cell fitness indirectly, via other cell variables. If these
  variables respond very slowly, they effectively depend on the
  average metabolic state; if they respond very fast, they depend on
  the metabolic state in each moment. \co{das in naechste FN?:} If
  they respond on an intermediate time scale, the resulting fitness
  functions should be defined by convolution integrals.}. There are
also more general fitness functionals that average state variables on
a time scale $\tau$ before evaluating the fitness\footnote{To obtain fitness functionals that
  score
  metabolic states on specific time scales, we can consider  convolution integrals
  $\bar f(t) = \int K(t-t')\,f(t')\, \md t'$. The kernel function $K$
  with temporal width $\tau$  accounts for the fact that metabolic
  outputs may be  smoothened in time (e.g.~in a wave through
  additional pathways) before the fitness effect is realised. By
  setting $\tau \rightarrow \infty$ or $\tau=0$, we reobtain our two
  simple functionals as limiting cases.}.  \co{klarer sagen, was die
  erwartetem AUSWIRKUNGEn verschiedener fitnessfunktionen sind; zb
  kopplung durch kosten (zwischen enzymen) ja oder nein? oder
  auswirkung von festem mittelwert auf die strategien}

\myparagraph{Assumption of small sine-wave enzyme rhythms} Biological
rhythms can be described as a sequence of discrete phases like sleep
and wake. Similarly, metabolic rhythms have been modelled as a
sequence of steady states with different enzyme profiles
\cite{pbbh:15}.  However, the real dynamics of metabolic cycles is
continuous rather than abrupt: if metabolic rhythms are driven by gene
expression, they are smooth on metabolic timescales\footnote{Protein
  synthesis acts as a low-pass filter, resulting in smooth enzyme
  profiles with low amplitudes at higher frequencies. Protein profiles
  with short peaks yield strong allosynchrony effects, but can only be
  realised posttranscriptionally (i.e.~at an extra cost).}  (see SI
\ref{sec:geneexpression}), and even sudden external perturbations are
smoothened while propagating through the metabolic network.  \co{FN:
  in modelling, fine time slices can be used to make this almost
  smooth (hoping for numerical convergence!); mention dynamic /
  periodic FBA again; BUT: ignoring kinetics, and less natural than a
  smooth description} Accordingly, the theory describes smooth
metabolic rhythms with small amplitudes. The focus on sine-wave
functions has a number of reasons. In control theory, optimal control
profiles in time can be computed by solving the Riccati equation, a
differential equation that is integrated backwards in time, starting
from the final system state. To describe periodic behaviour, the
initial and final states would have to be matched, which is difficult.
However, an elegant way to do this is to Fourier-expand all curves and
to convert the problem into an optimisation of the Fourier
coefficients\footnote{Other basis functions, e.g.~orthogonal
  polynomials, could be used to formulate non-periodic control
  problems.}. This is what I did here: rhythms are described by
smooth, small-amplitude sine-wave oscillations around a (possibly
shifted) static reference state, with amplitudes, phases, and average
shifts to be optimised\footnote{Using complex-valued frequencies, the
  same mathematical approach can also describe exponentially damped
  oscillations, which may occur in models with dilution.}.  Eqs
(\ref{eq:optstaticvector}) and (\ref{eq:optperiodic}) show how an
orchestrated enzyme behaviour can emerge from fitness synergies
between the sine-wave profiles of single enzymes. Finally, in models
without active constraints, Fourier synthesis also enables us to
handle periodic perturbations other than sine waves.

\co{WRITE AS FN: instead of large text, just briefly explain
  formula. If oscillations are described by sine-wave amplitudes, it
  is also relatively easy to handle differential-delay equations (by
  changing the phase shifts, depending on frequency). it would be
  interesting to further explored this mathematically. // could Fuu
  usw also come from delay-equation model?  (or simple whole-cell
  model?)  for this, delay equations would be needed (to get realistic
  phase shifts between processes // kommentar zu komplexer
  stöchiometrischer matrix einbauen (direkt verwenden in code?)
  describe how delay equations could yield synergy matrices; ab dann
  gleich?  -> diskutieren (auch in derivation of periodic lagrange
  mult) // idee zu ganzzell-osczillationsmodel (dougie) benutze
  diff-delay-equations (für richtige zeitskalen bei wenig globalen
  variablen); evtl: zeige dass die lösungen äquivalent sind zu
  normalen dgl auf einer schnelleren zeitskala; dh rechne einfach die
  zeitskala für enzymprodukton runter! denke an effektive gesamtkosten
  für pathways; versuche dann, die investments in pathways wieder
  aufzulösen in investments für einzeenzyme (modell wieder modular
  verfeinern)}

\co{the idea is that ..}
\co{vergleich mit system zur planung von bauprojekten (in anticipation: vorbereitungshypothese):\\
  o bauen: wenn A NOTWENDIGE bedingung fuer B ist (zb A mauern, B dach), dann kommt A vor B\\
  o evtl weitere ``regularisierungsbedingung'': nichts zu lange ungenutzt stehen lassen\\
  o modifikationen: 1. denken in KREISEN
  2. nicht unbedingt NOTWENDIG, sondern HILFREICH\\
  o myopic: like capitalist system (companies search profit NOW) }

\myparagraph{Future prospects} A theory of optimal rhythms in networks
can be useful in biological and medical modelling.  In the human body,
physiological states vary considerably between day and night,
entailing substantial changes in enzyme expression.  To better
understand these changes -- for example, glucose management by the
liver -- one may study how periodic enzyme activities affect metabolic
dynamics and what quantitative benefits result from them.
\co{konkreter!  mehr ausfuehren} Similar models may be used to find
optimal schedules for drug administration, or to plan combination
therapies in which several drugs are given at different times,
creating periodic blood profiles, to exploit allosynchrony.  For
example, a first drug could make cells more susceptible to a second
drug, which is administered with a phase delay. To model this, we may
use a mechanistic pharmacokinetics/pharmacodynamics model that
describes the distribution and effects of drugs in the body. The
control variables in this case would be drug dosages as time curves
(instead of enzyme activities), and the aim would be to maximise an
intended effect while minimising the side effects.  \co{or cancer
  cells could first be synchronised in their cell cycle or in a
  metabolic cycle before a second drug hits them in the right phase.}
Of course, with a simulation model at hand, optimal interventions can
be determined by numerical optimisation of time curves.  However, with
a large number of control variables this may be hard, and starting from
a good initial solutions (obtained from our perturbation theory) may
be helpful. Moreover, general formulae may provide additional
insights, for example about the roles of synergies and how they can be
arranged in beneficial cycles.  Such an abstract perspective may
reveal characteristic patterns, or a ``style'' of optimal cycles.
\co{move the rest to abstract?}  A theory of optimal rhythms can also
clarify the relation between network structure, biochemical dynamics,
and external interventions, from small pathways to larger networks.
It shows how network-wide behaviour emerges from local fitness synergies, it
thereby reveals general design principles behind enzyme rhythms, for
example, the general conditions for self-promoting beneficial
oscillations.

\section*{Acknowledgements}
I thank Rainer Machn\'e, Douglas B.~Murray, and Hermann-Georg
Holzh\"utter for thinking with me, and Matthias K\"onig and Mariapaola
Gritti for comments on the manuscript. Tàijíquán practice was a main
inspiration for this work, and I am very much indebted to my teachers
Edith Kock, Lee Hong Thay and Mizue Watanabe. This work was funded by
the German Research Foundation (Ll 1676/2-1 and Ll 1676/2-2).

\co{acknowledgements: mention that reinhart proposed ``flexibility''
  as a possible advantage of biological rhythms (das kommt eingentlich
  noch genauer in teil 2 ..).  here i tried to spell out this idea, in
  a way. but my main inspiration comes from taiji ..}

\bibliographystyle{unsrt}
\bibliography{files/biology}


\clearpage

\begin{appendix}


\section{Forced metabolic oscillations}
\label{sec:forcedOscillations}

\begin{figure*}[t!]
\begin{center}
\fbox{\parbox{16.5cm}{
\begin{center}
  \parbox{15.5cm}{\textbf{Box 3: Oscillations and complex amplitudes}\\[4mm]
    \centerline{
      \includegraphics[width=15cm]{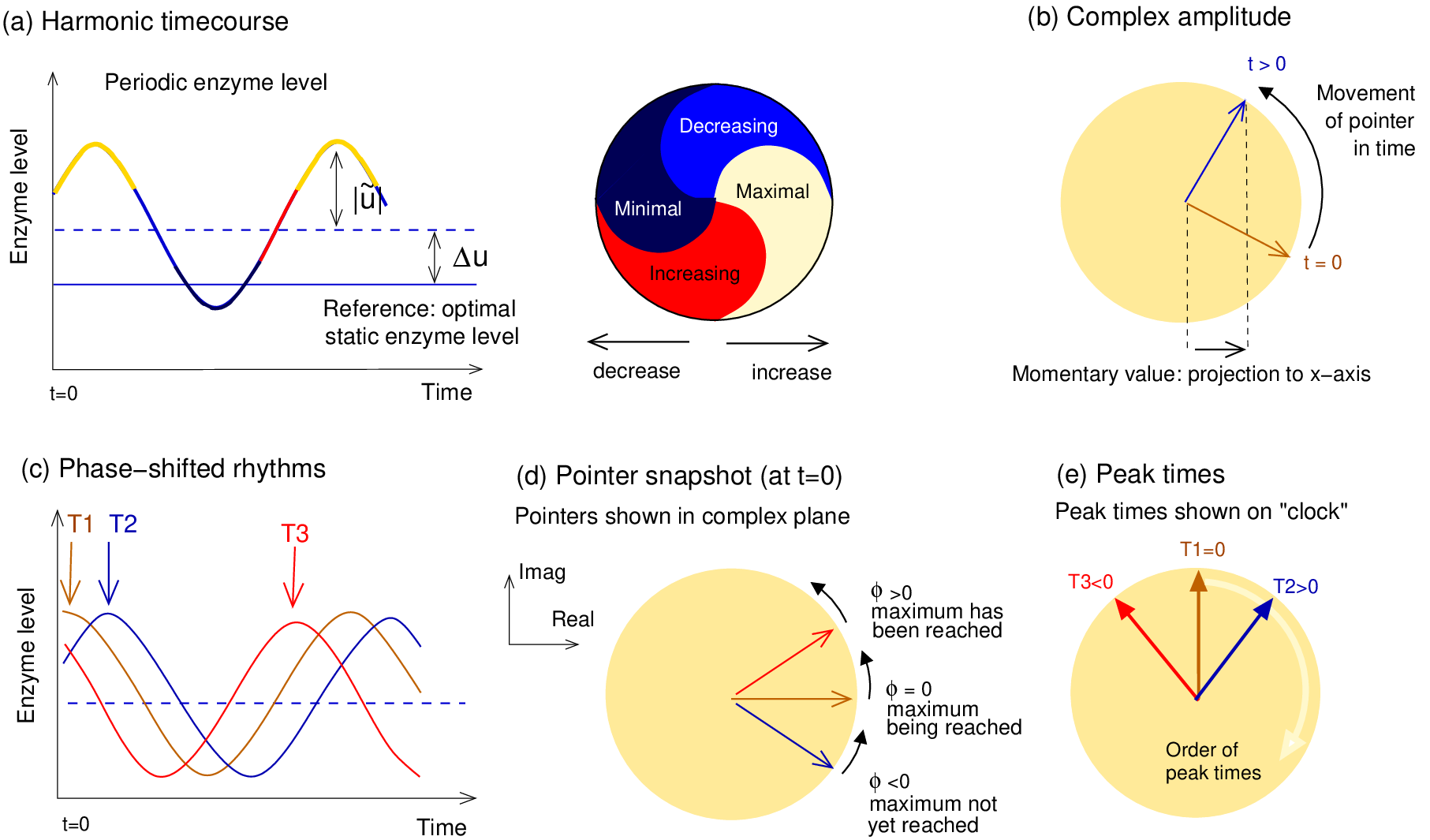}}
Enzyme rhythms and their complex
    amplitudes. (a) Sine-wave oscillation of an enzyme level.
    Starting from a steady reference state (straight solid line), we
    shift the enzyme level by $\Delta \ub$ and add a sine-wave
    oscillation with complex amplitude $\tilde u$ and circular
    frequency $\omega = 2\,\pi/T$ (with period length $T$).  In
    analogy to the notion of ``cell cycle phases'', the rhythm can be
    divided into phases (maximal, decreasing, minimal, increasing),
    but here discrete phases are used only for illustration and have
    little importance in the present approach. Instead, we describe
    sine-wave oscillations by complex exponential functions
    $\et\,\e^{i \,\omega \,t}$ and visualise them by rotating
    pointers. (b) The complex amplitude $\et$ encodes oscillation
    amplitude and phase shift (phase at time $t=0$). The real part
    (projection to x-axis) describes the periodic deviation
    $\Delta \ub(t)$.  (c) Oscillations with different phase shifts
    (peak times shown by arrows). (d) The same oscillations, shown by
    pointers at time point $t=0$. (e) Peak times displayed on a
    ``clock'': the vertical pointer corresponds to the curve that
    peaks at $t=0$; pointers for the other two curves follow in
    clockwise direction.
 \label{fig:OptimalityProblem}
}
 \end{center}}}
\end{center}
\end{figure*}

To understand how enzyme rhythms can provide benefits, we first need
to see how their effects propagate in metabolic networks.  Here I
summarise formulae from \cite{lksh:04} and \cite{lieb:2005}, using
complex-valued vectors as explained in SI section
\ref{sec:SInotation}.  A kinetic model describes metabolic dynamics by
rate equations
$\md \cv/\md t = \Nmat\, \ratev(\cv,\pv) - \lambda\,\cv$ with a
concentration vector $\cv$ (for internal metabolites), a
stoichiometric matrix $\Nmat$, and a flux vector $\vv=\ratev(\cv,\pv)$
containing the rate laws
$\rate_{l}(\esymbolv,\cv,\pv) = u_{l}\,\ratelaw_{l}(\cv,\pv)$. The term
$\lambda\,\cv$ describes dilution in growing cells with cell growth
rate $\lambda$ (if dilution can be ignored, we set or $\lambda = 0$).
In models with conserved moieties \cite{rede:88}, we can split the
stoichiometric matrix $\Nmat$ into a product $\Nmat = \Lmat\, \NRmat$,
where $\NRmat$ consists of linearly independent rows of $\Nmat$,
corresponding to independent internal metabolites.  The vector $\pv$
can contain two types of model parameters: externally defined
parameters in a subvector $\xv$ (e.g., external metabolite
concentrations) and control variables in a subvector $\esymbolv$ (e.g.~enzyme
activities).  A metabolic state with concentration vector $\cvsteady$,
satisfying the stationarity condition $0 = \Nmat \,\vv(\cvsteady)$, is
called a stationary state or steady state.  For growing cells, the
steady-state condition reads
$0 = \Nmat \,\vv(\cvsteady)-\lambda\,\cvsteady$.  To define a steady
reference state, we consider a reference parameter vector $\pv_{0}$
and require that all parameter sets $\pv$ close to $\pv_{0}$ lead to
similar stable steady states $\cvsteady(\pv)$ (i.e.~$p$ must not be
close to a bifurcation point).  Periodic parameter variations (i.e.~of
enzyme activities or external metabolite concentrations) are described
by a time-dependent parameter vector
\begin{eqnarray}
\label{eq:periodicparameter}
\pv(t)&=& \pv_{0} + \real(\e^{i \omega t}\,\pvt)
\end{eqnarray}
with a complex-valued amplitude vector $\pvt$ and circular frequency
$\omega$ (see Figure \ref{fig:OptimalityProblem}).  The parameter
rhythm leads to periodic concentration and flux changes that propagate
through the network as damped waves.  In a linearised model, the
concentration and flux curves will be sine waves with frequency
$\omega$.  In nonlinear models, sine-wave perturbations can evoke
higher harmonics and shift the average state.  Here we neglect higher
harmonics and focus on shifts of the average metabolite
concentrations, fluxes, and metabolic fitness, which we study in a
second-order approximation \cite{lieb:2005} (see Figure
\ref{fig:forced}). These are the effects that matter for our fitness
functions.

Oscillating metabolite and enzyme levels  have direct effects on
adjacent reaction rates. In a single reaction, if we  assume that
enzyme and metabolite concentrations are directly controllable, these effects
can be described by spectral elasticities \cite{lieb:2005}.  Since
enzyme activities appear as prefactors in the rate laws
($v_{l} = u_{l}\, \ratelaw_{l}(\cv)$), an enzyme rhythm alone (at
constant metabolite concentrations) cannot cause direct second-order effects
(such as higher harmonics or shifts of the average flux).  In contrast,
metabolite rhythms or combined metabolite and enzyme rhythms can have
such direct second-order effects (see SI Figure
\ref{fig:periodicelasticities}).  To see their effects on the average fluxes, we now consider two periodic parameters $a$ and $b$
(representing enzyme levels, metabolite concentrations, or other quantities
affecting the rate directly). Their time profiles read
\begin{eqnarray}
 \label{eq:twoparameters}
a(t)= \ab + \real(\e^{i \omega t}\,\at), \qquad
b(t)= \bb + \real(\e^{i \omega t}\,\bt)
\end{eqnarray}
with complex amplitudes $\at$ and $\bt$, frequency $\omega$, and phase
difference $\Delta \varphi = \varphi(\at)-\varphi(\bt)$. If $a$ peaks
before $b$, the phase shift $\Delta \varphi$ is small and positive.

In metabolic control theory, the effects of small perturbations on
reaction rates or steady states are described by sensitivities called
elasticity, response, and control coefficients. In this article,
unscaled sensitivities are used throughout.  Let us first consider a
single reaction. For small oscillations, the reaction rate reads
approximately
\begin{eqnarray}
\label{eq:approxrateelasticitiesall}
 v(t) \approx \bar v + E_{a} \,\Delta a(t) + E_{b} \,\Delta b(t) 
+ \half E_{a a} \,\Delta a(t)^{2} +  E_{a b} \,\Delta a(t)\,\Delta b(t) 
+ \half E_{b  b} \,\Delta b(t)^{2}
\end{eqnarray}
with  reference flux $\bar v$ and reaction elasticities $E_{a}, E_{b},
E_{a a}, E_{a b}$, and $ E_{b b}$.  The average flux shift  due to the rhythms is given by 
\begin{eqnarray}
\label{eq:approxrateelasticitiesalla}
\Delta \langle v \rangle_{t} \approx \half E_{\tilde a \tilde a} \,|\tilde a|^{2} 
+  E_{\tilde a \tilde b} \,|\tilde a|\,|\tilde b| \cos(\Delta \varphi) 
+ \half E_{\tilde b \tilde  b} \,|\tilde b|^{2},
\end{eqnarray}
with periodic second-order elasticities
$E_{\tilde a \tilde a} = \half E_{aa}$,
$E_{\tilde a \tilde b} = \half E_{ab}$, and
$E_{\tilde b \tilde b} = \half E_{bb}$. If parameter $a$ is an enzyme
activity $u$ and parameter $b$ is a metabolite concentration $x$, the
second-order elasticities are given by
$E_{\tilde a \tilde a} = E_{\tilde u \tilde u} = 0$,
$E_{\tilde a \tilde b} = E_{\tilde u \tilde x} = \frac{1}{2\,u}
E_{x}$, and $E_{\tilde b \tilde b} = \half E_{xx}$, and we obtain a
flux shift
\begin{eqnarray}
\label{eq:approxrateelasticitiesall2}
\Delta \langle v \rangle_{t} \approx 
E_{\tilde u \tilde x} \,|\tilde u|\,|\tilde x| \cos(\Delta \varphi) 
= \frac{E_{x}}{2\,u} \,|\tilde u|\,|\tilde x| \cos(\Delta \varphi).
\end{eqnarray}
This formula resembles Eq.~(\ref{eq:fluxsplitting2}) (with
$E_{\tilde u \tilde x}$ instead of the prefactor $k/2$).  The
second-order spectral elasticities describe synergisms between
metabolite and enzyme rhythms, relating two rhythmic profiles to the
resulting time average shift in reaction rate. Now we consider the
entire network. Network-wide rhythms (caused by enzyme and external
metabolite rhythms) can be described similarly, but with response
coefficients instead of elasticities as expansion
coefficients. Periodic response coefficients relate a sine-wave
perturbation (of an enzyme or external metabolite) to
the resulting flux and concentration rhythms (see SI \ref{sec:ResponseFormulae}).  If a state variable
$\ys$ (an  internal concentration or metabolic flux) is influenced by
oscillating parameters $a(t)$ and $b(t)$, a second-order expansion for
$\ys(t)$ yields three kinds of effects: forced oscillations with
frequencies $\omega$, forced oscillations with frequency $2 \omega$,
and a shift $\Delta \langle \ys \rangle_{t}$ of the average value. The
shift consists of three terms:
\begin{eqnarray}
\label{eq:deltayexpansion}
 \Delta \langle \ys \rangle_{t} &\approx& \frac{1}{2} ~
 R^{\rm \ys}_{\rm \at \at}(\omega) ~\at^{2}
+ \frac{1}{2} R^{\rm \ys}_{\rm \bt \bt}(\omega) ~ \bt^{2} 
+ \frac{1}{2} \re( \e^{-i \Delta \varphi} R^{\rm \ys}_{\rm \at \bt}(\omega))~ |\at| |\bt|.
\end{eqnarray}
The first two terms represent self-synergies of $a$ and of $b$
(describing, e.g.~the dynamic self-inhibition of an enzyme); the third
term describes a synergy between the parameters. \co{SAY: for symmetry reasons ..:} Its magnitude and
sign depend on the phase shift $\Delta \varphi$.
Eq.~(\ref{eq:deltayexpansion}) contains no first-order terms because
the reference state (where $\at = \bt = 0$) is an  extremum point with
respect to state shifts $\Delta \langle \ys \rangle_{t}$; it can
be a minimum, a maximum, or a saddle point.  In a model with many
periodic parameters (amplitude vector $\pvt$), the
total shift results from a sum over all pairwise synergies. We can
generalise Eq.~(\ref{eq:deltayexpansion}) and obtain
\begin{eqnarray}
 \label{eq:secondorderexpansion}
 \Delta \langle \ys \rangle_{t} &\approx& \frac{1}{2} ~ \pvt^\dag\, \RYmat_{\rm \pt \pt}(\omega) ~\pvt.
\end{eqnarray}
The symbol $\pvt^{\dag}$ denotes the adjoint vector (i.e.~the complex
conjugate transpose).  The synergy matrix
$\RYmat_{\rm \pt \pt}(\omega)$ can be computed from the stoichiometric
matrix $\Nmat$ and from the first-and second-order elasticities (see
SI \ref{sec:periodicdynamics}).  It is Hermitian and its matrix
elements, called second-order periodic response coefficients (see SI
\ref{sec:ResponseFormulae}), are equal to the second-order spectral
control coefficients in \cite{lieb:2005} except for a scaling factor.
In our second-order approximation, rhythms of different frequencies
have no fitness synergies between them.

\coout{allgemein anderes symbol statt $\Deltar \evb$? etwas analog zu
  $\evt$ und "ref"? zb $\evb + \check\esymbolv$ dann auch $\check{\langle
    \yv}\rangle_{t}$} \co{lieber $\RYmat_{\rm \pt \pt}(\omega)(\pvt
  \otimes
  \pvt)$? am besten nur manchmal, wenn vektorielles ergebnis, sagen
  ``for convenience''}

\begin{figure*}[t!]
 \begin{center}
   \parbox{16.5cm}{\parbox{9cm}{
       \includegraphics[width=8.5cm]{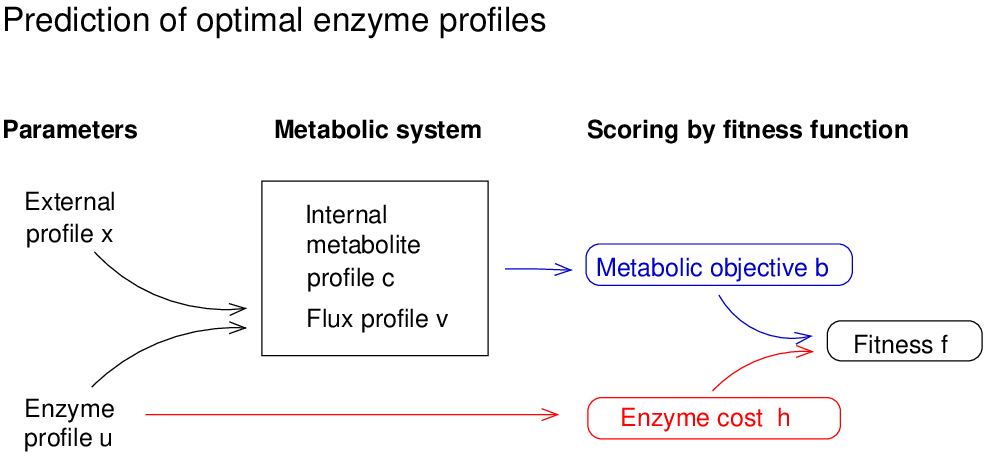}\\[5mm]}
     \parbox{7.5cm}{\caption{Optimality problem for enzyme rhythms. A metabolic
         state (with fluxes $\vv$ and metabolite concentrations $\cv$
         as state variables) is controlled by external parameters and
         enzyme levels $\esymbolv$. The fitness function $\ffit$ is defined
         as the difference of {\metabolicobjective} $\yy$ and
         enzyme cost $\hminus$. The need to maximise fitness creates
         an incentive for enzyme profiles with high benefit and low
         cost. The same scheme applies to optimal steady states (with
         static external and enzyme profiles) and optimal dynamic
         behaviour (with external and enzyme rhythms).}
 \label{fig:optimality}}}
 \end{center}
\end{figure*}

\section{Fitness functions for dynamic metabolic states}
\label{sec:AppFitness}

To define what we mean by optimal enzyme profiles, we need to refer to a given fitness
objective. Our fitness objectives consist of two terms: a cost
for enzyme levels and a benefit term for  metabolite concentrations
and fluxes (see Figure \ref{fig:optimality}).  For a single reaction,
we consider a {\metabolicobjective} function $\yy(u,x) = \gplus(v(u,x))$
and a cost function $\hminus(u)$ and define the fitness
$\ffit(u,x) = \gplus(u,x)-\hminus(u)$.  For a metabolic network, we
consider a fitness function \co{instead of separate costs and
  benefits, we could also assume a single fitness function
  $\ffit(\esymbolv) = \ffit'(\ysv(\esymbolv),\esymbolv)$. In metabolic value theory, this
  usually does not make a difference (because we only consider
  first-order effects). Here, in contrast, second-order effect play a
  role, and the difference between the two fitness functions would be
  in the mixed second-order derivatives between $\ysv$ and $\esymbolv$.}
\begin{eqnarray}
\label{eq:fitness0}
 \ffit(\esymbolv) = \underbrace{\yy(\esymbolv)}_{\gplus(\ysv(\esymbolv))} - \hminus(\esymbolv)
\end{eqnarray}
with state variables in a vector $\ysv$ (containing flux vector $\vv$
and concentration vector $\cv$), scored by a benefit $\yy(\ysv)$, and
a control vector $\esymbolv$ scored by a cost $\hminus(\esymbolv)$. The choice
variables $u_{l}$ may comprise enzyme activities and other variables,
e.g.~the dilution rate in growing cells \cite{lksh:04}.  In a pathway
model, the functions $\yy(\ysv)$ and $\hminus(\esymbolv)$ may describe how
the pathway contributes to cell fitness as assumed by the modeller.
Empirical protein cost functions can be obtained experimentally
\cite{deal:05,szad:10} from the growth deficits after a forced
expression of idle protein.  In our models, the enzyme cost function
must be an increasing (linear or nonlinear) function of the enzyme
activities. In our perturbation theory, {\metabolicobjective} and cost
functions are approximated by a second-order expansion
\begin{eqnarray}
 \label{eq:fitnessStationary2}
 \gplus(\ysv + \ysvd) &\approx& \gplus(\ysv) + \qz \cdot \ysvd + \half \ysvd\trans \Qzz \ysvd
 \nonumber\\ 
 \hminus(\esymbolv + \devb) &\approx& \hminus(\esymbolv) + \hu \cdot \devb + \half \devb\trans \Huu \devb,
\end{eqnarray}
where $\qz$, $\Qzz$, $\hu$, and $\Huu$ are  gradients and curvature
matrices of  $\gplus$ and $h$.  In models with external
parameters $x_{j}$ (e.g.~external metabolite concentrations), the steady-state
fitness
\begin{equation}
 \label{eq:fitness1}
 \ffit(\esymbolv,\xv)=\underbrace{\yy(\esymbolv,\xv)}_{\gplus(\ysv(\esymbolv,\xv))} - \hminus(\esymbolv)
\end{equation}
contains $\xv$ as an additional argument.  In static optimality
problems, the control variables $u_{l}$ must be chosen to maximise
$f(\esymbolv,\xv)$ at given external parameters $x_{j}$ and under all given
constraints.  The synergy matrices can be computed from the curvatures
of the fitness function Eq.~(\ref{eq:fitness1})
\begin{eqnarray}
 \label{eq:statichessians}
 \Fex &=& 
 \qz\trans ~\RYbmat_{\rm x e}
 + {\RYmat_{\rm x}}\trans \Qzz \RYmat_{\rm e} \nonumber \\
 \Fee &=& \qz\trans ~\RYbmat_{\rm x x}
 + {\RYmat_{\rm x}}\trans \Qzz \RYmat_{\rm x} - \Hee.
\end{eqnarray}
In  enzyme-optimal states, the fitness gradient
$\ffitu = (\partial \ffit/\partial e_{l})$ must vanish for all
active enzymes $e_{l}$ and the synergy matrix $\Fee$ must have
negative eigenvalues. 

\myparagraph{Fitness effects of periodic perturbations} Now we
consider such an enzyme-optimal state $(\xvref,\evref)$ as our
reference state, apply periodic perturbations, and search for optimal
adaptations of our enzyme profiles. We assume that all parameters show
sine-wave oscillations around their reference values:
\begin{eqnarray}
\xv(t) &=& \xvref+\real(\e^{i \omega t}\,\xvt) \nonumber\\
\esymbolv(t) &=& \evref+\real(\e^{i \omega t}\,\evt).
\end{eqnarray}
In a second-order approximation, these perturbations will shift the
average state and   the average fitness.  The time average of
a state variable $\ys$ can be expanded, using
Eq.~(\ref{eq:deltayexpansion}), as
\begin{eqnarray}
\label{eq:fitnessOscillatoryA1}
 \langle \ys \rangle_{t} &\approx& \ys(\xvref,\evref) + \frac{1}{2}~
 {\xvt \choose \evt}^{\dag}
\left( \begin{array}{ll}
\RYbsecmat_{\rm \xt \xt}(\omega) & \RYbsecmat_{\rm \xt \et} (\omega)\\
\RYbsecmat_{\rm \et \xt}(\omega) & \RYbsecmat_{\rm \xt \et} (\omega)
 \end{array} \right)
 {\xvt \choose \evt}.
\end{eqnarray}
To characterise   possible enzyme profiles $\esymbolv(t)$ by a  fitness score, we define a
fitness functional. In \emph{state-average fitness functionals}, we 
assume a slow realisation of fitness effects: the fitness function
(\ref{eq:fitness0}) is applied to the time-averaged state:
\begin{eqnarray}
  \label{eq:averagingmethod2}
  \Ftemp^{\rm (S)} = \gplus(\langle\ysv(t)\rangle_t) - \hminus(\langle\esymbolv(t)\rangle_t).
\end{eqnarray}
In \emph{fitness-average fitness functionals}, we assume an immediate
realisation of fitness effects: we evaluate the fitness function
(\ref{eq:fitness0}) in each moment and  take the time average:
\begin{eqnarray}
  \label{eq:averagingmethod1}
  \Ftemp^{\rm (F)} =\langle \gplus(\ysv(t)) - \hminus(\esymbolv(t))\rangle_t.
\end{eqnarray}
There also exist more general functionals which evaluate fitness on a
specific time scale.  In a functional
$F=\langle f(\int K(t-t')\,\ysv(t')\,\md t') \rangle_t$, the state
variables $\ys(t$) are convolved with a kernel function (temporal
width $\tau$), a fitness function is applied, and the resulting
fitness values are averaged over time.  State-averaged and
fitness-averaged fitness functions are special  cases of this general
 functional.  Given our fitness function (\ref{eq:fitness0})
and  type of functional ($\Ftemp^{\rm (S)}$ or $\Ftemp^{\rm (F)}$),
we can compute the fitness effects of parameter oscillations.   Periodic perturbations of external
concentrations $x_{j}$ and enzyme activities $u_{l}$ lead to
fitness changes that can be 
approximated by a quadratic function of $\xvt$ and $\evt$. The synergy
matrices (containing curvatures of  this quadratic function) can be computed from
periodic response coefficients and fitness derivatives.  For the
state-average fitness $\Ftemp^{\rm (S)}$,
Eq.~(\ref{eq:averagingmethod2}), they read
\begin{eqnarray}
 \label{eq:fitnessOscillatoryB3Text2}
 \Fetxt^{\rm (S)} &=& \qz\trans ~\RYbmat_{\rm \et \xt}(\omega) \nonumber \\
 \Fetet^{\rm (S)} &=& \qz\trans ~\RYbmat_{\rm \et \et}(\omega).
\end{eqnarray}
For the fitness-average fitness $\Ftemp^{\rm (F)}$,
Eq.~(\ref{eq:averagingmethod1}), they contain  additional terms:
\begin{eqnarray}
 \label{eq:fitnessOscillatoryB3Text}
 \Fetxt^{\rm (F)} &=& 
 \qz\trans ~\RYbmat_{\rm \et \xt}(\omega) 
 + \half\, (\RYtmat_{\rm \et}(\omega))^{\dag} \Qzz \RYtmat_{\rm \xt}(\omega) \nonumber \\
 \Fetet^{\rm (F)} &=& \qz\trans ~\RYbmat_{\rm \et \et}(\omega) 
 +  \half\, (\RYtmat_{\rm \et}(\omega))^{\dag} \Qzz \RYtmat_{\rm \et}(\omega)
 - \half \Hee.
\end{eqnarray}
For slow oscillations ($\omega \approx 0$) we can approximate
$\RYtmat_{\rm \xt} \approx \RYmat_{\rm x}$,
$\RYtmat_{\rm \et} \approx \RYmat_{\rm e}$,
$\RYmat_{\rm \et \et} \approx \half \RYmat_{\rm ee}$, and
$\RYmat_{\rm \xt \et} \approx \half \RYmat_{\rm xe}$.
Eq.~(\ref{eq:fitnessOscillatoryB3Text}) yields the same synergy
matrices as Eq.~(\ref{eq:statichessians}), but with a prefactor of
1/2.  This prefactor has a simple explanation: periodic synergy
matrices do not describe static parameter shifts with a fixed sign,
but alternating perturbations which realise only $\sqrt{1/2}$ of the
maximal amplitude on average.  This also holds in the theoretical
limit $\omega \rightarrow 0$ (where oscillations are infinitely slow).

\section{Optimal enzyme profiles}
\label{sec:computingoptimal}

To compute optimal enzyme profiles, we first need to know how small
(static or periodic) variations of enzyme activities or external
parameters affect fitness.  According to formula
(\ref{eq:deltafexpansion}), all relevant information is contained in
the synergy matrices (\ref{eq:fitnessOscillatoryB3Text2}) and
(\ref{eq:fitnessOscillatoryB3Text}).  Based on these matrices and
assuming no further constraints on the enzyme profile, we obtain
formulae for optimal enzyme profiles for a number of cases (see Figure
\ref{fig:Guulandscape} and Box 4).

\begin{figure*}[t!]
\begin{center}
  \fbox{\parbox{16.5cm}{\begin{center}
        \parbox{15.5cm}{\textbf{Box 4: Mathematical scenarios for optimal enzyme rhythms}\\[4mm]
         The diagrams
            show fitness landscapes and constraints for  enzymes
            in a larger metabolic system. Axes show the average enzyme
            level $\ub$ (x-axis) and amplitude $|\et|$
            (y-axis). The phase angle (not shown) is assumed to be
            optimised.\\
            
            \centerline{\includegraphics[width=15.5cm]{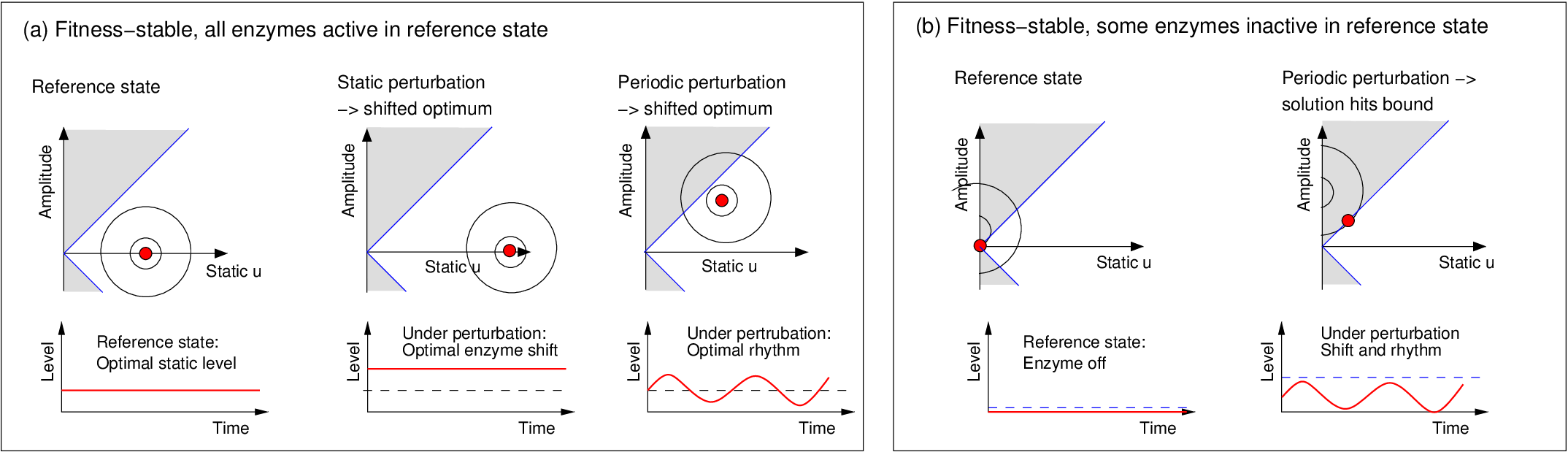}}\ \\[-5mm]

            \co{erlaubte bereiche lieber gelb?} (a) Unperturbed
            reference state and systems with static or periodic
            external perturbations.  Optimum points are shown by red
            dots.  Circles represent contour lines of the fitness
            function.  Inaccessible regions are shaded in
            grey. Perturbations shift the fitness function, and the
            enzyme must be optimally adaptated to reach follow the new
            optimum.  (b) A system in which an enzyme becomes active
            under periodic perturbations.  In the reference state,
            constraints (diagonal lines) force the enzyme to be
            inactive. Upon a periodic perturbation (right), an enzyme
            rhythm provides an advantage despite the cost of
            increasing of the average
            level.\\

            \centerline{\includegraphics[width=15.5cm]{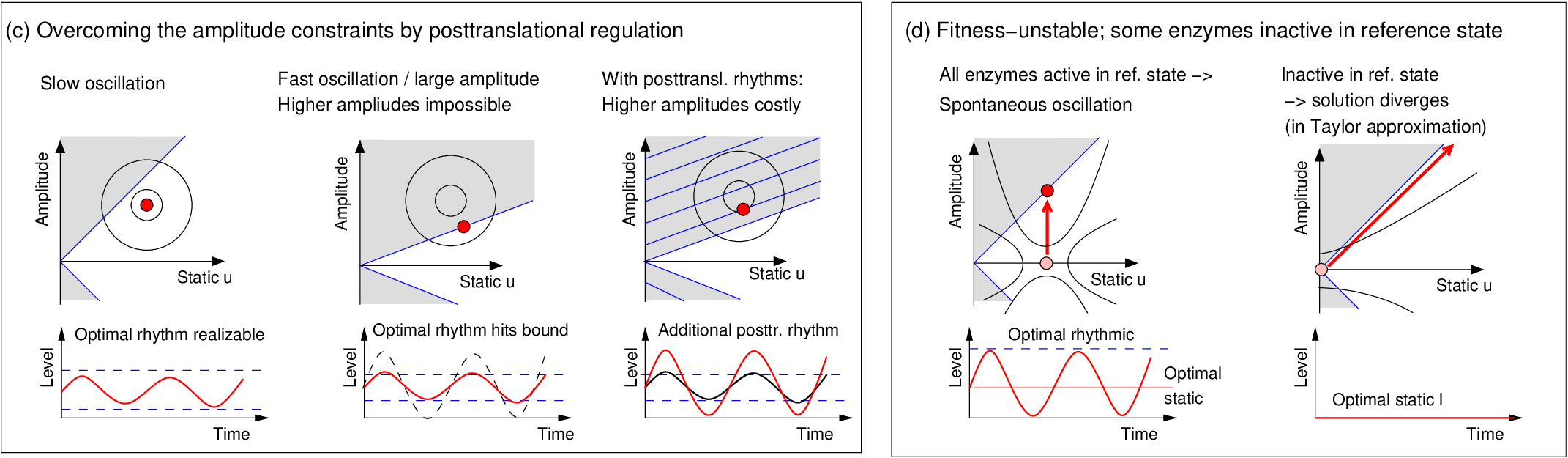}}\ \\[-5mm]

            (c) Enzyme rhythms realised by a joint differential
            expression and posttranslational modification.  The x and
            y axis refer to expression levels. Diagonal lines depict
            constraints on protein profiles. On the left, the optimal
            rhythm can be realised by expression changes alone (same
            as in Fig (a), right).  In the centre diagram (with a
            tighter constraint, e.g.~due to higher frequency), only a
            suboptimal rhythm (outside the centre point of the fitness
            landscape) can be realised. On the right, we assume that
            the cell can modulate its enzyme activities by
            posttranslational modification. Now any enzyme amplitude
            can be realised, but at a certain cost (function depicted
            by diagonal lines); this is why the optimum deviates from
            the nominal maximum of the fitness landscape (centre of
            the circle). (d) Cases in which the reference state is
            fitness-unstable against enzyme rhythms. In the case on
            the left, the amplitude will increase until it hits a
            bound, leading to spontaneous oscillations. In the case on
            the right, the solution diverges. In reality, large enzyme
            levels would be costly. In a model, this may be captured
            by additional bounds on maximal enzyme activities (not
            shown).
 \label{fig:scenarios}} \end{center}}}
\end{center}
\end{figure*}

\begin{enumerate}[leftmargin=5mm]
\item \textbf{Criterion for optimal steady states.} In a
  model with constant external parameters $x_{j}$,  a static enzyme
  profile $\esymbolv$ is an \emph{interior optimum} of the fitness function
  $f$ if all enzyme activities $u_{l}$ are positive, the fitness
  gradient $\fv_{\rm u}(\xv,\esymbolv)$ vanishes, and the synergy matrix
  $\Fuu(\xv,\esymbolv)$ is negative definite (i.e.~all eigenvalues are
  strictly negative). In this case, the parameter set $(\xv, \esymbolv)$ is called
  enzyme-optimal.  In contrast, an enzyme profile is a \emph{boundary
    optimum} if some of the enzyme activities vanish and have negative
  fitness slopes $\partial \ffit/\partial u_{l}<0$, while the others
  form an internal optimum.
\item \textbf{Optimal adaptation to static external perturbations.}
  Let $(\xvref, \evref)$ be an enzyme-optimal parameter set and
  $\dxvb$ be a static perturbation of $\xvref$.  A static enzyme
  change $\devb\opt$ is called an \emph{optimal adaptation} to $\dxvb$
  if it leads to a new enzyme-balanced state
  $(\xvref + \dxvb, \evref+\devb\opt)$.  For small perturbations
  $\dxvb$, the optimal adaptation can be approximated by
  \cite{lksh:04}
 \begin{eqnarray}
\label{eq:OptimalRespSimple}
 \devb\opt(\dxvb) &\approx& \Aux~ \dxvb
 \end{eqnarray}
 with the adaptation matrix $\Aux = -\Fee\inv\, \Fex$. The synergy
 matrices $\Fee$ and $\Fex$ are given by
 Eq.~(\ref{eq:statichessians}), and $\Fee$ is invertible because it
 stems from an enzyme-optimal reference
 state\footnote{Eq.~(\ref{eq:OptimalRespSimple}) can be proven as
   follows: after an optimal adaptation, the fitness gradient must
   vanish before the perturbation ($\fv_{\rm e}(\xvref,\evref)=0$) and
   after the adaptation ($\fv_{\rm e}(\xvref+\dxvb,\evref+\devb)=0$).
   After a first-order expansion, and equating the two expressions, we
   obtain the condition
   $\Delta \fv_{\rm e} = \Fee \devb + \Fex \dxvb = 0$. Solving for
   $\devb$ yields Eq.~(\ref{eq:OptimalRespSimple}).}.  \co{FN: what
   about non-strict optimum, with vanishing eigenvalues?}  Consider a
 single reaction as an example. The fitness function reads
 $\ffit(e,x) = \gplus(e,x)-\hminus(e)$. The optimality problem leads
 to a non-zero enzyme activity\footnote{With a very steep cost
   functions, the enzyme may be inactive.}
 $\eb^{\rm opt}(x) = \mbox{argmax}_{e}\, \ffit(e,\xb)$ (see SI
 \ref{sec:ex1} for an example). Since the fitness function is
 negatively curved, any spontaneous variation around the optimal point
 would lead to a fitness loss. When the substrate level increases from
 $\xb$ to $\xb+\Delta \xb$, the optimal enzyme activity also
 increases.  The optimal adaptation profile
\begin{eqnarray}
 \label{eq:optstatic}
 \Delta \eb^{\rm opt} &\approx& - \Fee\inv\, \ffit_{\rm
   ex}\, \Delta \xb
\end{eqnarray}
depends on the negative synergy
$\Fee = \frac{\partial^2 \ffit}{\partial \e^2}$ and positive synergy
$\Fex = \frac{\partial^2 \ffit}{\partial \e\, \partial x}$
in the reference  state.  An  increase
$\Delta \xb$ would lead to a positive adaptation $\Delta \eb^{\rm opt}$
and the fitness increases by
\begin{eqnarray}
\label{eq:AdaptationFormula1}
\Delta \ffit = \Fex\, \Delta {\eb_{\rm opt}}\, \Delta \xb +
\half \Fee\, \Delta \eb_{\rm opt}^{2}= - \half \ffit_{\rm
  ee}\inv\, \ffit^{2}_{\rm ex} \Delta \xb^{2} > 0.
\end{eqnarray}

\item \textbf{Optimal adaptation to external oscillations.}  The
  optimal adaptation to an external rhythm can be computed
  similarly. With a perturbation frequency $\omega$ and amplitude
  vector $\xvt$, the optimal enzyme  amplitude vector reads
 \begin{eqnarray}
   \evt\opt(\xvt) &\approx& \tilde{\bf{A}}^{\rm e}_{\rm x}~ \xvt.
 \end{eqnarray}
 The complex-valued, frequency-dependent adaptation matrix
 $\tilde{\bf{A}}^{\rm e}_{\rm x} = -\Fetet\inv\,\Fetxt$ follows from
 the synergy matrices $\Fetet$ and $\Fetxt$ for periodic perturbations
 (Eqs (\ref{eq:fitnessOscillatoryB3Text2}) and
 (\ref{eq:fitnessOscillatoryB3Text})).
\item \textbf{Static fitness changes} By inserting the adaptation
  vectors $\devb^{\rm opt}$ or $ \evt^{\rm opt}$ into
  Eq.~(\ref{eq:deltafexpansion}), we obtain the first-order fitness
  changes resulting from enzyme adaptation to static or periodic
  perturbations. \co{erst: static perturbations!}  For periodic
  perturbations $\xvt$, the fitness change is given by
  $\Delta \Ftemp = {\evt^{\rm opt\,\dag}}\, \Fetxt\, \xvt = -
  \xvt^{\dag}\, \Fetxt^{\dag}\, \Fetet\inv\, \Fetxt \xvt$.
\item \textbf{Benefit from  self-promoting oscillations.} An enzyme rhythm
  $\evt$ alone can change the fitness by  
  \begin{eqnarray}
    \label{eq:self-promotingbenefit}
    \Delta \Ftemp &\approx&  \frac{1}{2} \,\evt^\dag\, \Fetet\, \evt.
  \end{eqnarray}
  Any vector $\evt$ with a positive fitness change $\Delta \Ftemp$
  represents a beneficial, ``self-promoting'' enzyme rhythm. This
  holds, for example, for all eigenvectors of $\Fetet$ with positive
  eigenvalues.  Since $\Fetet$ is Hermitian, its eigenvalues are
  real-valued and its eigenvectors span the space of possible
  amplitude vectors.  A self-promoting enzyme rhythm requires that the
  reference state be fitness-stable against static enzyme changes, but
  fitness-unstable against certain enzyme rhythms. This means: all
  eigenvalues of $\Fee$ must be negative, but at least one eigenvalue
  of $\Fetet(\omega)$, for some non-zero frequency $\omega$, must be
  positive. What is the most profitable enzyme rhythm at a given
  frequency?  The eigenvector $\evt^{\rm opt}$ with the largest
  eigenvalue (called principal fitness synergy $\sigma(\omega)$)
  represents the best enzyme rhythm with a given norm
  $||\evt||$. \co{this will be the best solution for small
    oscillations. what if constraints make this vector impossible?
    include again lagrange multipliers? nee geht glaube ich nicht ..} Its elements define
  amplitudes and phases of all enzymes.  If there exist beneficial
  enzyme rhythms at different frequencies, those with the highest
  principal synergy $\sigma$ have the biggest selection advantage and
  are most likely to emerge in evolution. If the benefit is maximal at
  $\omega =0$, oscillations of finite frequency are dispreferred. If
  we consider large-amplitude oscillations, the optimal enzyme rhythms
  are not simply given by eigenvectors, but may also shaped by the
  amplitude constraints.
\end{enumerate}

 Optimal enzyme profiles can also be shaped by constraints.

\begin{enumerate}[leftmargin=5mm]
\item \textbf{Bounds for enzyme amplitudes}
\label{sec: amplitudeconstraints}
The amplitudes $|\pt|$ of protein profiles
$\pb + \real(\pt\, \e^{i \omega t})$ are constrained for several
reasons (see Figure \ref{fig:amplitudeConstraint}). (i) A protein
amplitude cannot be larger than the average protein level
(i.e.~$|\pt| \le \pb$) because protein levels would become negative
otherwise.  (ii) A dynamic model of protein production and degradation
(with degradation constant $\kappa$) yields the tighter constraint
$|\pt| \le \sqrt{\frac{\kappa}{\kappa+\omega}}\, \pb$ (see SI
\ref{sec:geneexpression}), which coincides with the previous
constraint at frequency $\omega=0$ (see Figure
\ref{fig:amplitudeConstraint} (a)).  (iii) The time derivative of a
protein concentration, $\md p/\md t = -\pt\,\omega\,\sin(\omega t)$,
reaches its higherst negative value $-\pt\,\omega$ when the protein
level is $\pb$ and a degradation rate is $\kappa\,\pb$.  Since the
rate of decrease cannot be bigger than the degradation rate, we obtain
the constraint $\pt \le \frac{\kappa}{\omega} \pb$. (iv) Due to space
restriction in cells, we can put a bound on the maximal protein level
$\pb + |\pt|$ and on the average level $\pb$.  With all these
constraints, each protein amplitude $|\pt|$ has an upper bound
$\pt^{\rm max}(\pb,\omega,\kappa) =
\mbox{min}(\sqrt{\frac{\kappa}{\kappa+\omega}},
\frac{\kappa}{\omega})\,\pb$ or briefly $\gamma(\omega,\kappa) \pb$.
The maximal relative amplitude
$\gamma(\omega,\kappa) = \pt^{\rm max}/\pb$ depends on frequency
$\omega$ and effective degradation constant $\kappa$; at frequencies
$\omega \gg \kappa$, it becomes very small.

\item \textbf{The role of constraints in enzyme optimisation} The
  fitness expansion formula Eq.~(\ref{eq:deltafexpansion}) contains no synergy
  terms between enzyme rhythms and static changes, and no synergy terms  between enzyme
  rhythms of different frequencies. If a temporal parameter
  perturbation consists of sine-wave oscillations of different
  frequencies, and if Eq.~\ref{eq:optstaticvector} applies, the
  optimal adaptations to these components should therefore  be additive. However,
  this only holds if there are no constraints on the enzyme activities
  (i.e.~if Eq.~(\ref{eq:OptimalityProblem}) applies), or if the
  overall solution (i.e.~optimal enzyme rhythms, summed over all
  frequencies) respects all constraints (e.g.~on maximal enzyme
  activities). In practice, we may proceed by first computing a preliminary
  solution by summing the solutions for individual frequencies. If this solution
  violates some constraints, we need to discard this solution and
  instead  determine a solution containing all frequencies by
  applying a single joint optimisation with constraints.

\item \textbf{Enzymes that only become active in periodic states}
  \label{sec:inactive} If an enzyme is inactive in the reference
  state, then it cannot start oscillating unless its mean level also
  increases: the mean value must be at least as large as the
  amplitude, and the resulting cost can be seen as an effective
  ``oscillation cost'' proportional to the amplitude.  In this case, our  two optimality
  problems -- static and periodic adaptation -- cannot be treated
  separately: instead of Eqs (\ref{eq:optstaticvector}) and
  (\ref{eq:optperiodic}), a constrained optimality problem
  Eq.~(\ref{eq:OptimalityProblem}) must be solved, even if the
  oscillations are very small.  \todo{There are two possible cases:
    \co{gute bezeichnungen! uea verwenden} (i) The reference state is
    fitness-stable against enzyme oscillations and enzyme rhythms are
    promoted by external oscillations. (ii) The reference state is
    fitness-unstable against enzyme oscillations despite the
    effective oscillation cost; then, in our second-order expansion,
    due to the linear increase in {\metabolicobjective}, the resulting
    oscillations will increase until they hit the bounds for maximal
    enzyme activities or amplitudes.}

\item \textbf{Non-optimal enzyme profiles and the resulting fitness
    loss.}  If a cell fails to use an optimal enzyme profile,
  this leads to a fitness loss.  The fitness loss
  can be quantified by Eq.~(\ref{eq:deltafexpansion}) (see
  Figure \ref{fig:Guulandscape} (c)). With a mismatch 
  $\evt^{\rm mis} = \evt - \evt^{\rm opt}$  between 
  actual and desired amplitude vector, the fitness difference reads
  \begin{eqnarray}
\label{eq:fitnesLoss}
 \Delta \Ftemp^{\rm mis} &\approx& \real(\xvt^{\dag}\,\Fxtet\,\evt^{\rm mis} +
  \evt_{\rm opt}^{\dag}\,\Fetet \evt^{\rm mis} + \evt^{*\dag}\,\Fetet
  \evt^{\rm mis})=\real(\evt^{*\dag}\,\Fetet \evt^{\rm mis})
\end{eqnarray}
(note that $\xvt^{\dag}\,\Fxtet\ + \evt_{\rm opt}^{\dag}\,\Fetet = 0 $
in the reference state).  If a metabolic state is fitness-stable
against enzyme rhythms, the synergy matrix $\Fetet$ is negative
definite, so Eq.~(\ref{eq:fitnesLoss}) describes an actual loss.

\end{enumerate}

Matlab code and models for optimal enzyme rhythms are available at
\url{www.metabolic-economics.de/enzyme-rhythms/}.

\section{Periodic economic potentials}
\label{sec:AppPeriodicEconomicPotentials}

The shapes of optimal enzyme profiles can be  understood using 
concepts from metabolic value theory \cite{lieb:18theory,
  lieb:18lagrange,lieb:14a,lieb:14b}.  Following 
Eq.~(\ref{eq:deltafexpansion}), we first expand the fitness function
quadratically around a steady reference state. Assuming that  our solution does
not hit any constraints, the optimality conditions read
\begin{eqnarray}
\label{eq:optimalityConditionEconomic}
 0 &=&  \underbrace{\gev+ \Gex\,\dxvb + \Gee\,\devb}_{\pergev}
       -\underbrace{(\hev+\Hee\,\devb)}_{\perhev}  \qquad \mbox{Static changes}\,\dxvb, \devb\nonumber  \\
 0 &=&  \underbrace{\gevt+\Getxt\,\dxvt + \Getet\,\devt}_{\pergevt}
       -\underbrace{(\hevt+\Hetet\,\devt)}_{\perhevt} \qquad \mbox{Periodic changes}\,\dxvt, \devt.
\end{eqnarray}
The terms \co{wu statt yu ueberall? zb in $\gev$!}  $\gev, \Gex, \Gee$, .., denote
first and second derivatives of our {\metabolicobjective} 
$\gplus$ and cost function $\hminus$ in the unperturbed reference
state.  The bracket terms $\pergev$ or $\pergevt$ (called static or
periodic enzyme {\myvalue}s) and terms $\perhev$ or $\perhevt$
(called static or periodic enzyme {\price}s), denoted by circle
$\circ$ can be seen as  derivatives too. Unlike the usual
gradients $\gev$ and $\hev$ (or $\gevt$ and $\hevt$) defined in the
reference state, they are gradients describing variations around an
existing periodic state. With these enzyme {\myvalue}s and {\price}s,
the optimality conditions in a periodic state,
Eq.~(\ref{eq:optimalityConditionEconomic}), can be simply written as
\begin{eqnarray}
\label{eq:optimalityConditionEconomic1}
\pergev=\perhev, \qquad \pergevt=\perhevt.
\end{eqnarray}
Enzyme {\myvalue}s and {\price}s in periodic states can be  defined
more generally outside our quadratic approximation: for general
periodic states, with {\metabolicobjective} and cost functionals $g(\evb,\evt)$
and $h(\evb,\evt)$, we define them as
\begin{eqnarray}
\label{eq:optimalityConditionEconomic2}
\pergev  = \frac{\partial g}{\partial \evb},\qquad
\pergevt = \frac{\partial g}{\partial \evt},\qquad
\perhevt = \frac{\partial h}{\partial \evb},\qquad
\perhev  = \frac{\partial h}{\partial \evt}.
\end{eqnarray}
Again, the optimality conditions
$\frac{\partial f}{\partial \evb}= \frac{\partial g}{\partial
  \evb}-\frac{\partial h}{\partial \evb}= 0$ and
$\frac{\partial f}{\partial \evt}= \frac{\partial g}{\partial
  \evt}-\frac{\partial h}{\partial \evt}= 0$ \todo{are given by
  Eq.~(\ref{eq:optimalityConditionEconomic})}\co{oder nochmal
  schreiben mit $\pergev, \pergevt, \perhevt, \perhev$?}. What can we
learn from these optimality conditions?  If our expansion point is an
enzyme-optimal steady reference state, we know that $\gev=\hev$; and
if our fitness functional is not explicitly time-dependent, we also
obtain $\gevt=\hevt=0$.  In this case, the first-order terms in
Eq.~(\ref{eq:optimalityConditionEconomic}) cancel out and can be
ignored. Then,  by solving for $\devb$ and $\devt$ we reobtain Eqs
(\ref{eq:optstaticvector}) and (\ref{eq:optperiodic}).  If, in
addition, the cost functional is linear (i.e.~$\Huu=0$), then
$\perhevt$ will vanish and we obtain $\pergevt=0$ as a simple
optimality condition.  In other words: in the state with optimal
enzyme rhythms, the periodic {\myvalue} of each enzyme must vanish.
Intuitively, this means that no small variation of enzyme amplitudes
or phases can improve (even change) the metabolic objective to first
order.

What else can we learn from optimality condition
(\ref{eq:optimalityConditionEconomic})? If enzyme {\myvalue}s are
displayed on the network, their amplitudes and phases reflect the
network structure: for example, enzyme {\myvalue}s are related between
adjacent reactions. To see this, we introduce similar economic
variable for metabolite production, called economic potentials, with
the following definition. The function $g$ describes the overall
{\metabolicobjective}, evaluated in the periodic state of the model.
Now, for each independent metabolite $i$, we consider a small virtual
exchange flux $\prodrate_i^{\rm virt}$ producing this metabolite.  The
static and periodic economic potentials are defined as
$\wri = \frac{\partial g}{\partial \prodrate_i^{\rm virt}}$ and
$\wrti = \frac{\partial g}{\partial \tilde{\prodrate}_i^{\rm virt}}$,
for static and periodic variations of the exchange flux.  As shown in
SI \ref{sec:SIProofPeriodicbalanceequations}, the (static and
periodic) {\price}s and the (static and periodic) economic potentials
of their metabolites are linked by the balance equation
\begin{eqnarray}
\label{eq:EconomicBalanceStaticPeriodic}
{\perhev \choose \perhevt} &=& 
{\pergev \choose \pergevt} = 
\left(\begin{array}{ll}
\Eper^v_u & \Eper^v_\et\\
\Eper^\vt_u & \Eper^\vt_\et
\end{array}\right)^\dag\,
{\Deltar \wrv+\bv_v \choose \Deltar \wrvt+\bv_\vt}
\end{eqnarray}
with effective elasticities $\Eper^x_y$ defined for the periodic state. Here
$\dwrl$ denotes the economic potential difference along reaction $l$,
and the (static and periodic) flux gains $\bv_v$ and $\bv_\vt$ are
direct derivatives of the {\metabolicobjective} functional with respect
to metabolic fluxes. If we expand the elasticity coefficients around
our steady reference state, we obtain the formula
\begin{eqnarray}
\label{eq:EconomicBalanceStaticPeriodic2}
  {\diag(\evb)\,\perhev \choose \diag(\evb)\,\perhevt } 
  &=&  
  {\diag(\evb)\,\pergev \choose \diag(\evb)\,\pergevt } 
  =
\left(\begin{array}{lr}
\diag(\mathring{\vvb}) & \diag(\mathring{\vvt}) \\
\diag(\mathring{\vvt}) & \diag(\mathring{\vvb})
\end{array} \right)
^\dag
  {\Deltar \perwrv + \bvtot\choose  \Deltar \perwrvt +  {\bv_\vt}}
\end{eqnarray}
where $\evb=\evref+\devb$ contains the average enzyme activities,
$\mathring{\vvb} = \vv_{\rm ref} + \Emat^{\rm v}_{\rm c}\,\Delta
\cintvb + \Emat^{\rm v}_{\rm x}\,\dxvb$ is the vector of fluxes in the
average state, and
$\mathring{\vvt} = \Emat^{\rm v}_{\rm c}\, \cintvt + \Emat^{\rm
  v}_{\rm x}\,\xvt$.  \co{hadamard erklaeren und verwenden; unten
  genauso /} Eq.~(\ref{eq:EconomicBalanceStaticPeriodic}), for
periodic states, shows that static and periodic economic variables may
affect each other.  Applying this formula to the reference state
itself, we can set $\mathring{\vvb} = \vv$ and $\mathring{\vvt} = 0$
and obtain separate balance equations for static and periodic fitness
values
\begin{eqnarray}
\hev  &=& \gev  = \diag(\vv/\evb)\,[\dwrv+\bv_v] \nonumber \\
\hevt &=& \gevt = \diag(\vv/\evb)\,[\dwrv+\bv_\vt].
\end{eqnarray}
If the cost function is not explicitly time-dependent, all periodic
enzyme {\price}s must vanish ($\hevt = 0$).  For details, see SI
\ref{sec:SIProofPeriodicbalanceequations}.

\section{Extensions of  the theory}
\label{sec:extending}

The theory of optimal enzyme rhythms can be extended to cover more general
cases.

\begin{enumerate}[leftmargin=5mm]
\item \textbf{Other types of constraints.}  In
  Eq.~(\ref{eq:OptimalityProblem}) for  optimal profiles $\devb$
  and $\evt$, we may consider extra constraints on metabolite
  concentrations, fluxes, the sum of enzyme activities, or time
  averages of these quantities\footnote{Negative enzyme activities are
    always excluded by positivity constraints.  To also exclude
    negative metabolite concentrations, in a first-order expansion we may
    employ the approximative constraint
    $\cv + \Rmat^{\rm \tilde c}_{\rm \tilde u} \evt + \Rmat^{\rm
      \tilde c}_{\rm \tilde x} \xvt \ge 0$. This constraint excludes
    enzyme profiles for which the model yields meaningless results,
    predicting negative concentrations within the range of it Taylor
    expansion.}. Some of the periodic flux or metabolite profiles may
  even be predefined\footnote{Here is an example. If amplitude and
    phase of the cellular ATP/ADP ratio are known from experiments, we
    can require that a model realises this ratio. Using a linear
    approximation, we would constrain
    $\tilde R_{\rm ATP/ADP} \tilde u$ to yield a given complex
    number.}.  In general, with constraints of the form
  $\Mmat\,\xvt= \tilde \av$, we can 
  predefine linear combinations of periodic fluxes or concentrations.
  Alternatively, soft constraints can be implemented by penalty terms
  in the fitness function.

\item \textbf{Other control variables.}  The control variables $u_l$ in
  our example models represent enzyme activities, whereas state
  variables $\cint_{i}$ represent metabolite concentrations.  In other
  models, however, the variables can have different meanings.  Enzyme
  levels may also be described as dynamic variables, while other
  quantities, such as mRNA levels, transcription factor activities,
  drug dosages, or cell growth rate (which determines dilution in
  metabolism) could be treated as control variables.

\item \textbf{Frequency-dependent cost.}  Apparent enzyme costs in
  bacteria can be defined as the observable growth deficits after an
  overexpression of protein.  In experiments, protein overexpression
  evokes a strong initial growth deficit, which decreases after some
  hours \cite{szad:10}: possibly, ribosome levels or other cellular
  resources get out of balance and the cell needs time to return to a
  balanced state.  What if the same adjustments occur during slow
  metabolic oscillations? This would imply higher costs whenever
  enzyme levels are changing. Enzymes oscillations would be costly,
  and fast oscillations even more, because ribosome adaptation will be
  harder to achieve at high frequencies.  To model this extra cost, we
  may assume frequency-dependent enzyme cost functions. However, when
  biological rhythms are slow and predictable (e.g.~day-night cycles),
  cells may be able to anticipate and regulate ribosome demands, and
  no frequency-dependent cost needs to be assumed.

\item \textbf{Periodic fitness functions} So far we assumed fitness
  functions that do not explicitly depend on time: time dependencies
  are caused by parameter oscillations only. However, time-dependent
  fitness functions may be a reasonable assumption.  Consider a
  pathway within in a large metabolic network and assume that the cell
  lives in a periodic environment. With optimally adapted enzyme
  profiles, all cellular subsystems will oscillate, and the
  \emph{requirements} for our pathway product change periodically.  To
  model our pathway alone, while capturing these periodic external
  demands, we need an effective \emph{time-dependent} fitness function
  for our pathway.  This is a general principle: whenever we like to
  use models to ``zoom'' into a system, we need be describe subsystems
  by separate effective models. Within these models, we need to allow
  for periodic fitness functions, which breaks the time-shift
  invariance of our fitness functionals, and new terms may appear in
  the fitness expansion Eq.~(\ref{eq:deltafexpansion}): there can be
  first-order terms scoring periodic enzyme activities as well as
  synergy terms linking static shifts and amplitudes, or linking
  rhythms of different frequencies.

\item \textbf{Rhythms in protein modification and
    allosteric regulation}.  Enzyme rhythms driven by periodic gene
  expression have limited amplitudes, and at high frequencies the
  possible amplitudes become very small (Figure
  \ref{fig:amplitudeConstraint} (a)). To obtain larger amplitudes,
  enzyme activities can be modulated by enzyme phosphorylation,
  which can be steered much more quickly than gene expression
  (Figure \ref{fig:amplitudeConstraint} (b)-(d)).  However, this comes at a cost. When an
  enzyme  is periodically inhibited (or incompletely
  activated), its efficiency decreases on average. To keep the average enzyme activity at  its
  original value, the average protein level must be increased, wich entails a cost. To
  account for this cost in the model, we distinguish between an enzyme's
  \co{fix the words!}  concentration $p(t)$ (concentration of enzyme
  molecules, appearing in the enzyme cost function) and its activity
  $u(t)$ (concentration of enzyme molecules in the active state, which
  appears in the rate laws).  To write the enzyme cost as a function of
  $u$ (instead of $p$), we follow a simple  logic. If a desired
  enzyme amplitude can be realised by protein expression alone, the 
  posttranslational regulation is not used, enzyme and protein curves are
  identical, and the enzyme cost is given by protein cost (setting
  $p= u$).  In contrast, if the desired enzyme amplitude is too large
  and hits the constraints, we introduce an auxiliary variable
  $q_l = \max(0,|\et_l|-\pt^{\max}(\pb,\omega))$, describing the
  difference between the desired \todo{activity\co{anderes wort: ``effective conc''?}}
  amplitude $\et_l$ and the
  maximal protein amplitude $\pt^{\rm max}_l$ (see Figure
  \ref{fig:amplitudeConstraint}). Then, we reformulate the
  optimisation problem with $\evb,\evt, \pvb,$ and $\qv$ (see SI
  \ref{sec:PosttranslationalCost}).  The resulting extra cost is
  proportional to the extra amplitude achieved by inhibition and may
  briefly be called ``cost of posttranslational inhibition''.

\item \textbf{Amplitude constraints and resulting Lagrange terms}
  Optimal enzyme rhythms under constraints are described by
  Eq.~(\ref{eq:OptimalityProblem}).  Active constraints are treated by
  Langrange multipliers, and instead of Eqs (\ref{eq:optstaticvector})
  and (\ref{eq:optperiodic}) we obtain (proof in SI
  \ref{sec:RhythmConstraintsLagrange})
\begin{eqnarray}
\label{eq:SolutionWithConstraints}
\devb &=& - \Fee\inv \, \left[\Fex \,\dxvb - \betav - \diag(\pvt^{\rm max}_{\rm rel}(\omega))\, \gammav \right] \nonumber \\
\evt &=& -\Fetet\inv\, \left[ 
\Fetxt \,\xvt -\diag\left(\frac{\evt}{|\evt|}\right))\, \gammav \right].
\end{eqnarray}
The signs of the Lagrange multipliers (in vectors $\betav$ for the
upper bounds on $\eb_{l}$ and $\gammav$ for the upper bounds on
$|\et_{l}|$) reflect different types of constraints: $\beta_l<0$ for
inactive enzymes (with $\eb_l=0$), $\beta_l>0$ for enzymes whose
average value hits an upper bound, and $\beta_l=0$
otherwise. Likewise, $\gamma_l>0$ holds for enzymes with active
amplitude constraints, and $\gamma_l=0$ otherwise.  If no constraints
are hit, all Lagrange multipliers vanish and we reobtain our solution
formulae (\ref{eq:optstaticvector}) and (\ref{eq:optperiodic}).
Otherwise, each active constraint yields a new term, and each of these
terms affects the profiles of all enzymes \todo{through the matrix
  multiplication}. If an enzyme hits its amplitude constraint, there
is a nonzero vector component in $\gammav$ that couples $\devb$ and
$\evt$. At high frequencies the extra term in the first equation goes
to zero and the coupling can be neglected. In the second equation, the
vector $\evt$ appears on both sides.  Thus, instead of computing
$\evt$ directly from the equation, a self-consistent solution needs to
be found. \co{wie? iterate?  would an iteration converge?}

\co{check in proof!! sind die Lagrangevektoren wirklich reell?  gelten die
  kuhn-tucker-bedingungen fuer komplexe groessen? ich glaube ja, weil
  man das problem ja auch mit reellen real-iund imaginaerteilen
  formulieren koennte und das die L.M. nicht betreffen sollte.}
\end{enumerate}

\co{
\section{Perspectives on optimal biological rhythms}
\label{sec:APPperspectives}
\co{ref in main text!}

\co{The style of optimal movements in cycles} \todo{Optimal enzyme
  rhythms and biological cycles  general, have a special
  ``style'', one  that also arises with other complex planning tasks and even
  in our daily experience. What   ``strategies''
  will be  optimal if a behaviour needs to  repeat itself   periodically?  We can understand
  these strategies in different ways: we may
   see them as a sequence of (discrete or continuous) actions, or
   we describe them as one ``flowing'' state that is modulated in time.
   Altogether, we obtain a number of
  complementary
  perpectives: some require thinking in time,
  others require thinking in terms of oscillation modes; some are
  based on discrete actions, some on continuous changes of variables. They also
  suggest  different description of possible control
  strategies.  

  \myparagraph{Perspective 1: A series of interwoven \co{oefters
      ``interwoven''?} actions} \co{WO? FN: The topics in this section
    do not concern science, but are meant as food for thoughts. They
    are about perspectives rather than about facts. However,
    discussing this here is important because control of
    high-dimensional, cyclic processes is complicated: it is difficult
    to get an intuition, and changing between different complementary
    perspectives may improve our understanding.} One perspective on
  rhythmic control considers discrete actions in time. In the cell, an
  action such as ``DNA replication'' involves some of the cell
  variables and happens in a specific phase of the cycle.  Different
  actions enable or require each other. For an analogy, we may think
  of a building \co{construction?} project (or of the assembly of a
  machine \co{ref like the bacterial flagella system}) that involves
  steps, where some steps are preconditions for others (e.g.~building
  the walls before adding a roof), others should be performed
  together, and some should not be performed at the same time.  In
  cells, additionally, some processes will lead to others
  automatically for physical reasons (e.g.~a reaction producing free
  radicals will lead to subsequent damaging reactions). In the case of
  cycles, all steps need to repeat themselves periodically: to close
  the cycle, the last process creates the right conditions for the
  first.  How can we put all processes in a meaningful order, such
  that all these \todo{conditions} are satisfied?  We could do that by
  thinking in terms of anticipation, and then closing the
  cycle. \co{FN: ref max-plus algebra; evtl das mal tatsaechlich
    verwenden, mit rainer? oder abfolge von przessen mit
    max-plus-leuten in paris? kann man kreise schliessen mit (einer erweiterten form der)
    max-plus algebra?} Or we start with a preliminary cyclic
  arrangement and then move processes in time until everything fits
  together. In reality, the problem is even more complicated, because
  actions have different durations and may overlap or be run simultaneously. \co{the next step
    would be to see the actions not as discrete phases, but as
    continuous, complex movements (like movements in taiji)}

  \co{ZU BIFURKATION: (bezieht sich auf alle punkte 1-4) sagen, dass
    das triggern manche prozesse zu manchen zeiten bevorzug und diese
    spaltung hervorbringt; aber dass sie auch von selbst entstehen
    kann; analogie magnetisierung von aussen; spontane magnetisierung
    (gute analogie, wg erniedrigung der energie und spontaner
    symmetriebrechung)}

  \myparagraph{Perspective 2: A \co{crystallised??}
    \todo{disintegrated} \co{WORT!} \co{eher sowas wie ``ausfaellen''}
    steady states} \todo{Thinking in cycles can help us understand
    steady states. In a cell, concentrations and fluxes are constantly
    affecting each other. However, in steady states, all variables are
    constant in time and these causal relations remain
    invisible. \co{Oscillations bring out causality nd control
      patterns that are less well visible in steady states.} If we see
    variables oscillate in time, causality becomes much more obvious:
    the temporal separation of processes shows clearly how processes
    are promoted by each other, and adapted to different external
    conditions.}  Thus, in our second perspective, we compare a
  rhythmic sequence in time (e.g.~subsequent peaks in molecule
  concentrations) to the sequence of cause and effect within steady
  states of the same system: a temporal cycle makes visible a cycle of
  causation and helps us understand steady \co{flux} states. In
  contrast to a building project (or the assembly of a machine), a
  production chain (e.g.~in a factory) runs continuously: since
  products are constantly produced, all production steps must be
  performed continuously and at the same time. Yet, if we trace an
  item long the production chain, the different steps are performed in
  a meaningful order (if process A is a precondition for B, then A
  must happen before B). If the same system runs in periodic cycles,
  this constant flow is ``broken'', but only to an extent,
  e.g.~certain steps become stronger in certain phases in time, and
  ``peaks of activity'' move similar like the products, along a
  sequence of preconditions. For example, if a causal chain consists
  of ribosomes producing proteins that catalyse a reaction, then in
  the corresponding cycle a peak in ribosome concentration would allow
  for a (later) peak in protein levels, which would then allow for a
  (later) peak in the flux.  Causal chains, which remain hidden within
  steady states, are broken into phases: this makes them better
  visible than in a continuous steady flow. \co{JA!  EDIT!  das kann
    aussehen wie just-in-time-production (oder nach-sich-ziehen), ist
    aber in wahrheit ein global optimales verhalten. damit ist der
    zyklus ein ``fenster'', durch das man abhaengigkeiten zwischen
    prozessen sehen kann. funktional oder mechanistisch koennte er aus
    dem stationaeren zustand entstehen durch eine fluktuation, die
    sich (mechanistisch oder funktional) selbst verstaerkt //
    splitting causality in steady states leads to time arrangement;
    reflection becomes more clearly visible!}  \co{DORT! Series
    expansion also holds for static optimisation.}  \co{JA, EDIT!
    Periodische Zyklen machen die verhältnisse im stationären zustand
    sichtbar!  Topologischer Zyklus -> stationärer Zustand: balance,
    stoff geht immer weiter -> Stoerung fuehrt zur propagation in Zeit
    (entsprechend elastizitaeten) .. auch rueckwaertswelle!
    geschlossener ring kann zu resonanz fuehren; oder spontaner
    osczillation?  die wellen folgen ungefaehr dem ablauf der prozesse
    (bzw den elastizitaeten); all das ist aehnlich bei oekonomischer
    logik; stabilisierung mit delay fuehrt zu oszillationen!

    vorgehensweise: understand steady states through cycles or cycles
    through steady states}
  
  \myparagraph{Perspective 3: An optimal usage of pairwise synergies}
  \co{bisschen teile aus anderem abschnitt hierher?}  Our third
  perpective focuses on the curves of individual control variables
  (e.g.~enzyme curves) and how these curves should be adjusted to each
  other and to the environment. We assume that each variable can be
  modulated in time, and ask how these modulations should be chosen to
  obtain beneficial synergies.  We already saw this logic above.
  \co{das in den text schieben oder hierlassen?}  Each variable in our
  models has its own amplitude and phase. For each pair of variables,
  there can be a synergies between the rhythms, characterised by a
  synergy strength \co{wort??} and a preferred phase difference.  The
  ``style'' of optimal rhythms requires that there are many (or
  stronger) positive synergy effects, and only few (or weaker)
  negative ones -- or more precisely, the sum of all synergy terms
  should be maximal.

  \myparagraph{Perpective 4: A reflection of metabolic dynamics, and
    possible resonance, in the controlling enzyme curves} \co{bisschen
    teile aus anderem abschnitt hierher?} In a fourth perspective, we
  consider the dynamic of a metabolic system and the system
  controlling it, and observe an analogy \co{FN about ``metonymic''
    (causal) vs ``metaphoric'' (reflection, promoting)?} relation
  between them.  If metabolism shows specific dynamics
  (e.g.~perturbations propagating like waves), its dynamics may be
  reflected in its own control systems (e.g.~in ``complementary''
  \co{uea?}  enzyme profiles that can be applied to evoke and exploit
  certain waves). We already saw this above. For example, if a
  metabolic system \todo{shows a dynamic resonance} \co{besser erklaeren}
  (like the repressilator
  or inducilator system), a control system may ``drive'' the system at
  the resonance frequency to improve its performance. Likewise, if
  metabolism shows a certain network structure (e.g.~containing linear
  pathways), this structure may be reflected in the regulation system
  (e.g.~in the presence of operons).  \co{eigentlich
    wiederspiegelungsprinzip, wie in text bschrieben // Dynamics and
    control strategies are entangled. Different understandings! Thus
    whatever an enzyme forward action can do dynamically, will imply
    the possibility of exploiting this for control, i.e.~create a
    backwards incentive. from each dynamic effect follows a possible
    strategy // ref to portrayal in text, which follows from this //
    ``adapt (not to situation, but) to the possibilities''}

  \myparagraph{Perspective 5: Achieving without effort} \co{ganzer
    absatz: sort and edit!}  Finally, systems under optimal periodic
  control tend to ``avoid unnatural effort'', an idea that is also
  advocated in Taoist philosophy \co{REF tao te king} and practiced in
  martial arts such as \todo{Taiji Chuan}. \co{Effortlessness requires
    smartness, taking advantage of opportunities, adapting to the
    affordances of the environment and of the moments, and
    anticipating future demands and and preparing for future
    opportunities, all of which makes such strategies more
    sustainable.} Such sustainable strategies are also put forward by
  the permaculture movement.  \todo{ which advocates an ecological
    thinking in economics}.  \co{ref buch von andreas} \co{mention
    synergy graph again} In all these cases, resource saving and
  sustainability require that outcomes are achieved with a minimal
  effort and considering, in every action, the preceding and following
  actions: in the ideal arrangement, all resources are used
  efficiently, material for each phase is prepared by previous
  processes, side products are reused in following processes, and
  little material remains unused. Specific principles related to such
  ``effortlessness'' are: (i) Separating different processes and
  alternating between them \co{(e.g. activity and repair; in yeast
    metabolic cycles, and in activity and recreating, see
    glucocorticoids during the day)}; (ii) Using synergies through
  coordinated or simultaneous actions; (iii) Performing actions that
  build on existing conditions (resulting from previous actions); (iv)
  Performing actions that create good conditions for following
  actions; (v) Performing actions not sequentially, but such that the
  end of one action is already the beginning of another one}.}

\end{appendix}

%
%
%

\end{document}